\documentclass[paper]{JHEP3}
\usepackage{epsfig}
\usepackage{amsmath}
\usepackage{amssymb}
\usepackage{amsbsy}
\usepackage{bbm}

\newcommand{\tmmathbf}[1]{\ensuremath{{\bf #1}}}
\newcommand{\tmtexttt}[1]{{\ttfamily{#1}}}

\newcommand{\tmop}[1]{\ensuremath{{\rm #1}}}

\newenvironment{enumeratenumeric}{\begin{enumerate}}{\end{enumerate}}
 
\def\beq{\begin{equation}}
\def\beqn{\begin{eqnarray}}
\def\eeq{\end{equation}}
\def\eeqn{\end{eqnarray}}
\def\abs#1{\left|#1\right|}

\def\FKSeq#1{eq.~({\bf FKS}.#1)}
\def\half{\frac{1}{2}}

\newcommand\FKS{Frixione, Kunszt and Signer}
\newcommand\CS{Catani and Seymour}

\newcommand\HERWIG{{\tt HERWIG}}

\newcommand\PYTHIA{{\tt PYTHIA}}
\newcommand\ARIADNE{{\tt ARIADNE}}

\newcommand\BASES{{\tt BASES}}
\newcommand\SPRING{{\tt SPRING}}

\def\lq{\left[} 
\def\rq{\right]} 
\def\rg{\right\}} 
\def\lg{\left\{} 
\def\({\left(} 
\def\){\right)}

\def\th{\theta}

\newcommand\sss{\mathchoice%
{\displaystyle}%
{\scriptstyle}%
{\scriptscriptstyle}%
{\scriptscriptstyle}%
}

\newcommand\nplus{\oplus}
\newcommand\nminus{\ominus}

\newcommand\splus{{\sss \nplus}}
\newcommand\sminus{{\sss \nminus}}

\newcommand\splusminus{{\mathchoice%
{\vplusminus\displaystyle}%
{\vplusminus\scriptstyle}%
{\vplusminus\scriptscriptstyle}%
{\vplusminus\scriptscriptstyle}%
}}

\newcommand\sminusplus{{\mathchoice%
{\vminusplus\displaystyle}%
{\vminusplus\scriptstyle}%
{\vminusplus\scriptscriptstyle}%
{\vminusplus\scriptscriptstyle}%
}}

\newdimen\hbigcirc
\newdimen\wbigcirc

\newcommand\vplusminus[1]{%
\settoheight{\hbigcirc}{$#1\bigcirc$}%
\settowidth{\wbigcirc}{$#1\bigcirc$}%
\makebox[\wbigcirc]{%
\makebox[0pt]{\rule[0.4\hbigcirc]{0.5\wbigcirc}{0.05\hbigcirc}}%
\makebox[0pt]{\rule[0.1\hbigcirc]{0.5\wbigcirc}{0.05\hbigcirc}}%
\makebox[0pt]{\rule[0.1\hbigcirc]{0.05\wbigcirc}{0.6\hbigcirc}}%
\makebox[0pt]{$#1\bigcirc$}}%
}

\newcommand\vminusplus[1]{%
\settoheight{\hbigcirc}{$#1\bigcirc$}%
\settowidth{\wbigcirc}{$#1\bigcirc$}%
\makebox[\wbigcirc]{%
\makebox[0pt]{\rule[0.2\hbigcirc]{0.5\wbigcirc}{0.05\hbigcirc}}%
\makebox[0pt]{\rule[0.5\hbigcirc]{0.5\wbigcirc}{0.05\hbigcirc}}%
\makebox[0pt]{\rule[-0.1\hbigcirc]{0.05\wbigcirc}{0.6\hbigcirc}}%
\makebox[0pt]{$#1\bigcirc$}}%
}

\newcommand\xplus{x_\splus}
\newcommand\xminus{x_\sminus}
\newcommand\xplusminus{x_\splusminus}
\newcommand\kplus{k_\splus}
\newcommand\kminus{k_\sminus}
\newcommand\Kplus{K_\splus}
\newcommand\Kminus{K_\sminus}
\newcommand\bxplus{\bar{x}_\splus}
\newcommand\bxminus{\bar{x}_\sminus}
\newcommand\bxplusminus{\bar{x}_\splusminus}

\newcommand\fmo{{f_{\splus}}}
\newcommand\fmop{{f_{\splus}'}}
\newcommand\fmt{{f_{\sminus}}}

\newcommand\ep{\epsilon}

\newcommand\as{\alpha_{\sss\rm S}}
\newcommand\asotpi{\frac{\as}{2\pi}}
\newcommand\Lum{{\cal L}}
\newcommand\Lumt{\tilde{\cal L}}

\newcommand\pt{p_{\sss\rm T}}
\newcommand\ptmin{{\pt^{\min}}}
\newcommand\kt{k_{\sss\rm T}}

\newcommand\tildept{\tilde{p}_{\sss\rm T}}
\newcommand\tildekt{\tilde{k}_{\sss\rm T}}
\newcommand\tildektmaxsq{{\tilde{k}_{\sss\rm T,\,max}^2}}
\newcommand\boost{\mathbb B}

\newcommand\matB{{\cal B}}
\newcommand\matR{{\cal R}}

\newcommand\matVb{{\cal V}_{\bare}}

\newcommand\matI{{\cal I}}
\newcommand\matSV{{\cal V}}

\newcommand\bare{{\rm b}}
\newcommand\matCb{{\cal G}_\bare}
\newcommand\matCpb{{\cal G}_{\splus,\bare}}
\newcommand\matCmb{{\cal G}_{\sminus,\bare}}
\newcommand\matCpmb{{\cal G}_{\splusminus,\bare}}
\newcommand\matCp{{\cal G}_\splus}
\newcommand\matCm{{\cal G}_\sminus }
\newcommand\matCpm{{\cal G}_{\splusminus}}

\newcommand\ctindex{\alpha}
\newcommand\ctindr{{\alpha_{\sss\rm r}}}
\newcommand\ctindpm{{\alpha_{\splusminus}}}
\newcommand\ctindp{{\alpha_{\splus}}}
\newcommand\ctindm{{\alpha_{\sminus}}}
\newcommand\matct{{\cal C}}
\newcommand\matcti{{\cal C}^{(\ctindex)}}
\newcommand\matctiP{\bar{\cal C}^{(\ctindex)}}

\newcommand\PSnpo{\Phi_{n+1}}
\newcommand\PSn{\Phi_n}

\newcommand\MapNLOi{{\bf M}^{(\ctindex)}}

\newcommand\Kinnpo{{\bf \Phi}_{n+1}}
\newcommand\Kinn{{\bf \Phi}_n}
\newcommand\BKinn{{\bf \bar{\Phi}}_n}
\newcommand\BKinm{{\bf \bar{\Phi}}_m}

\newcommand\Kinncp{{\bf \Phi}_{n,\splus}}
\newcommand\Kinncm{{\bf \Phi}_{n,\sminus}}
\newcommand\Kinncpm{{\bf \Phi}_{n,\splusminus}}

\newcommand\Kinmpo{{\bf \Phi}_{m+1}}

\newcommand\Kinnpot{{\bf \tilde{\Phi}}_{n+1}}

\newcommand\stepf{\theta}
\newcommand\Dfun{{\cal D}}
\newcommand\Sfun{{\cal S}}

\newcommand\Sz{{\cal S}}

\newcommand\Szi{{\cal S}_i}
\newcommand\Szip{{\cal S}_i^\splus}
\newcommand\Szim{{\cal S}_i^\sminus}
\newcommand\Szipm{{\cal S}_i^\splusminus}
\newcommand\Szimp{{\cal S}_i^\sminusplus}
\newcommand\Soij{{\cal S}_{ij}}
\newcommand\Soji{{\cal S}_{ji}}
\newcommand\mydot{\!\cdot\!}
\newcommand\xiic{\left(\frac{1}{\xi_i}\right)_{\!\!\xi_c}}
\newcommand\xic{\left(\frac{1}{\xi}\right)_{\!\!\xi_c}}

\newcommand\omzc{\left(\frac{1}{1-z}\right)_{\!\!\xi_c}}
\newcommand\lomzc{\left(\frac{\log(1-z)}{1-z}\right)_{\!\!\xi_c}}
\newcommand\omyid{\left(\frac{1}{1-y_i}\right)_{\!\!\deltaI}}
\newcommand\opyid{\left(\frac{1}{1+y_i}\right)_{\!\!\deltaI}}
\newcommand\ompyid{\left(\frac{1}{1\mp y_i}\right)_{\!\!\deltaI}}

\newcommand\omyijd{\left(\frac{1}{1-y_{ij}}\right)_{\!\!\deltaO}}
\newcommand\ompyd{\left(\frac{1}{1\mp y}\right)_{\!\!\delta}}

\newcommand\Mrec{M_{\rm rec}}

\newcommand\mmod[1]{\underline{#1}}
\newcommand\MCatNLO{{\tt MC@NLO}}

\newcommand\xicut{\xi_{c}}
\newcommand\deltaI{\delta_{\sss\rm I}}
\newcommand\deltaO{\delta_{\sss\rm O}}
\newcommand\MSB{{\rm \overline{MS}}}

\newcommand\CA{C_{\sss\rm A}}

\newcommand\CF{C_{\sss\rm F}}
\newcommand\TF{T_{\sss\rm F}}
\newcommand\NC{N_{\rm c}}
\newcommand\NCt{N_{\rm c}^2}
\newcommand\NF{n_{\rm f}}

\newcommand\APreg{\hat{P}}

\newcommand\fb{{f_b}}

\newcommand\POWHEG{{POWHEG}}

\newcommand\Rad{\Phi_{\rm rad}}
\newcommand\Xrad{X_{\rm rad}}

\newcommand\muF{\mu_{\sss\rm F}}

\newcommand\frindsing{{\ctindr}}

\newcommand\qb{\bar{q}}

\def\ord#1{{\cal O}\(#1\)}

\newcommand\xpm{{x_{\splus\sminus}}}

\newcommand\vplus{{\tilde{v}_\splus}}
\newcommand\vminus{{\tilde{v}_\sminus}}
\newcommand\vpm{{\tilde{v}_\splusminus}}

\preprint{
Bicocca-FT-07-9\hfill\\
 GEF--TH--21/2007}
\title{Matching NLO QCD computations
with Parton Shower simulations: the \POWHEG{} method}
\author{Stefano Frixione\\
  INFN, Sezione di Genova,
  Via Dodecaneso 33, 16146 Genova, Italy\\
  E-mail: \email{Stefano.Frixione@cern.ch}}
\author{Paolo Nason\\
  INFN, Sezione di Milano-Bicocca,
  Piazza della Scienza 3, 20126 Milan, Italy\\
  E-mail: \email{Paolo.Nason@mib.infn.it}}
\author{Carlo Oleari\\
  Universit\`a di Milano-Bicocca and INFN, Sezione di Milano-Bicocca\\
  Piazza della Scienza 3, 20126 Milan, Italy\\
  E-mail: \email{Carlo.Oleari@mib.infn.it}}
\abstract{
The aim of this work is to describe in detail the \POWHEG{} method,
first suggested 
by one of the authors,
for interfacing parton-shower generators with NLO QCD computations.
We describe the method in its full generality, and then specify
its features in two subtraction frameworks for NLO calculations:
the Catani-Seymour and the Frixione-Kunszt-Signer
approach. Two examples are discussed in detail in both approaches:
the production of hadrons in $e^+e^-$ collisions,
and the Drell-Yan vector-boson production in hadronic collisions.
}
\keywords{QCD, Monte Carlo, NLO Computations, Resummation, Collider Physics}

\renewcommand\arraystretch{1.1}

\begin{document}

\section{Introduction}

In the past two decades, next-to-leading order (NLO) QCD computations have
become standard tools for phenomenological studies at lepton and hadron
colliders. QCD tests have been mainly performed by comparing NLO results with
experimental measurements, with the latter corrected for detector effects.

On the experimental side, leading order (LO) calculations, implemented in the
context of general purpose Shower Monte Carlo (SMC) programs, have been the
main tools used in the analysis. SMC programs include dominant QCD effects at
the leading logarithmic level, but do not enforce NLO accuracy. These
programs were routinely used to simulate background processes and signals in
physics searches. When a precision measurement was needed, to be compared
with an NLO calculation, one could not directly compare the experimental
output with the SMC output, since the SMC does not have the required
accuracy. The SMC output was used in this case to correct the measurement for
detector effects, and the corrected result was compared to the NLO
calculation.

In view of the positive experience with QCD tests at the NLO level, it has become
clear that SMC programs should be improved, when possible, with NLO
results. In this way a large amount of the acquired knowledge on QCD
corrections would be made directly available to the experimentalists in a
flexible form that they could easily use for simulations.

The problem of merging NLO calculations with parton shower simulations is
basically that of avoiding overcounting, since the SMC programs do implement
approximate NLO corrections already. Several proposals have appeared in the
literature~\cite{Dobbs:2001dq,Frixione:2002ik,Kurihara:2002ne,Nason:2004rx}
that can be applied to both
$e^+ e^-$ and hadronic collisions, and two
approaches~\cite{Nagy:2005aa,Kramer:2005hw} suitable for $e^+e^-$ annihilation.
Furthermore, proposals for new shower algorithms, that should be better
suited for merging with NLO results, have appeared in the literature
(see refs.~\cite{Bauer:2006mk,Bauer:2007ad,Nagy:2007ty,Giele:2007di,%
Dinsdale:2007mf,Schumann:2007mg}).

The \MCatNLO{}
proposal~\cite{Frixione:2002ik} was the first one to give an acceptable
solution to the overcounting problem.  The generality of the method has been
explicitly proven by its application to processes of increasing complexity,
such as heavy-flavour-pair~\cite{Frixione:2003ei} and
single-top~\cite{Frixione:2005vw} production.\footnote{ A complete list of
processes implemented in \MCatNLO{} can be found at\\\centerline{\tt
http://www.hep.phy.cam.ac.uk/theory/webber/MCatNLO.}}
  The basic idea of
\MCatNLO{} is that of avoiding the overcounting by subtracting from the exact
NLO cross section its approximation, as implemented in the SMC
program to which the NLO computation is matched. Such approximated cross
section (which is the sum of what have been denoted in \cite{Frixione:2002ik}
as MC subtraction terms) is
computed analytically, and is SMC dependent. On the other hand, the MC
subtraction terms are process-independent, and thus, for a given SMC, can be
computed once and for all. In the current version of the \MCatNLO{} code, the
MC subtraction terms have been computed for \HERWIG~\cite{Corcella:2000bw}.
In general, the exact NLO cross section minus the MC subtraction terms does
not need to be positive.  Therefore \MCatNLO{} can generate events with
negative weights. For the processes implemented so far, negative-weighted
events are about 10--15\% of the total. Their presence does not imply a
negative cross section, since at the end physical distributions must turn out
to be positive.

The features implemented in \MCatNLO\ can be summarized as follows:
\begin{itemize}
\item[-]
Infrared-safe observables have NLO accuracy.
\item[-]
Collinear emissions
are summed at the leading-logarithmic level.
\item[-]
The double
logarithmic region (i.e.\ soft and collinear gluon emission)
is treated correctly if the SMC code used for showering has
this capability.
\end{itemize}
In the case of \HERWIG{} this last requirement is satisfied, owing to the fact
that its shower is based upon an angular-ordered branching.

In ref.~\cite{Nason:2004rx} a method, to be called \POWHEG\ in the following
(for Positive Weight Hardest Emission Generator), was proposed that overcomes
the problem of negative weighted events, and that is not SMC specific. In the
\POWHEG{} method the hardest radiation is generated first, with a technique
that yields only positive-weighted events using the exact NLO matrix
elements.  The \POWHEG{} output can then be interfaced to any SMC program
that is either $\pt$-ordered, or allows the implementation of a $\pt$
veto.\footnote{All SMC programs compatible with the {\em Les Houches Interface
for User Processes}~\cite{Boos:2001cv} should comply with this requirement.}
However, when interfacing \POWHEG{} to angular-ordered SMC programs, the
double-log accuracy of the SMC is not sufficient to guarantee the double-log
accuracy of the whole result.  Some extra soft radiation (technically called
vetoed-truncated shower in ref.~\cite{Nason:2004rx}) must also be included in
order to recover double-log accuracy. In fact, angular ordered SMC programs
may generate soft radiation before generating the radiation with the largest
$\pt$, while \POWHEG{} generates it first. When \POWHEG{} is interfaced to
shower programs that use transverse-momentum ordering, the double-log
accuracy should be correctly retained if the SMC is double-log accurate. The
\ARIADNE{} program~\cite{Lonnblad:1992tz} and
\PYTHIA{}~6.4~\cite{Sjostrand:2006za} (when used with the new showering
formalism), both adopt transverse-momentum ordering, in the framework of
dipole-shower algorithm
\cite{Gustafson:1987rq,Pettersson:1988zu,Lonnblad:1988nh}, and aim to have
accurate soft resummation approaches, at least in the large $N_c$ limit
(where $N_c$ is the number of colours).

A proof of concept for the \POWHEG{} method has been given in
ref.~\cite{Nason:2006hf}, for $ZZ$ production in hadronic
collisions. In ref.~\cite{Frixione:2007nw} the method was also applied to $Q\bar{Q}$ hadroproduction.
Detailed comparisons have been carried out
between the \POWHEG{} and \MCatNLO{} results, and reasonable agreement has been found,
which nicely confirms the validity of both approaches. In
ref.~\cite{Latunde-Dada:2006gx} the \POWHEG{} method, interfaced to the
\HERWIG{} Monte Carlo, has been applied to $e^+e^-$ annihilation, and compared
to LEP data. The method yields better fits compared to \HERWIG{} with
matrix-element corrections. The authors of ref.~\cite{Latunde-Dada:2006gx}
have also provided an estimate of the effects of the
truncated shower, which turned out to be small.

In the present work we give a detailed description of the \POWHEG{}
method. Our aim is to provide all of the necessary formulae and procedures
for its application to general NLO calculations.  We first formulate
\POWHEG{} in a general subtraction scheme.  Then, we illustrate it in detail
in two such schemes: the Frixione, Kunszt and Signer
(FKS)~\cite{Frixione:1995ms,Frixione:1997np} and the Catani and Seymour
(CS)~\cite{Catani:1996vz} one.  The CS method has been widely used in the
literature. On the other hand, the FKS method has already been used
extensively in the \MCatNLO{} implementations.  Furthermore, the NLO cross
sections for vector-boson and heavy-quark pair production used in the
\POWHEG{} implementations of refs.~\cite{Nason:2006hf,Frixione:2007nw} have a
treatment of initial-state radiation that is essentially the same one used in
FKS.

Our paper is organized as follows. In section~\ref{sec:nlo} we summarize the
general features of the NLO computations and of the subtraction formalisms.
In section~\ref{sec:FKSsub} all the details of the FKS subtraction method are
given, and in section~\ref{sec:dipsub} the basic features of the CS approach
are summarized.

In section~\ref{sec:mcsim} a general discussion of the inclusion of NLO
corrections in a parton shower framework is given, together with a basic
introduction of the \POWHEG{} method.

In section~\ref{sec:powhegmth} we go through all the details of the \POWHEG{}
method.  The method is presented in general, and it is shown how to apply it
within any subtraction framework. Thus, this section does not refer in
particular to either the FKS or the CS method.

In section~\ref{sec:sudak} we discuss the accuracy of the \POWHEG{} approach
in the resummation of soft-gluons effects. We show that, with an appropriate
prescription for the evaluation of the running coupling used in \POWHEG{} for
the generation of radiation, one can easily obtain next-to-leading
logarithmic (NLL) accuracy in soft-gluon radiation, provided the process in
question has no more than three incoming or outgoing coloured partons at the
Born level.  If the Born process involves more than three coloured partons,
there are left-over soft terms that are not correctly represented by \POWHEG.
We show, however, that with a modest modification of the algorithm one can also
correctly resum these contributions at the level of the leading terms in a
large $N_c$ expansion.

In section~\ref{sec:powhegfks} the formulae needed for the explicit
construction of a \POWHEG{} in the FKS framework is given.  The same is done
in section~\ref{sec:powhegcs} for the CS method.

Finally, in section~\ref{sec:examples} two simple examples are discussed in
both the FKS and the CS framework, namely the production of hadrons in
$e^+e^-$ annihilation, and the production of a massive vector boson (or a
virtual photon) in hadronic collisions.

This paper is considerably long, and it involves many technical details.
The length is partly due to the fact that we deal with two subtraction methods.
The reader may skip the one she/he is not interested in. The example sections
are particularly long and pedantic. The reader may not be interested in
reading all of them. Section~\ref{sec:sudak} is technically complex,
but it may be almost completely skipped on a first reading.

\section{NLO computations}\label{sec:nlo}

\subsection{Generalities}\label{sec:NLOgen}
In this section, we describe the general features of an NLO calculation for
a generic hadron-hadron collision process. In lepton-hadron and lepton-lepton
collisions, the treatment is similar, but simpler. For example, in the case
of lepton-lepton collisions, the parton-distribution functions for the
incoming particles are replaced by delta functions.

We consider $2\to n$ processes, where the momenta of the particles
satisfy the momentum conservation
\begin{equation}
\xplus \Kplus+\xminus \Kminus = k_1+\ldots +k_n\,,
\label{eq:momconsn}
\end{equation}
where $x_{\splusminus}$ are the momentum fractions of the incoming partons,
and $K_{\splusminus}$ the
four-momenta of the incoming hadrons. In what follows, we also use the
notation
\begin{equation}
\kplus=\xplus \Kplus\,,\qquad\qquad
\kminus =\xminus \Kminus\,,
\end{equation}
to denote the momenta of the incoming partons. We define, as usual,
\begin{equation}
\label{eq:S_and_s}
S=(\Kplus+\Kminus)^2\,,\quad\quad s=(\kplus+\kminus)^2\,.
\end{equation}
We denote by $\Kinn$ the set of variables
\begin{equation}
\Kinn=\lg \xplus ,\xminus ,k_1,\ldots,k_n \rg
\end{equation}
constrained by momentum conservation (eq.~(\ref{eq:momconsn})), and by
the on-shell conditions for final-state particles.  We collectively
denote by $\matB$ the squared matrix elements\footnote{ We always assume spin
and colour sums and averages when needed, and the inclusion of the
appropriate flux factor.}  relevant to the LO contributions to our
process. The total cross section at leading order is given by
\begin{equation}
\label{eq:sigtotlo}
\sigma_{\sss\rm LO} =
\int d\Kinn\,\Lum \; \matB\!\(\Kinn\)
\end{equation}
where $\Lum$ is the parton luminosity\footnote{In this section we drop
the parton flavours and the scale dependence in the
luminosity, for ease of notation.}
\begin{equation}
\Lum=\Lum(\xplus ,\xminus )=f_\splus(\xplus ) \, f_\sminus (\xminus )\,,
\label{eq:Lum1}
\end{equation}
 and
\begin{equation}
d\Kinn=d\xplus \;d\xminus \;d\PSn\(\kplus+\kminus ; k_1,\ldots,k_n\)\,,
\label{eq:PSn}
\end{equation}
with $d\PSn$ the $n$-body phase space
\begin{equation}
d\PSn\(q;k_1,\ldots,k_n\)=(2\pi)^4\,\delta^4\!\(q-\sum_{i=1}^n k_i\)
\,\prod_{i=1}^n \frac{d^{3} k_i}{\(2\pi\)^{3} 2k^0_i }\,.
\end{equation}
In case of leptons in the initial state, the corresponding parton
distribution function $f(x)$ in eq.~(\ref{eq:Lum1}) is replaced by
$\delta(1-x)$.

The real contributions at the NLO arise from the
tree-level squared amplitudes for the $2 \to n+1$ parton process,
which we denote by $\matR$.  As before,
we denote by $\Kinnpo$ the corresponding set of variables
\begin{equation}
\Kinnpo=\lg \xplus ,\xminus ,k_1,\ldots,k_{n+1} \rg
\end{equation}
constrained by momentum conservation and on-shell conditions.

The virtual contributions arise from the interference of the one-loop
amplitudes times the LO amplitudes. We denote by $\matVb$ the renormalized
virtual corrections, that is, we assume that all ultraviolet divergences have
already been removed by renormalization.
These terms still contain infrared divergences.
Therefore, they are computed in $d=4-2\ep$ dimensions, and
the divergences appear as $1/\ep^2$ and $1/\ep$ poles.  The subscript $\bare$
(for ``bare'') reminds us of the presence of infrared divergences in the
amplitude.

In hadronic collisions, the complete cancellation of the initial-state
collinear singularities is achieved
by adding two counterterms, one for each of the incoming partons ($\nplus$,
$\nminus$), to the differential cross section. We denote them by
$\matCpb$ and $\matCmb$. The factorization counterterms are
infrared divergent in four dimensions. Therefore, they are computed in
$d=4-2\ep$ dimensions, and the divergences appear as $1/\ep$ poles. To remind
this fact, also in this case a subscript $\bare$ has been included in the
notation.

The total NLO cross section is given by\footnote{The $\matCpmb$ terms are
present only for incoming hadrons. If one or both the incoming particles are
leptons, the corresponding $\matCb$ is zero.}
\begin{eqnarray}
\label{eq:sigtot}
\sigma_{\sss\rm NLO}
 &=&\int d\Kinn\,\Lum \; \Big[\matB\!\(\Kinn\)+\matVb\!\(\Kinn\) \Big]
 + \int d\Kinnpo\,\Lum\;\matR\!\(\Kinnpo\)
\nonumber \\
&+& 
\int d\Kinncp\,\Lum\;\matCpb\!\(\Kinncp\) + 
\int d\Kinncm\,\Lum\;\matCmb\!\(\Kinncm\)\,,
\end{eqnarray}
where
\begin{equation}
d\Kinnpo=d\xplus \;d\xminus \;d\PSnpo\(\kplus+\kminus ; k_1,\ldots,k_{n+1}\)\;.
\label{eq:PSnpo}
\end{equation}
$\Kinncpm$ denotes configurations in which one of
the final-state partons is collinear to one of the incoming partons. Thus,
such configurations are effectively $n$-body final-state ones, except for the
energy degree of freedom of the parton collinear to the beam. We then write
\begin{eqnarray}
\Kinncp & = & \lg \xplus ,\xminus ,z,k_1,\ldots,k_n \rg\;,\quad
 z\,\xplus \Kplus+\xminus \Kminus  = \sum_{i=1}^n k_i\;,
\label{eq:procmomcp}
\\
\Kinncm & = & \lg \xplus ,\xminus ,z,k_1,\ldots,k_n \rg\;,\quad
 \xplus \Kplus+z\,\xminus \Kminus  = \sum_{i=1}^n k_i\;,
\label{eq:procmomcm}
\end{eqnarray}
where $z$ is the fraction of momentum of the incoming parton after radiation.
We can associate with the phase-space configuration $\Kinncpm$ an
{\em underlying $n$-body configuration} $\BKinn$ defined as
\begin{equation} \label{eq:barfromcpm}
\BKinn=\{\bar{x}_\splus,\bar{x}_\sminus,k_1,\ldots k_n\}\,,\qquad
 \bar{x}_\splusminus=zx_\splusminus\,,
\qquad\bar{x}_\sminusplus=x_\sminusplus\;.  
\end{equation}
Thus, the values of $\bar{x}_\splusminus$ in the underlying $n$-body
configuration are constrained by momentum conservation, and do not depend
upon $z$.  We also define
\begin{eqnarray}
d\Kinncp&=&d\xplus \;d\xminus \;dz\;
d\PSn\(z\,\kplus+\kminus ;k_1,\ldots,k_n\)\,,
\label{eq:PSqn1}
\\
d\Kinncm&=&d\xplus \;d\xminus \;dz\;
d\PSn\(\kplus+z\,\kminus ;k_1,\ldots,k_n\)\,.
\label{eq:PSqn2}
\end{eqnarray}

We now consider a generic observable $O$, function of the final-state
momenta. $O$ could be, for example,
a product of theta functions describing a
particular histogram bin for the distribution of some kinematic observable.
Its expected value is given by
\begin{eqnarray}
\label{eq:dsdO}
\langle O \rangle 
&=&
\int d\Kinn\,\Lum \;
O_n\!\(\Kinn\)\,\Big[\matB\!\(\Kinn\)+\matVb\!\(\Kinn\)\Big]
\nonumber \\*&+&
\int d\Kinnpo\,\Lum\; O_{n+1}\!\(\Kinnpo\)\,\matR\!\(\Kinnpo\)
\nonumber \\*&+&
\int d\Kinncp\,\Lum\;O_n\!\(\BKinn\)\,\matCpb\!\(\Kinncp\) + 
\int d\Kinncm\,\Lum\;O_n\!\(\BKinn\)\,\matCmb\!\(\Kinncm\),~
\end{eqnarray}
where $O_n$ and $O_{n+1}$ are the expressions of the observable $O$ in terms
of $n$ and $(n+1)$ final-state particle momenta, and $\Kinn$, in the
$\Kinncpm$ integrals, is the corresponding underlying $n$-body configuration.
We require that $O$ is an infrared-safe observable, and, furthermore,
we require that the Born contribution in eq.~(\ref{eq:dsdO})
(i.e.\ the term proportional to $\matB$) is infrared finite (thus, for example,
if our $n$-body process corresponds to $Z+{\rm jet}$ production, the observable
$O_n$ must suppress the region where the jet is emitted at low transverse
momentum).
Under these assumptions,
the real matrix elements contribution (i.e.\ the term proportional to $\matR$)
is finite in the whole phase space
$d\PSnpo$, except for the regions that correspond to soft and collinear
emissions. There, the divergences are integrable only in $d$ dimensions, and
yield $1/\ep^2$ and $1/\ep$ poles. Furthermore
the divergences of each term on the r.h.s. of eq.~(\ref{eq:dsdO}) cancel
in the sum, and the total cross section is finite.
Observe that the argument of $O$ in the last
two terms on the right hand side of eq.~(\ref{eq:dsdO}) is set equal to
$\BKinn$ rather than $\Kinncpm$, owing to the fact that $O$ is an infrared-safe
observable.
The integrals in eq.~(\ref{eq:dsdO}) are
usually too difficult to be performed analytically (because of the involved
functional form of $O$) and, being divergent, they are not suited for
numerical computations. For these reasons, different strategies have been
proposed for the computation of observables in QCD. One of the most
successful is the so-called subtraction method, pioneered in
ref.~\cite{Ellis:1980wv}, which we discuss in the next section.

\subsection{Subtraction formalism}
\label{sec:SubtractionFormalism}
The subtraction formalism requires the definition of a set of functions
$\matcti$, called real counterterms. Each $\ctindex$ is associated with a
particular singular region, i.e.\ with a configuration that has either a
final-state parton with four momentum equal to zero, or a final-state
massless parton with momentum proportional to an initial-state or to another
final-state massless parton.  Furthermore, for each $\ctindex$, a
mapping\footnote{In some approaches, the counterterms are different from zero
only in a finite neighborhood of the corresponding singular regions. In these
cases, the mapping needs to be defined only there.} $\MapNLOi$
\begin{equation}
\Kinnpot^{(\ctindex)}=\MapNLOi\left(\Kinnpo\right)
\label{eq:mapsNLO}\;,\quad
\Kinnpot^{(\ctindex)}= \lg \tilde{x}_\splus^{(\ctindex)},
\tilde{x}_\sminus^{(\ctindex)} ,
\tilde{k}_1^{(\ctindex)},\ldots,\tilde{k}_{n+1}^{(\ctindex)}
 \rg
\end{equation}
is defined that maps the $(n+1)$-body configuration into a singular
one.

The real counterterms and the mapping have the following property:
for any infrared-safe observable $O_{n+1}(\Kinnpo)$, that vanishes fast enough
if $\Kinnpo$ approaches
two singular regions at the same time, the function
\begin{equation}
\matR(\Kinnpo)\,O_{n+1}(\Kinnpo)-\sum_\ctindex\matcti(\Kinnpo)\,O_{n+1}\(\MapNLOi\left(\Kinnpo\right)\)
\label{eq:sub0}
\end{equation}
has at most integrable singularities
in the $\Kinnpo$ space. Observe that the above condition
does not always imply that
\begin{equation}
\matR(\Kinnpo)-\sum_\ctindex\matcti(\Kinnpo)
\label{eq:sub1}
\end{equation}
is also integrable. This is the case if the corresponding $n$-body
process has no singularities, like, for example, in $Z$ production in
hadronic collisions.

Each singular region
$\ctindex$ is characterized by a different mapping,
and, for this reason, we use the superscript $\ctindex$ on the
tilded variables.  For ease of notation, we use the following {\em context
convention}: if an expression is enclosed in the subscripted squared brackets
\begin{equation}
\label{eq:context}
\big[\ldots\big]_{\ctindex}\,,
\end{equation}
we mean that all variables appearing inside have, when applicable, the
superscripts corresponding to the subscript of the bracket.  Thus we write
\begin{equation}
\Kinnpot^{(\ctindex)}= \lq \lg \tilde{x}_\splus,
\tilde{x}_\sminus ,
\tilde{k}_1,\ldots,\tilde{k}_{n+1} \rg \rq_\ctindex\,.
\end{equation}
The form of the singular configurations $\Kinnpot^{(\ctindex)}$ differs
according to the nature of the singular region.  More specifically:
\begin{itemize}
\item If $\ctindex$ is associated with a soft (S) region,
the singular configuration has a final-state parton with null
four-momentum.
\item If $\ctindex$ is associated with a final-state collinear
singularity (FSC),
the singular configuration has two massless final-state partons
with parallel three-momenta.
\item If $\ctindex$ is associated with an initial-state collinear
(ISC) singularity,
the singular configuration has a massless outgoing parton with three-momentum
parallel to the momentum of one incoming parton.
\end{itemize}
The mapping~(\ref{eq:mapsNLO}) must be smooth near the singular region, and
it must become the identity there.  In other words if, for instance,
$\ctindex$ is associated with the FSC region where the particles $i$ and $j$
become collinear, we must have $\Kinnpot^{(\ctindex)}=\Kinnpo$ for
$\vec{k}_i\parallel \vec{k}_j$.  Notice that also the $x$'s of the
configurations $\Kinnpot^{(\ctindex)}$ do not necessarily coincide with
$\xplus $ and $\xminus $ for all $(n+1)$-body configurations $\Kinnpo$.  On
the other hand, they do coincide in the singular limit.

As in the case of the collinear configurations
(eqs.~(\ref{eq:procmomcp}) and (\ref{eq:procmomcm})),
we associate with each $\Kinnpot^{(\ctindex)}$ configuration
an $n$-body configuration $\BKinn^{(\ctindex)}$, that we will call
the \emph{underlying $n$-body configuration}
\begin{equation}
\BKinn^{(\ctindex)}= \lq \lg \bar{x}_\splus,
\bar{x}_\sminus ,
\bar{k}_1,\ldots,\bar{k}_n
 \rg \rq_\ctindex \;.
\end{equation}
$\BKinn^{(\ctindex)}$ is obtained as follows:
\begin{itemize}
\item
If $\alpha \in  {\rm S}$ (i.e.\ it is a soft region), $\BKinn^{(\ctindex)}$
is obtained by deleting the zero momentum parton.
\item
If $\alpha \in  {\rm FSC}$ (i.e.\ it is a final-state collinear region),  $\BKinn^{(\ctindex)}$
is obtained by replacing the momenta of the two collinear partons with their sum.
\item
If $\alpha \in  {\rm ISC}$ (i.e.\ it is an initial state collinear region),  $\BKinn^{(\ctindex)}$
is obtained by deleting the radiated collinear parton, and by replacing the momentum fraction of the
initial-state radiating parton with its momentum fraction after radiation.
\end{itemize}
In all the above cases, the final-state momenta are relabelled with an index that
takes values in the range $1,\ldots,n$.
Observe that, as a consequence of the procedure itemized above,
the variables in $\BKinn^{(\ctindex)}$ are constrained by momentum conservation
\begin{equation}
\bar{x}_\splus \Kplus + \bar{x}_\sminus \Kminus = \sum_{j=1}^n \bar{k}_j\;.
\end{equation}
Furthermore, for S or FSC regions, we have
\begin{equation}
\bar{x}_{\splusminus} = \tilde{x}_{\splusminus}\,.
\end{equation}
This does not hold for ISC regions: in the $\nplus$ direction, for example,
we have
\begin{equation}
\bar{x}_\splus<\tilde{x}_\splus\;,\qquad \bar{x}_\sminus=\tilde{x}_\sminus\;,
\end{equation}
and the analogous one for the case of ISC in the $\nminus$ direction.

We stress that the difference in our notation between the
$\Kinnpot^{(\ctindex)}$ and $\BKinn^{(\ctindex)}$ is a minor one: the
former has an unresolved parton, while in the latter all partons are
resolved. On the other hand, it is necessary to introduce
$\BKinn^{(\ctindex)}$ (together with the concept of underlying
$n$-body configuration), since it is formally the argument of Born-like matrix
elements, and will play a central role in the development of the \POWHEG{}
formalism.

In the subtraction method one rewrites the contribution to any observable $O$
coming from real radiation in the following way
\begin{eqnarray}
&&\int d\Kinnpo\,
\Lum\,O_{n+1}\!\(\Kinnpo\)\,\matR\!\(\Kinnpo\)
= \sum_\ctindex \int d\Kinnpo\, \lq \Lumt\,
 O_n\!\(\BKinn\)\,\matct\!\(\Kinnpo\)\rq_\ctindex \;+
\label{eq:sub2}
\nonumber\\ 
&&\hspace{2cm}
\int d\Kinnpo\,\Big\{
\Lum\,O_{n+1}\!\(\Kinnpo\)\,\matR\!\(\Kinnpo\) -
\sum_\ctindex \lq \Lumt\,
O_n\!\(\BKinn\)\,\matct\!\(\Kinnpo\)\rq_\ctindex \Big\},
\phantom{aaa}
\end{eqnarray}
where $\Lumt=\Lum(\tilde{x}_\splus,\tilde{x}_\sminus )$.  In this way, under
the assumptions we have made about the counterterms, and the assumption that
$O$ is an infrared-safe observable, the second term on the r.h.s.\ of
eq.~(\ref{eq:sub2}) is integrable in $d=4$ dimensions.

The first term on the r.h.s.\ of eq.~(\ref{eq:sub2}) is divergent. In order
to deal with it, we introduce, for each $\ctindex$, the $(n+1)$-phase space
parametrization
\begin{equation}
\Kinnpo \; \stackrel{(\ctindex)}{\Longleftrightarrow} \;
 \lg \BKinn^{(\ctindex)}, \Rad^{(\ctindex)} \rg\,,
\label{eq:phspfact}
\end{equation}
and the corresponding phase-space element
\begin{equation}
d\Kinnpo = d\BKinn^{(\ctindex)}\,d \Rad^{(\ctindex)}\;.
\label{eq:emfacphspe0}
\end{equation}
In words, we parametrize the $(n+1)$-phase space in terms of an $n$-body
phase space (obtained as described earlier), plus (three) more variables that
describe the radiation process.  The left-right arrow in
eq.~(\ref{eq:phspfact}) indicates that the correspondence is one to
one.\footnote{The correspondence needs only to be defined where the
corresponding counterterm is non-vanishing.}  The range of the radiation
variables in $\Rad^{(\ctindex)}$ may depend upon $\BKinn^{(\ctindex)}$.
Furthermore, eq.~(\ref{eq:emfacphspe0}) implicitly defines a Jacobian,
possibly dependent upon $\BKinn^{(\ctindex)}$, that we conventionally include into
$d \Rad^{(\ctindex)}$.  We call the parametrization~(\ref{eq:phspfact}) the
{\em emission factorization}.

We now distinguish two cases: the FSC+S case and the ISC one.  In the former
case we have
\begin{equation}
 \Lumt=\Lum(\tilde{x}_\splus,\tilde{x}_\sminus
)=\Lum(\bar{x}_\splus,\bar{x}_\sminus )\,.
\end{equation}
Defining
\begin{equation}\label{eq:cdef1}
\lq \bar{\cal C}\(\BKinn\)= \int d \Rad\, \matct\!\(\Kinnpo\)
\rq_{\ctindex\in\lg {\rm\sss FSC, S}\rg},
\end{equation}
we can write the generic
term in the first sum on the r.h.s.\ of eq.~(\ref{eq:sub2}) as follows
\begin{equation}
\lq \int d\Kinnpo\, \Lumt\,
 O_n\!\(\BKinn\)\,\matct\!\(\Kinnpo\)
=\int d\BKinn \,\Lumt\, O_n\!\(\BKinn\) \bar{\cal C}\(\BKinn\)
\rq_{\ctindex\in\lg {\rm\sss FSC, S}\rg}.
\end{equation}
In the ISC case, we cannot factor out the luminosity so easily, since
$\tilde{x}_{\splusminus} \ne \bar{x}_{\splusminus}$.
We define
\begin{equation}\label{eq:cdef2}
\lq \bar{\cal C}\(\BKinn,z\)= \int d \Rad\, \matct\!\(\Kinnpo\)\;
z\, \delta\(z-\bar{x}_{\splusminus}/\tilde{x}_{\splusminus}\)
\rq_{\ctindex\in\lg {\sss\rm ISC}_\splusminus\rg},
\end{equation}
which formally introduces the momentum fraction $z$,
and write
\begin{equation}\label{eq:ccontrib}
\lq \int d\Kinnpo\, \Lumt\,
 O_n\!\(\BKinn\)\,\matct\!\(\Kinnpo\)
=\int d\BKinn \frac{dz}{z} \Lumt\, O_n\!\(\BKinn\) \bar{\cal C}\(\BKinn,z\)
\rq_{\ctindex\in\lg {\sss\rm ISC}\rg}.
\end{equation}
Notice that, owing to the delta function in eq.~(\ref{eq:cdef2})
we have
\begin{equation}
\Lumt=\Lum(\tilde{x}_\splus,\tilde{x}_\sminus)=
\left\{\begin{array}{l} \Lum(\bar{x}_\splus/z,\bar{x}_\sminus)
\quad\mbox{for}\quad\alpha\in{{\rm ISC}_\splus} \\
 \Lum(\bar{x}_\splus,\bar{x}_\sminus/z)
\quad\mbox{for}\quad\alpha\in{{\rm ISC}_\sminus}\end{array}\right.\;\;.
\end{equation}
We also notice that the variables
$\lg \tilde{x}_\splus,\tilde{x}_\sminus, z,\bar{k}_1,\ldots,
\bar{k}_n \rg$ in the ISC regions can be identified with the
$\Kinncpm$ variables in eqs.~(\ref{eq:procmomcp}) and (\ref{eq:procmomcm}).
In fact, the $\bar{k}_i$ are integration variables, and can be identify with
the $k_i$'s in eqs.~(\ref{eq:procmomcp}) and (\ref{eq:procmomcm}).
Furthermore, the $\tilde{x}_\splusminus$ variables are identical
to the $x_\splusminus$  in eqs.~(\ref{eq:procmomcp}) and (\ref{eq:procmomcm}),
since those equations refer to a singular configuration, and (as we have remarked
earlier) the mapping of eq.~(\ref{eq:mapsNLO}) is the identity in the singular
region. It follows that the $z$ variables of eqs.~(\ref{eq:barfromcpm})
and (\ref{eq:cdef2}) are identical.  Hence, from eqs.~(\ref{eq:PSqn1})
and (\ref{eq:PSqn2}), we obtain
\begin{equation}
 d \Kinncpm = d \BKinn \frac{dz}{z}\,
\end{equation}
(the $1/z$ factor in the second equation being due to the Jacobian for the
change of variables $\bar{x}_{\splusminus} \to \tilde{x}_{\splusminus}$).

The choice of the counterterms in eq.~(\ref{eq:sub0}) and of the mapping
(\ref{eq:mapsNLO}) should be such that the integrals in eqs.~(\ref{eq:cdef1})
and~(\ref{eq:cdef2}) are easily performed analytically in $d$
dimensions. In this way, the $\bar{\cal C}$ terms contain explicitly the
divergences as poles in $\epsilon$.

We now write eq.~(\ref{eq:dsdO}) as
\begin{eqnarray}\label{eq:dsdOsub}
\langle O \rangle &=&
\int d\Kinn\,\Lum\,
O_n\!\(\Kinn\)\,\Big[\matB\!\(\Kinn\)+\matVb\!\(\Kinn\)\Big]
\nonumber
 \\*&+&
\int d\Kinnpo\,\Big\{
\Lum\,O_{n+1}\!\(\Kinnpo\)\,\matR\!\(\Kinnpo\) -
\sum_\ctindex \lq \Lumt\,
O_n\!\(\BKinn\)\,\matct\!\(\Kinnpo\) \rq_\ctindex \Big\}
\nonumber \\*&+&
\sum_{\ctindex\in\lg {\rm\sss FSC, S}\rg} \lq \int d\BKinn\,\Lumt\,
O_n\!\(\BKinn\)\,
\bar{\cal C}\(\BKinn\)\rq_\ctindex
+ \sum_{\ctindex\in\lg {\sss\rm ISC}_\splusminus\rg} \lq \int
d\Kinncpm\,\Lumt\, O_n\!\(\BKinn\)\,
\bar{\cal C}\(\Kinncpm\)\rq_\ctindex \nonumber
\\*&+&
\int d\Kinncp\,\Lumt\,O_n\!\(\BKinn\)\,\matCpb\!\(\Kinncp\) + 
\int d\Kinncm\,\Lumt\,O_n\!\(\BKinn\)\,\matCmb\!\(\Kinncm\) \,.
\end{eqnarray}
Notice that, in the last line, we have substituted $\Lum \to \Lumt$ for
uniformity of notation. This is correct, since, as pointed out earlier,
in the phase space of the collinear counterterms we have
$x_\splusminus=\tilde{x}_\splusminus$.

It turns out that it is always possible to write
\begin{equation}\label{eq:collcanc}
\matCpmb\!\(\Kinncpm\) + \sum_{\ctindex\in\lg {\sss\rm ISC}_{\splusminus}\rg}
\matctiP\!\(\Kinncpm\)=\matCpm\!\(\Kinncpm\) + \delta(1-z)\, \matCpm^{\rm
  div}\(\BKinn\)\;, 
\end{equation}
where $\matCpm\!\(\Kinncpm\)$ is finite in $d=4$ dimensions.\footnote{We point
out that $\matCpm$, although finite, may contain distributions associated
with the soft region $z\to 1$.}  The only remaining poles in $\epsilon$ are
included in the last term of eq.~(\ref{eq:collcanc}), and have soft origin.
It also turns out that in the quantity
\begin{equation}
\matSV\!\(\Kinn\) = \matVb\!\(\Kinn\)+\lq \sum_{\ctindex\in\lg {\sss\rm
    FSC, S}\rg} 
\matctiP\!\(\BKinn\)+\matCp^{\rm div}\(\BKinn\)
+\matCm^{\rm div}\(\BKinn\)\rq^{\BKinn=\Kinn},
\end{equation}
all poles in $\epsilon$ cancel.
With the notation
\begin{equation}
\big[ \ldots \big]^{\BKinn=\Kinn},
\end{equation}
we mean that the argument between the brackets is evaluated for values of the
phase-space variables $\BKinn$ equal to $\Kinn$.  Notice that the
identification $\BKinn=\Kinn$ is possible, since $\BKinn$ refers to the
underlying $n$-body configuration,
that must correspond to the Born term.  Defining
now the following abbreviations
\begin{equation}
R=\Lum\,\matR,\qquad 
C^{(\ctindex)}=\Lumt^{(\ctindex)}\,\matcti,\qquad
G_{\splusminus}=\Lumt \,\matCpm,\qquad
B=\Lum\, \matB,\qquad V=\Lum\,\matSV\;,
\end{equation}
equation~(\ref{eq:dsdOsub}) becomes
\begin{eqnarray}\nonumber
\langle O \rangle &=&
\int d\Kinn \;
O_n\!\(\Kinn\)\,\Big[B\!\(\Kinn\)+V\!\(\Kinn\)\Big]
 \\*&+&
\int d\Kinnpo\,\Big\{
O_{n+1}\!\(\Kinnpo\)\,R\!\(\Kinnpo\) -
\sum_\ctindex \lq
O_n\!\(\BKinn\)\,C \!\(\Kinnpo\) \rq_\ctindex \Big\}
\nonumber
\nonumber \\*&+&
\int d\Kinncp\;O_n\!\(\BKinn\)\,G_\splus\!\(\Kinncp\) + 
\int d\Kinncm\;O_n\!\(\BKinn\)\,G_\sminus \!\(\Kinncm\) \;,
\label{eq:dsdOsub2}
\end{eqnarray}
and it is now suited to be integrated numerically, since all the integrals
that appear in it are finite and can be evaluated in 4 dimensions.

\subsection{Subtraction formalism using the ``plus'' distributions}
\label{subsec:plus}
The subtraction method naturally arises when results of NLO computations are
expressed in terms of distributions in final-state variables.
In order to illustrate this issue, we assume now for simplicity
that there is just one singular region,
and, in 4 dimensions, we describe the kinematics of the emitted parton (with
momentum $k$) in the centre-of-mass (CM) frame of the incoming partons with
the following variables
\begin{equation}
\label{eq:FKSvar}
\xi=2 k^0/\sqrt{s}\,,\qquad y=\cos\theta\,,\qquad \phi\,,
\end{equation}
where $s=(\kplus+\kminus)^2$, $\theta$ is the angle of the emitted parton
relative to a reference direction (typically another parton), and $\phi$ is
an azimuthal variable around the same reference direction.  The singular
regions (soft and collinear) are associated with $\xi\,\to\,0$ and
$y\,\to\,1$ respectively.  More generally, in $d=4-2\epsilon$ dimensions, we
can write
\begin{equation} 
\label{eq:PS_d_dim}
\frac{d^{d-1} k}{2k^0(2\pi)^{d-1}}=
\frac{s^{1-\epsilon}}{(4\pi)^{d-1}}\,\,\xi^{1-2\epsilon}\,
(1-y^2)^{-\epsilon}\,\,d\xi\,dy\, d\Omega^{d-2}\;,
\end{equation}
where
\begin{equation}
d\Omega^{d-2}= (\sin\phi)^{-2\epsilon}\,d\phi\,d\Omega^{d-3}\,,
\qquad \int d\Omega^{d-3} =
\frac{2\pi^{\frac{d-3}{2}}}{\Gamma\(\frac{d-3}{2}\)}\,. 
\end{equation}
If we write
\begin{equation}
\matR = \frac{1}{\xi^2}\frac{1}{1-y} \lq \xi^2\,(1-y)\,\matR \rq\,,
\end{equation}
then $[ \xi^2\,(1-y)\,\matR ]$ is regular for $\xi \to 0$ and $y\to 1$. The
phase-space integral of $\matR$ is infrared divergent, and in $d=4-2\epsilon$
dimensions (with $\epsilon<0$) the singular part of the integration is
proportional to (see eq.~(\ref{eq:PS_d_dim}))
\begin{equation}
\label{eq:example_plus}
 \int_{-1}^1 dy \, (1-y)^{-1-\epsilon}\int_0^1 d\xi\, \xi^{-1-2\epsilon}
\; [\xi^2\,(1-y)\,\matR]\,.
\end{equation}
In order to deal with the singularities, one uses the expansions
\begin{eqnarray}
\label{eq:xiepplus}
\xi^{-1-2\ep}&=&-\frac{1}{2\ep}\delta(\xi) + \left(\frac{1}{\xi}\right)_+
-2\ep \left(\frac{\log \xi}{\xi}\right)_+ + {\cal O}(\ep^2)\,,
\\
\label{eq:yepplus}
(1-y)^{-1-\epsilon}&=&-\frac{2^{-\epsilon}}{\epsilon}\delta(1-y) +
\left(\frac{1}{1-y}\right)_+ + {\cal O}(\ep)\,, 
\end{eqnarray}
with the usual definition of the $+$ prescription
\begin{eqnarray}
\label{eq:xiplusdef1}
\int_0^1 d\xi\, \(\frac{1}{\xi}\)_+ f(\xi) 
&=&\int_0^1 d\xi\, \frac{f(\xi)-f(0)}{\xi}\;,
\\
\label{eq:xiplusdef2}
\int_0^1 d\xi\, \( \frac{\log \xi}{\xi} \)_+ f(\xi) 
&=&\int_0^1 d\xi\, \log \xi \, \frac{f(\xi)-f(0)}{\xi}\;,
\\
\label{eq:yplusdef}
\int_{-1}^1 d y\, \(\frac{1}{1-y}\)_+ f(y) 
&=&\int_{-1}^1 d y\, \frac{f(y)-f(1)}{1-y} \;.
\end{eqnarray}
Inserting eqs.~(\ref{eq:xiepplus}) and~(\ref{eq:yepplus})
into~(\ref{eq:example_plus}), and defining $g(\xi,y)\equiv
\xi^2\,(1-y)\,\matR$, for ease of notation, we have
\begin{eqnarray}
\label{eq:example_expans}
&&
\int_{-1}^1 dy \, (1-y)^{-1-\epsilon}\!\!\int_0^1 d\xi\, \xi^{-1-2\epsilon}\,
g(\xi,y) =\! -\frac{1}{2\ep}\int_{-1}^1\!\! dy\, (1-y)^{-1-\ep} g(0,y)
\nonumber\\ 
&& \phantom{aaaaaaaaa} -\int_0^1 d\xi
 \lq \frac{2^{-\ep}}{\ep}  \(\frac{1}{\xi}\)_+\!\! - 2 \( \frac{\log \xi}{\xi}
\)_+\rq g(\xi,1)
\nonumber\\
&& \phantom{aaaaaaaaa} +\int_{-1}^1 dy\int_0^1 d\xi
\(\frac{1}{\xi}\)_+  \(\frac{1}{1-y}\)_+ g(\xi,y) + \ord{\ep}.
\end{eqnarray}
The first term on the r.h.s.\ of eq.~(\ref{eq:example_expans}), after
integration in $y$ and $\phi$, gives a contribution with the
same structure as the virtual term, with which it is combined. Since this
term arises from the $\delta(\xi)$ factor, it can be easily obtained
using the eikonal approximation for soft emissions in $d$ dimensions.  The
second term on the r.h.s.\ of eq.~(\ref{eq:example_expans}) is the
contribution to $\matR$ proportional to the delta-function term in
eq.~(\ref{eq:yepplus}), multiplied by the second and third term of
eq.~(\ref{eq:xiepplus}). It gives rise to terms that, in the case of
final-state singularities, can also be integrated in $\xi$ and in $\phi$,
and yields terms of the same form of the virtual
terms, with which it is combined.  Also for this term it is not necessary
to know ${\cal R}$ in $d$ dimensions, since one can use the collinear
approximation in the $y\to 1$ limit in order to obtain it.
In the case of initial-state
singularities, the same procedure gives terms of the same form of the
collinear counterterms, with which they are combined.  Finally, a term of the
form
\begin{equation}\label{eq:doubleplus0}
\int_{-1}^1 dy\int_0^1 d\xi
\(\frac{1}{\xi}\)_+  \(\frac{1}{1-y}\)_+ g(\xi,y)
=\int_{-1}^1 dy\int_0^1 d\xi \, \xi\; \hat\matR
\end{equation}
remains, where
\begin{equation}\label{eq:doubleplus}
\hat\matR\equiv \frac{1}{\xi} \left\{
\(\frac{1}{\xi}\)_+  \(\frac{1}{1-y}\)_+\lq \xi^2(1-y)\,\matR \rq\right\}\;.
\end{equation}
Observe that the $1/\xi$ factor in front of
eq.~(\ref{eq:doubleplus}) cancels against the phase-space $\xi$ factor
in eq.~(\ref{eq:doubleplus0}),
and that $[\xi^2(1-y)\,\matR]$ has no singularities
at $\xi=0$ and $y=1$, so that distributions in $\xi$ and
$y$ act on a regular function.

The procedure outlined in this section is fully general. It can be shown that,
defining $\hat{R}=\Lum\,\hat\matR$, one can rewrite eq.~(\ref{eq:dsdOsub2}) in the form
\begin{eqnarray}
\langle O \rangle &=&
\int d\Kinn \;
O_n\!\(\Kinn\)\,\Big[B\!\(\Kinn\)+V\!\(\Kinn\)\Big]
\nonumber \\*&+&
\int d\Kinnpo\,
O_{n+1}\!\(\Kinnpo\)\,\hat{R}\!\(\Kinnpo\)
\nonumber \\*&+&
\int d\Kinncp\;O_n\!\(\BKinn\)\,G_\splus\!\(\Kinncp\) + 
\int d\Kinncm\;O_n\!\(\BKinn\)\,G_\sminus \!\(\Kinncm\) \;.
\label{eq:dsdOsub2hat}
\end{eqnarray}
By handling the $+$ distributions in $\hat\matR$ according to the
prescriptions~(\ref{eq:xiplusdef1}), (\ref{eq:xiplusdef2})
and~(\ref{eq:yplusdef}), one automatically generates the real counterterms,
provided the variables $\xi$ and $y$ appear in the phase-space
parametrization.  If more than one singular region is present, the real cross
section is decomposed into a sum of terms, each of them having
singularities in no more than one singular region. For each term, the phase
space is parametrized in such a way that the variables $\xi$ and $y$,
appropriate to that particular singular region, are present.

The expression of a cross section in terms of distributions has sometimes the
advantage that the associated projections are not uniquely fixed.
In fact, one has the freedom to chose the integration variables other than
$y$ and $\xi$ at one's convenience. This amounts to choosing a different
projection.

\subsection{\FKS{} subtraction}
\label{sec:FKSsub}
In this section, we briefly review the FKS general subtraction formalism, proposed
in refs.~\cite{Frixione:1995ms,Frixione:1997np},
including a few modifications that have been
introduced recently (see ref.~\cite{Frixione:2005vw}).

In FKS one expresses the cross section for the real-emission
contribution as a sum of terms, each of them having at most one collinear and
one soft singularity associated with one parton (called the FKS parton).  The
singular region associated with the final-state parton $i$ becoming soft or
collinear to one of the incoming partons are labeled by $i$,
while those associated
with final-state parton $i$ becoming soft or collinear to a final-state
parton $j$ are labeled by the pair $ij$.  For each singular region one
introduces certain non-negative functions\footnote{The notation of
refs.~\cite{Frixione:1995ms,Frixione:1997np} has been slightly changed here
in order to simplify the discussion. Functions $\Szi$ and $\Soij$ of the present
paper play the same role as
$\Theta^{(0)}_i$ and $\Theta^{(1)}_{ij}\theta(k_{jT}^2-k_{iT}^2)$
of  ref.~\cite{Frixione:1997np} respectively.}
$\Sz$ of the $(n+1)$-body phase space such that
\begin{equation}
\sum_{i} \Szi + \sum_{ij} \Soij=1\,.
\label{Sident}
\end{equation}
We have two options for the range of the indices in the sums:
we can let them range from 1 to $(n+1)$ (excluding only
the $i=j$ possibility in the second sum), or we can assume that
the $\Szi$ and $\Soij$ are zero (i.e.\ they are excluded from the
sum) if the corresponding regions are not singular.
For example, if $ij$ refer to a quark and an antiquark of different
flavour there cannot be a FSC singularity in this region, and we can
set $\Soij=0$. Also, if $i$ is a gluon and $j$
is a quark, we may set to zero $\Soji$, since
there is no soft singularity associated to $j$ becoming soft.
Notice that if $i$ and $j$ are both gluons, both terms $\Soij$ and $\Soji$
appear in the sum, since there is a soft singularity for either parton
becoming soft.

The $\Sz$ function have the following properties
\begin{eqnarray}
&&\lim_{k^0_m\to 0}\left(\Szi+\sum_j \Soij\right)=\delta_{im}\,,
\label{Slimsf}
\\
&&\lim_{\vec{k}_m\parallel \vec{k}_\splusminus}\Szi=\delta_{im}\,,
\label{Slimcl}
\\
&&\lim_{\vec{k}_m\parallel \vec{k}_l}\(\Soij+\Soji\)=\delta_{im}\delta_{jl}
+ \delta_{il}\delta_{jm}\,,
\label{Slimclf}
\\
&&\lim_{\vec{k}_m\parallel \vec{k}_{\splusminus}}\Soij=0\,.
\label{Slimcl0}
\end{eqnarray}
Thus, in a given soft region, (i.e.\ if parton $m$ is soft), all
$\Szi$ and $\Soij$ with $i\ne m$ vanish, and eq.~(\ref{Sident}) is
consistent with eq.~(\ref{Slimsf}).  In a given initial-state
collinear region, i.e.\ parton $m$ becomes collinear to an initial
state parton, only ${\cal S}_m$ is non-zero and equal to one,
consistently with eq.~(\ref{Sident}).  In a given final-state
collinear region, i.e.\ if partons $i$ and $j$ are collinear, only
$\Soij$ and $\Soji$ can differ from zero, their sum being 1, again
consistently with eq.~(\ref{Sident}).

We now write
\begin{equation}
\matR=\sum_i \matR_i + \sum_{ij} \matR_{ij}\,,
\label{FKSpart}
\end{equation}
where
\begin{equation}
\matR_i= \Szi\matR\,,\qquad  \matR_{ij}= \Soij\matR\,.
\label{FKSpart1}
\end{equation}
The $\matR_i$ terms give a divergent contribution (i.e.\ a contribution which
has to be subtracted) only in the regions in which parton $i$ is soft and/or
collinear to one of the initial-state partons, and the $\matR_{ij}$ terms are
divergent only in the regions in which parton $i$ is soft and/or collinear to
final-state parton $j$. Notice that if we have chosen the option of keeping
all the $\Sz$ functions different from zero, the  $\matR_i$ and  $\matR_{ij}$
functions corresponding to non-singular regions are non-zero but finite.

Equations~(\ref{Slimsf})--(\ref{Slimcl0}) are the only properties of the
$\Sfun$ functions used in the analytical computations of
refs.~\cite{Frixione:1995ms,Frixione:1997np}. Their actual functional forms,
away from the infrared limits, are only relevant to numerical integrations.

After the $(n+1)$-body cross section is decomposed as in
eq.~(\ref{FKSpart}), FKS chooses a different parametrization of the
$(n+1)$-body phase space for each term, such that one can perform the
necessary analytical and numerical integrations in an easy way.  The key
variables in the phase-space parametrization associated with $\Szi$ are the
energy of parton $i$ (directly related to soft singularities), and the angle
between parton $i$ and one of the initial-state partons (directly related to
initial-state collinear singularities). For $\Soij$, the energy of parton $i$
and the angle between parton $i$ and $j$ (related to a final-state
collinear singularity) are chosen instead. Therefore, there are only two
independent functional forms for phase spaces in FKS, one for initial- and
one for final-state emissions.

In the following, we will need a further refinement of the FKS
decomposition (eq.~(\ref{Sident})). This is because, in FKS, the
$\splus$ and $\sminus$ collinear regions are both singled out by
the $\Szi$ functions. In \POWHEG{} we will need sometimes to treat
the two collinear regions separately. We thus introduce the notation
\begin{equation}
\Szi=\Szip+\Szim\,,
\end{equation}
with the properties
\begin{equation}
\lim_{\vec{k}_m\parallel \vec{k}_\splusminus}\Szipm=\delta_{im}\,,
\quad\qquad
\lim_{\vec{k}_m\parallel \vec{k}_\splusminus}\Szimp=0\,,
\label{Slimclpm}
\end{equation}
that refine eq.~(\ref{Slimcl}). Eqs.~(\ref{FKSpart}) and
(\ref{FKSpart1}) are modified accordingly.

\subsubsection{The $\boldsymbol{\Sfun}$ functions}
\label{sec:Sfun}
In the original formulation of the FKS subtraction, the $\Sfun$ functions
were defined as sets of $\stepf$ functions. The different contributions to
the real cross section, separated out in this way, corresponded to a
partition of the phase space into non-overlapping regions. In the more recent
calculation of single-top hadroproduction in \MCatNLO~\cite{Frixione:2005vw},
the $\Sfun$ functions were instead defined as smooth functions. In view of
Monte Carlo implementations, step functions should be avoided as much as
possible, and therefore we consider here the latter approach.  We introduce
the functions $d_i$ and $d_{ij}$, where $i,j=1,\ldots,n+1$, with the
following properties
\begin{equation}
\label{dijdef}
\begin{array}{lllllll}
 d_i= 0  \ \ &\mbox{if and only if} &
\ \  E_i= 0\ &\mbox{or}&\  \vec{k}_i\parallel \vec{k}_\splus\ &\mbox{or}&\
\vec{k}_i\parallel \vec{k}_\sminus \,, \\ 
 d_{ij}=0 \ \ &
\mbox{if and only if}& \ \  E_i=0\ &\mbox{or}&\ E_j= 0\ &\mbox{or}&\ 
\vec{k}_i\parallel \vec{k}_j\,,
\end{array}
\end{equation}
where energies and three-momenta are computed in the centre-of-mass frame of
the incoming partons.  A possible definition of the $d$'s is
\begin{eqnarray}
d_i&=&\left(\frac{\sqrt{s}}{2}E_i\right)^a\left(1-\cos^2\theta_i\right)^b\,,
\label{di1}
\\
d_{ij}&=&\Big(E_i E_j\Big)^a\Big(1-\cos\theta_{ij}\Big)^b\,,
\label{eq:di2}
\end{eqnarray}
where $\theta_i$ is the angle between $\vec{k}_i$ and $\vec{k}_\splus$,
$\theta_{ij}$ the angle between $\vec{k}_i$ and $\vec{k}_j$, $s=\(\kplus+
\kminus\)^2$, and $a$, $b$ are positive real numbers.  Equations~(\ref{di1})
and~(\ref{eq:di2}) can be easily expressed in terms of invariants using
\begin{equation}
\kplus=\frac{\sqrt{s}}{2}(1,0,0,1)\,,\;\;\;\;
\kminus =\frac{\sqrt{s}}{2}(1,0,0,-1)\,,
\end{equation}
which imply
\begin{eqnarray}
E_i&=&\frac{1}{\sqrt{s}}\,\(\kplus+\kminus \)\mydot k_i\,,
\label{eq:Evsinv}
\\
\cos\theta_i&=&1-\frac{2k_i\mydot \kplus}{E_i\sqrt{s}}\,,
\\
\cos\theta_{ij}&=&1-\frac{k_i\mydot k_j}{E_i E_j}\,.
\end{eqnarray}
We now introduce the quantity
\begin{equation}
\Dfun=\sum_k\,\frac{1}{d_k}+
\sum_{kl}\,\frac{1}{d_{kl}}\,,
\label{Ddef}
\end{equation}
and define
\begin{eqnarray}
\Szi&=&\frac{1}{\Dfun\,d_i}\,,
\label{Szidef}
\\
\Soij&=&\frac{1}{\Dfun\,d_{ij}}\;
h\!\left(\frac{E_i}{E_i+E_j}\right)\,,
\label{Soijdef}
\end{eqnarray}
where $h$ is a function such that
\begin{equation}
\lim_{z\to 0}h(z)=1\,,\;\;\;\;\;\;
\lim_{z\to 1}h(z)=0\,,\;\;\;\;\;\;
h(z)+h(1-z)=1\,.
\label{hdef}
\end{equation}
For example, one can define\footnote{ In the original FKS formulation
$h(z)=\stepf(1/2-z)$, which implies that $\Soij$ defined in
eq.~(\ref{Soijdef}) vanishes if $E_j<E_i$. With such a choice, a proof was
given in ref.~\cite{Frixione:1995ms} that all the infrared singularities
are canceled by the FKS subtraction. We have checked that, if a
smooth form for $h$ is adopted, this proof goes through unaltered by
replacing $\stepf(z-1/2)$ in eq.~(4.84) and~(4.85) of
ref.~\cite{Frixione:1995ms} with $h(1-z)$.}
\begin{equation}
h(z)=\frac{(1-z)^c}{z^c+(1-z)^c}
\end{equation}
for some positive $c$. Notice that the $h$ factor is necessary only if
one considers both functions $\Soij$ and $\Soji$ (which is strictly necessary
only if both $i$ and $j$ are gluons).

It is manifest that eqs.~(\ref{Szidef})
and~(\ref{Soijdef}) fulfill eqs.~(\ref{Slimsf})--(\ref{Slimcl0}).

As in section~\ref{sec:FKSsub}, in order to separate the $\splus$ and $\sminus$
collinear regions, we modify the previous formulae as follows.
We introduce
\begin{equation}\label{eq:dispm}
d_i^\splusminus=\left(\frac{\sqrt{s}}{2}E_i\right)^a 2^b (1\mp \cos \theta_i)^b\,,
\end{equation}
instead of $d_i$ of eq.~(\ref{di1}). The definition of $\Dfun$ in eq.~(\ref{Ddef})
becomes
\begin{equation}
\Dfun=\sum_k\,\left(\frac{1}{d_k^\splus}+\frac{1}{d_k^\sminus}\right)+
\sum_{kl}\,\frac{1}{d_{kl}}\,.
\label{Ddefpm}
\end{equation}
We then define
\begin{equation}\label{eq:szipmdef}
\Szipm=\frac{1}{\Dfun\,d_i^\splusminus}\,.
\end{equation}

\subsubsection{Contributions to the cross section}
\label{sec:xsec}
We introduce, for the FKS parton $i$, the following variables
\begin{equation}
\label{eq:csi_yij_def}
\xi_i=\frac{2k^0_i}{\sqrt{s}}\,,\qquad y_i=\cos \theta_i\,,\qquad
y_{ij}=\cos \theta_{ij}\,,
\end{equation}
where $\theta_i$ is the angle of parton $i$ with the incoming parton
$\nplus$, and $\theta_{ij}$ is the angle of parton $i$ with parton $j$.  All
variables are computed in the centre-of-mass frame of the incoming partons.
The phase space for the $\matR_i$ and $\matR_{ij}$ contributions is written
in $d=4-2\epsilon$ dimensions as
\begin{eqnarray}
d\Phi_{n+1} &=& (2\pi)^d\,\delta^d\!\!\(\kplus+\kminus -\sum_{i=1}^{n+1} k_i\)\!
\left[\prod_{l\ne i} \frac{d^{d-1} k_l}{(2\pi)^{d-1} \,2k^0_l}\right]
\nonumber \\
&\times&
\frac{s^{1-\epsilon}}{(4\pi)^{d-1}}\,\xi_i^{1-2\epsilon}
\(1-y^2\)^{-\epsilon}d\xi_i\,dy\, d\Omega^{d-2},
\label{eq:ddimphasespace}
\end{eqnarray}
where $y,\Omega$ stands for either $y_i$, $\Omega_i$ or $y_{ij}$,
$\Omega_{ij}$.  The transverse angular variables $d\Omega_{i}^{d-2}$ are
relative to the collision axis, while $d\Omega_{ij}^{d-2}$ are relative to
the direction of parton $j$.  The singularities for $\xi \to 0$, $y_i \to \pm
1$ or $y_{ij}\to 1$ are treated along the lines of section~\ref{subsec:plus}.
The final expression in the FKS formalism results from the cancellation of
the infrared singularities which emerge in the intermediate steps of the
computation. It thus involves only non-divergent terms.

The contributions to the real-emission cross section, in the notation of
eq.~(\ref{eq:dsdOsub2hat}), are
\begin{equation}
\label{eq:FKS0}
\hat{\matR}=\sum_i \hat{\matR}_i + \sum_{ij}\hat{\matR}_{ij}\,, 
\end{equation}
where
\begin{eqnarray}
\label{eq:FKSrin}
\hat{\matR}_i&=& \frac{1}{\xi_i} \left\{ \half\xiic\left[\omyid+\opyid\right]
\lq \(1-y_i^2\) \, \xi_i^2 \, \matR_i\rq\right\} \,,
\phantom{aaa}
\\
\label{eq:FKSrout}
\hat{\matR}_{ij}&=& \frac{1}{\xi_i} \left\{ \xiic\omyijd
\lq\(1-y_{ij}\)\,\xi_i^2 \,\matR_{ij}\rq\right\} \,.
\end{eqnarray}
If we need to separate the $\splus$ and $\sminus$ collinear regions, as discussed
at the end of sec.~\ref{sec:FKSsub}, we have
\begin{eqnarray}
\label{eq:FKS0p}
\hat{\matR}&=&\sum_i \left(\hat{\matR}_i^\splus+\hat{\matR}_i^\sminus\right)
 + \sum_{ij}\hat{\matR}_{ij}\,, 
\\
\label{eq:FKSrinp}
\hat{\matR}_i^\splusminus&=& \frac{1}{\xi_i} \left\{ \xiic \ompyid
\lq \(1\mp y_i\) \, \xi_i^2 \, \matR_i^\splusminus \rq\right\} \,.
\end{eqnarray}

The distributions that appear in eqs.~(\ref{eq:FKSrin})
and~(\ref{eq:FKSrout}) are defined as follows
\begin{eqnarray}
\label{eq:uoxidef}
\int_0^1 d\xi \, f(\xi)\xic &=& \int_0^1 d\xi\,
\frac{f(\xi)-f(0)\,\stepf\!\(\xicut-\xi\)}{\xi}\,,
\\
\label{eq:uoyipmdef}
\int_{-1}^{1}dy \, f(y)\ompyd &=& \int_{-1}^{1}dy\,
\frac{f(y)-f(\pm 1)\,\stepf\!\(\pm y-1+\delta\)}{1\mp y}\,.
\end{eqnarray}
The parameters $\xicut$, $\deltaI$ and $\deltaO$ must be chosen in the ranges
$0<\xicut\le 1$ and $0<\delta_{\sss I,O}\le 2$.  The dependence they
introduce in the $(n+1)$-body contribution is exactly compensated by the same
dependence in the $n$-body contribution. In the \POWHEG{} framework it is
often convenient to use the maximal range of integration, and we will thus
also use the notation
\begin{equation}\label{eq:fksplusdef}
\left(\frac{1}{\xi}\right)_+=\xic\;\;\mbox{with}\;\;\xicut=1\,,\quad\quad
\left(\frac{1}{1\mp y}\right)_+=\ompyd\;\;\mbox{with}\;\;\delta=2\,.
\end{equation}
The parametrization of the $(n+1)$-body phase space, appropriate to the
integration of $\matR_i$ and $\matR_{ij}$, can be chosen as the
$d=4$ version of equation~(\ref{eq:ddimphasespace}), as suggested in
the original FKS papers.  This is however not necessary. Any parametrization
of the phase space that allows a simple handling of the distributions is
acceptable. This freedom is exploited in the present work, in order to
simplify the formulation of the \POWHEG{} method in the case of the
$\matR_{ij}$ contributions, where we make a choice of the azimuthal variables
different from that of eq.~(\ref{eq:ddimphasespace}).

We now consider the soft-virtual term $V$ in eq.~(\ref{eq:dsdOsub2hat}).  We
define the set of all the $n+2$ parton labels for an $n$-body process
\begin{equation}
  \matI = \lg \nplus, \nminus, 1, \ldots, n \rg.
\end{equation}
The virtual contribution $\matVb$ of eq.~(\ref{eq:dsdOsub}) is given
by\footnote{We stress that $\matVb$ is the contribution to the cross section
due to the
interference of the virtual amplitude with the Born term. It thus includes the
corresponding factor of 2.
}
\begin{equation}
\matVb={\cal N}\,\asotpi
\Bigg[-\sum_{i \in \matI} \left(\frac{1}{\ep^2} \,C_{f_i}+
\frac{1}{\ep} \,\gamma_{f_i}\right)\matB
+\frac{1}{\ep}\sum_{\stackrel{i,j \in \matI}{i\ne j}}\log\frac{2k_i\mydot
  k_j}{Q^2}\, 
\matB_{ij}+\matSV_{\sss\rm fin}\Bigg]\,,
\label{eq:FKSoneloop}
\end{equation}
where
\begin{equation}\label{eq:virtnorm}
{\cal N}=
\frac{(4\pi)^\ep}{\Gamma(1-\ep)}
\left(\frac{\mu^2}{Q^2}\right)^\ep.
\end{equation}
Notice that, in the second sum on the r.h.s.\ of
eq.~(\ref{eq:FKSoneloop}), we sum over $i\ne j$, and thus,
since (as we show later) $\matB_{ij}$ is symmetric,
each term appears twice in the sum.
The definition of the finite part $\matSV_{\sss\rm fin}$ depends upon the
definition of the normalization factor ${\cal N}$, for which we have adopted
the common choice of eq.~(\ref{eq:virtnorm}), and from the regularization
scheme, that we assume to be the standard conventional dimensional
regularization (CDR).\footnote{If we instead use the dimensional reduction
(DR) scheme, we have $\matSV_{\sss\rm fin} = \matSV_{\sss\rm fin}^{\rm DR}
-\as/(2\pi)\, \matB \sum_{i \in \matI} \tilde{\gamma}(f_i)$, where,
$\tilde{\gamma}(g)=\NC/6$ and $\tilde{\gamma}(q)=(\NCt-1)/(4\NC)$, where
$\NC=3$ is the number of colours.}
Finally, $\mu^2$ is the renormalization scale, and $Q^2$ is an (arbitrary)
physical scale that is factored out from the virtual amplitude in order to
make the normalization ${\cal N}$ dimensionless (thus $\matSV_{\sss\rm fin}$
depends upon $\mu^2$ and $Q^2$).

The symbol $f_i$ denotes the flavour of parton $i$, i.e.\ $g$ for a
gluon, $q$ for a quark and $\bar{q}$ for an antiquark. We define
\begin{eqnarray}
&&C_g=\CA\,,\phantom{aaaaaaaaaaaaaaaaaaaaaaa\!}
C_q=C_{\bar{q}} = \CF\,,
\\
&&\gamma_g=\frac{11\CA-4\TF\, \NF}{6}\,,\phantom{aaaaaaaaaaaaa}
\gamma_q=\gamma_{\bar q}= \frac{3}{2}\CF\,,
\\
&&\gamma^\prime_g=\( \frac{67}{9}-\frac{2\pi^2}{3}\)\CA
-\frac{23}{9}\TF \,\NF\,,\;\;\;\;\;\;
\gamma^\prime_q=\gamma^\prime_{\bar{q}}=
\(\frac{13}{2}-\frac{2\pi^2}{3}\)\CF\,.
\phantom{aaaa}
\end{eqnarray}
In case $i$ is a colorless particle all the above quantities are zero.

The quantities $\matB_{ij}$, commonly referred to as the colour-correlated
Born amplitudes, are defined in the following way
\begin{equation}
\label{eq:colourcorr}
\matB_{ij}=-\frac{1}{2s}\frac{1}{N_{\rm sym} D_\splus \,D_\sminus\,
S_\splus\,S_\sminus}
\sum_{\stackrel{{\mbox{\rm
      \scriptsize 
      spins}}}{{\mbox{\rm \scriptsize colours}}}}
{\cal M}_{\{c_k\}}\, \left( {\cal M}^\dagger_{\{c_k\}} \right)_{
\stackrel{\scriptstyle c_i \to c'_i}{c_j \to c'_j}}\;
 T^a_{c_i,c'_i} \, T^a_{c_j,c'_j}\,.
\end{equation}
Here ${\cal M}_{\{c_k\}}$ is the Born amplitude, and $\{c_k\}$ stands for
the colour indices of all
partons in $\matI$.  The suffix on the parentheses that enclose ${\cal
M}^\dagger_{\{c_k\}}$ indicates that the colour indices of partons $i,j \in \matI$ are
substituted with primed indices in ${\cal M}^\dagger_{\{c_k\}}$.  Furthermore $N_{\rm sym}$
is the symmetry factor for identical particles in the final state,
$D_{\splusminus}$ are the dimension of the colour representations of the
incoming partons (3 for quarks and 8 for gluons), and $S_{\splusminus}$ are
the number of spin states.  The factor $1/(2s)$
is the flux factor.  We assume summation over repeated colour indexes ($c_k$
for $k\in \matI$, $c'_i$, $c'_j$ and $a$) and spin indices.  For gluons
$T^a_{cb}=if_{cab}$, where $f_{abc}$ are the structure constants of the
$SU(3)$ algebra. For incoming quarks $T^a_{\alpha\beta}=t^a_{\alpha\beta}$,
where $t$ are the colour matrices in the fundamental representation
(normalized as ${\rm Tr}[t\,t]=1/2$). For
antiquarks $T^a_{\alpha\beta} =-t^a_{\beta\alpha}$. It follows from colour
conservation that $\matB_{ij}$ satisfy
\begin{equation}\label{eq:colcons}
\sum_{i\in \matI, i\ne j}\matB_{ij}=C_{f_j}\matB\,.
\end{equation}
The soft-virtual term in eq.~(\ref{eq:dsdOsub2hat}) is given by
\begin{equation}
V=\Lum\,\matSV\,,\qquad \matSV=\asotpi\Bigg( {\cal Q}\,\matB
+\sum_{\stackrel{i,j \in \matI}{i\ne j}}\,{\cal I}_{ij}\,
\matB_{ij}+
\matSV_{\sss\rm fin}\Bigg)\,.
\label{eq:FKSsv}
\end{equation}
The quantities ${\cal Q}$ and ${\cal I}_{ij}$ depend on the flavours and
momenta of the incoming and outgoing partons. They are defined as follows
\begin{eqnarray}
{\cal Q}&=&\sum_{i=1}^{n}\Bigg[\gamma^\prime_{f_i}
-\log\frac{s\deltaO}{2Q^2}\left(\gamma_{f_i}
-2C_{f_i}\log\frac{2E_i}{\xicut\sqrt{s}}\right)
\nonumber \\*&&\phantom{\sum_{j=1}^{n}}
+2C_{f_i}\left(\log^2\frac{2E_i}{\sqrt{s}}-\log^2\xicut\right)
-2\gamma_{f_i}\log\frac{2E_i}{\sqrt{s}}\Bigg]
\nonumber \\*&&-
\log\frac{\muF^2}{Q^2}
\lq \gamma_\fmo + 2C_\fmo\log\xicut
+\gamma_\fmt + 2C_\fmt\log\xicut \rq\,,\phantom{aaaa}
\label{eq:Qdef}
\\
{\cal I}_{ij}&=&
\frac{1}{2}\log^2\frac{\xicut^2 s}{Q^2}
+\log\frac{\xicut^2 s}{Q^2}\log\frac{k_j\mydot k_i}{2E_j E_i}
-{\rm Li}_2\left(\frac{k_j\mydot k_i}{2E_j E_i}\right)
\nonumber \\*
&&
   +\frac{1}{2}\log^2\frac{k_j\mydot k_i}{2E_j E_i}
   -\log\left(1-\frac{k_j\mydot k_i}{2 E_j E_i}\right)
   \log\frac{k_j\mydot k_i}{2E_j E_i}\,,\phantom{aaaaa}
   \label{eq:Iijreg}
\end{eqnarray}
where $E_i$ is the energy of parton
$i$ in the partonic centre-of-mass frame.

We finally report the expressions for the initial-state collinear remnants
that appear in eq.~(\ref{eq:dsdOsub2hat}). For each collinear singular
configuration, relevant to the process, we have a term
$G_\splus=\Lumt\,\matCp$ (and a corresponding one for
$G_\sminus=\Lumt\,\matCm$), where
\begin{eqnarray}
\matCp^{\fmo\fmt}(z) &=&
\asotpi \sum_\fmop \Bigg\{ (1-z) \, P^{\fmo\fmop}(z,0) 
\Bigg[\omzc\log\frac{s\deltaI}{2\muF^2}
+2\lomzc\Bigg] \phantom{AAA}
\nonumber \\*&&\phantom{\asotpi\sum_{d}}
-\lq \frac{\partial P^{\fmo\fmop}(z,\ep)}{\partial \ep}\rq_{\ep=0} 
-K^{\fmo\fmop}(z)\Bigg\} \, \matB^{\fmop\fmt}(z) \,,
\label{eq:FKScrp}
\end{eqnarray}
for a
process in which an incoming parton $\nplus$ of flavour $\fmo$ splits into a
parton $\fmop$ (with fraction $z$ of the incoming momentum) that enters the
Born process $\matB$.  The superscripts on $\matB$ denote the flavours of the
incoming parton, and the $z$ dependence is to remind that the incoming
$\nplus$ momentum is rescaled.  The distributions are defined as in
eq.~(\ref{eq:uoxidef}), with $\xi=1-z$.  The functions $P(z,\epsilon)$ are
the leading order Altarelli-Parisi splitting functions in $d=4-2\epsilon$ dimensions,
given by
\begin{eqnarray}
P^{qq}(z,\ep)
&=& \CF
\;\left[ \frac{1 + z^2}{1-z} - \ep (1-z) \right] \;\;,
\\
P^{qg}(z,\ep) &=& \CF
\;\left[ \frac{1 + (1-z)^2}{z} - \ep z \right] \;\;,
\\
P^{gq}(z,\ep) &=& \TF
\left[ 1 - \frac{2 z(1-z)}{1-\ep} \right] \;\;,
\\
P^{gg}(z,\ep) &=& 2\,\CA
\left[ \frac{z}{1-z} + \frac{1-z}{z}
+ z(1-z) \right] \;\;.
\end{eqnarray}
The distributions $K^{ff'}$ control the change of scheme in the evolution of
parton distribution functions. They are
defined in ref.~\cite{Frixione:1995ms}, and
equivalently, with the notation $K^{ff'}_{\rm F.S.}$ in
ref.~\cite{Catani:1996vz}. They are identically zero in $\MSB$.

\subsection{\CS{} subtraction}
\label{sec:dipsub}
In this section we briefly review the  general subtraction formalism
proposed in ref.~\cite{Catani:1996vz}, called {\em dipole
subtraction}.  A dipole is defined by three partons: the
emitted, the emitter and the spectator parton (the last two forming
the dipole).  In the dipole formulation, one given singular region receives,
in general, contributions by several dipoles, differing among each other by
the spectator parton. Thus, the counterterms $\matcti$ are associated with
dipoles, rather than singular regions. The maps $\MapNLOi$ of eq.~(\ref{eq:mapsNLO})
in the dipole formulation (that are summarized in section~\ref{sec:powhegcs}) are
constructed in such a way that, in most cases, they affect only the
momenta of the dipole partons, and all other momenta remain
unchanged.\footnote{The only exception is when the emitter and the spectator
are the two incoming partons.}  The maps $\MapNLOi$ appropriate to
the dipole subtraction will be discussed in
section~\ref{sec:powhegcs}.  These definitions, together with the definitions
of the relative dipole counterterms (to be found in the original
reference~\cite{Catani:1996vz}) are necessary to define a \POWHEG{}
implementation. The last two missing ingredients are the soft-virtual term
$\matSV$, and the collinear remnants $\matCpm$. In this section we report
explicitly the form of these terms, expressed in our notation.

For the soft-virtual contribution $\matSV$ we obtain the following
\begin{equation}\label{eq:svcs}
\matSV=\frac{\as}{2\pi}\Bigg\{\!
\matSV_{\sss\rm fin}-
\sum_{\stackrel{i,j \in \matI}{i\ne j}}\lq\frac{1}{2} 
\log^2\frac{Q^2}{2k_i\cdot k_j} 
+\frac{\gamma_{f_i}}{C_{f_i}}\log\frac{Q^2}{2k_i\cdot k_j}\rq \matB_{ij}
+\sum_{i \in \matI} \lq-\frac{\pi^2}{3}C_{f_i}+\gamma_{f_i}+K_{f_i}\rq
\! \matB \!\Bigg\}.
\end{equation}
Equation~(\ref{eq:svcs}) has been obtained by a suitable manipulation of
eqs.~(C.27) and~(C.28) of ref.~\cite{Catani:1996vz}.  The virtual term
$\matSV_{\sss\rm fin}$ coincides with that of our eq.~(\ref{eq:FKSoneloop}).

The collinear remnant is given by
\begin{eqnarray}
\matCp^{\fmo\fmt}(z) &=& \frac{\as}{2\pi} \sum_{\fmop} 
\Bigg\{
\lq  \overline{K}^{\fmo\fmop}(z)  - K_{\rm F.S.}^{\fmo\fmop}(z)\rq
\matB^{\fmop\fmt}(z)
\nonumber\\
&-& \delta^{\fmo\fmop} \sum_{i=1}^{n}
\frac{\gamma_{f_i}}{C_{f_i}} \lq \(\frac{1}{1-z}\)_{\!\!+}\!\!\! +\delta(1-z)
\rq  
\matB_{i\splus}^{\fmop\fmt}(z) + \frac{1}{C_{\fmop}} \tilde{K}^{\fmo\fmop}(z) 
\,\matB_{\sminus\splus}^{\fmop\fmt}(z)
\nonumber\\
&-&
P^{\fmo\fmop}(z)\frac{1}{C_{\fmop}}
\sum_{\stackrel{i \in \matI}{i\ne \splus}} 
\matB_{i\splus}^{\fmop\fmt}(z) \log\frac{\muF^2}{2\,z\,\kplus\cdot \kminus}
\Bigg\}\,,
\label{eq:cremcs}
\end{eqnarray}
and the analogous one for $\matCm$. This formula is the translation in our notation
of eq.~(10.30) of ref.~\cite{Catani:1996vz}.  The definition of the functions
$\overline{K}$, $\tilde{K}$, $K_{\rm F.S.}$ and $P$ are given in appendix~C
of ref.~\cite{Catani:1996vz}.

In eq.~(\ref{eq:cremcs}), $\matCp^{\fmo\fmt}$ is the collinear remnant
contribution for flavours $\fmo,\fmt$ of the incoming partons.  Analogously,
the superscripts in $\matB$ (and $\matB_{ij}$) single out a given flavour
combination for the incoming partons in the Born amplitudes and its
colour-correlated components.

\section{NLO with Parton Showers}\label{sec:mcsim}
\subsection{Parton Shower Monte Carlo programs}
A detailed discussion of the ideas upon which Shower Monte Carlo programs are
based is beyond the scope of the present paper. The interested reader can
find some pedagogical introductions e.g.\ in refs.~\cite{Webber:1986mc,Sjostrand:2006su}.
Here, we simply need to recall few basic features of SMC programs.

First, an MC starts from a kinematic configuration (``hard'') which is
generated according to an exact LO computation. Usually such configuration is
that of a $2\to 2$ partonic process. The final-state multiplicity is then iteratively
increased, by letting each initial- and final-state parton branch into a
couple of partons with a probability related to a Sudakov form factor. Thus,
if at a given stage of the shower, the scattering process is described by $m$
partons, the algorithm decides with a certain probability whether branching
is over at this stage, or further branchings will take place.  In the latter
case, one of the $m$ partons splits into a pair, generating an $m+1$ body
final state. Thus, the algorithm defines a mapping
\begin{equation}
\Kinmpo \; \stackrel{(\ctindex)}{\Longleftrightarrow} \;
 \lg \BKinm^{(\ctindex)}, \Rad^{(\ctindex)} \rg\,,
\label{eq:phspfactmc}
\end{equation}
that is fully analogous to the mapping (\ref{eq:phspfact}). Also in this case
there is one mapping for each singular region, where the singular region is
associated with the parton that undergoes the splitting.  Observe that
mappings defined in eq.~(\ref{eq:phspfactmc}) act non-trivially also on the
momenta of the partons that do not undergo any splitting process. This is due
to the fact that momentum conservation must be restored after branching, an
operation that is usually referred to as ``momentum reshuffling'' in the SMC
jargon. Also the value of the momentum fraction of the incoming partons may
require readjustment, which leads to the fact that the value of the
luminosity used for the cross-section computation does not correspond exactly
to what one would have used if the $(m+1)$-particle matrix element had been
computed with standard methods. These readjustments usually amount to
corrections beyond the (leading log) Monte Carlo accuracy, but should be
analyzed carefully if one aims at NLO accuracy.

\subsection{Including NLO corrections into Monte Carlo programs}
\label{subsec:matching}
The embedding of a NLO computation into an MC framework, as first clarified
in ref.~\cite{Frixione:2002ik}, aims at reaching NLO accuracy for inclusive
observable, maintaining the leading logarithmic accuracy of the shower
approach. This requires that the hardest emission that is generated has the
correct distribution also far from the collinear directions, and that
integrated quantities around the soft and collinear directions have NLO
accuracy. This requirements are met in the \MCatNLO{} approach by carefully
tracing the differences of the MC simulation relative to the exact NLO
one. The similarity of the mappings (\ref{eq:phspfactmc})
and~(\ref{eq:phspfact}) are the starting point for this task. The shower
algorithm is analyzed to determine its own approximate NLO structure in the
subtraction framework, in order to determine unambiguously the difference
with the exact NLO formulae.

In the \POWHEG{} approach, one performs
the generation of the hardest event with NLO accuracy, in a framework that
does not depend upon the SMC's shower algorithm. This is why it is fully
independent from the SMC.
Furthermore, the subsequent showers takes place at softer transverse momenta,
and thus affects infrared-safe observables only at the
next-to-next-to-leading order (NNLO). Thus,
the matching problem considerably simplifies, since it no longer requires
a detailed examination of the properties of the SMC. 

\subsection{\POWHEG}
\label{subsec:POWHEG}
In the \POWHEG{} formalism, the generation of the hardest emission is
performed first, using full NLO accuracy, and using the SMC to generate
subsequent radiation.  We give here a simple illustration of the method,
ignoring, for the moment, the complications due to the presence of several
singular regions in the NLO cross section.  We begin by defining
\begin{eqnarray}
\bar{B}(\Kinn)&=&B(\Kinn)+V(\Kinn)
\nonumber \\
&+&\lq \int d\Rad\lq R(\Kinnpo)-C(\Kinnpo) \rq
+\int \frac{dz}{z} \lq G_\splus(\Kinncp)+ G_\sminus (\Kinncm) \rq 
\rq^{\BKinn=\Kinn}\mbox{\hskip -1cm},\mbox{\hskip 1cm}
\end{eqnarray}
where we have assumed that all the $\Kinnpo$, $\Kinncpm$ are expressed in
terms of the barred variables.  Next we introduce the Sudakov form
factor\footnote{Torbj\"orn Sj\"ostrand has pointed out to us that a similar
Sudakov form factor is also used in \PYTHIA{} for weak vector-bosons decay and
production, in order to implement a matrix-element matching for the first
emission in the shower, see refs.~\cite{Sjostrand:2006su,Bengtsson:1986hr}.}
\begin{equation}
\Delta\(\Kinn, \pt\)=\exp\lg
- \int  \frac{\big[d\Rad\, R(\Kinnpo) \; \stepf\!\(\kt\(\Kinnpo\)-\pt\)
\big]^{\BKinn=\Kinn}}{B(\Kinn)}\,
\rg.
\end{equation}
The function $\kt\(\Kinnpo\)$ should be equal, near the singular limit, to the
transverse momentum of the emitted parton relative to the emitting one.
The \POWHEG{} cross section for the generation of the hardest event
is then
\begin{equation}
\label{eq:POWHEGsigmasimple}
d\sigma= \bar{B}(\Kinn)\, d \Kinn
 \Bigg\{ \Delta\(\Kinn,\ptmin\)
+
\Delta\(\Kinn,\kt\(\Kinnpo\)\)\, \frac{
 R\(\Kinnpo\)}{ B\!\(\Kinn\)} d\Rad
\Bigg\}_{\BKinn=\Kinn},\phantom{aa}
\end{equation}
where it is assumed that $\Kinnpo$ is parametrized in terms of $\Rad$ and $\Kinn$,
and that values of $\kt\(\Kinnpo\)<\ptmin$ are not allowed. The
cross section~(\ref{eq:POWHEGsigmasimple}) has
the following properties:
\begin{itemize}
\item At large $\kt$ it coincides with the NLO cross section up to NNLO terms.
\item It reproduces correctly the value of infrared safe observables at the NLO.
      Thus, also its integral around the small $\kt$ region has NLO accuracy.
\item At small $\kt$ it behaves no worse than standard Shower Monte Carlo generators.
\end{itemize}
Thus, it fulfills the requirement of the previous subsection for the inclusion
of NLO corrections in an SMC.

As it stands, the \POWHEG{} formula (\ref{eq:POWHEGsigmasimple}) can be used
to feed a SMC program, that will perform all subsequent (softer) showers and
hadronization.
If the SMC is ordered in $\pt$, we simply require that the shower is started
with an upper limit on the scale equal to the $\kt$ of the \POWHEG{} event.
In case the SMC uses a different ordering variable, a problem
arises, since the \POWHEG{} cross section requires the emissions with
higher $\kt$ to be suppressed in the SMC. This problem typically arises
when interfacing \POWHEG{} to angular ordered SMC's. It is dealt with
by vetoing emissions with larger $\kt$ in the shower, and by introducing vetoed
truncated showers (see ref.~\cite{Nason:2004rx}),
that compensate for the fact that in angular ordered
shower the hardest emission may not be the first.

Modern SMC programs, such as \HERWIG{} and \PYTHIA{}, have
the capability of generating a vetoed shower. This is not the case
for the vetoed truncated showers. We point out, however, that the
need of vetoed truncated showers
is not specific to the \POWHEG{} method. As discussed in ref.~\cite{Nason:2004rx},
it also emerges naturally when interfacing standard
matrix element calculations with parton shower, as in the approach
of ref.~\cite{Catani:2001cc}.
At present, there is no evidence that the effect of vetoed truncated showers may
have any practical important.

\section{The \POWHEG{} method}
\label{sec:powhegmth}
In order to implement the \POWHEG{} method one must specify
the separation of the singular regions, and the kinematics that associates a
given $(n+1)$-body singular region with an $n$-body one. 
We discuss \POWHEG{} in the framework of a generic subtraction formalism.

When one aims at the construction of an event generator, flavour should be
carefully tracked, since different flavour structures always give rise to
different events. We thus distinguish the contributions to the cross section
also by their flavour structures, which are determined by the flavours of the
incoming and outgoing partons. We call equivalent two flavour structures that
differ only by a permutation of final-state partons. In particular, we label
with the index $\fb$ the flavour structure of the $n$-body processes and
write $B^\fb$ and $V^\fb$ for the various Born and soft-virtual
contributions.

We label with $\ctindr$ a particular contribution to the real cross section
that is singular in only one singular region of integration and has a
specific flavour structure. Thus, to each $\ctindr$ corresponds one and only
one singular region.  We then write
\begin{equation}
R =\sum_\ctindr R^\ctindr\,.
\end{equation}
A similar separation also holds for the counterterms, so that they are also
labelled by an index $\ctindr$.

In the FKS case, for example,
the $\ctindr$ contributions are obtained by first separating the real contribution
$R$ into the sum of all its flavour components. For each flavour component,
one constructs the $\Sfun$ functions, according to the procedure of
section~\ref{sec:Sfun}, and then multiplies it by the
factors $\Szi$ or $\Soij$.  In the CS case, for each
flavour component of the real contribution, one defines
\begin{equation}
\mathcal{S}_\alpha=\frac{{\cal D}_\alpha}{\sum_\beta {\cal D}_\beta}\;,
\end{equation}
where $\alpha$ ranges in the set of dipoles ${\cal D}$ with the same
flavour structure.

To each contribution $\ctindr$ we can associate an underlying $n$-body
process, with a specific flavour structure.  The association is performed as
follows. If the singular region is collinear, the two collinear particles are
merged into a single particle in such a way that flavour is conserved.  In
particular, a $gg$ pair is merged into a $g$, a $q\bar{q}$ pair is merged
into a $g$, and a $qg$ ($\bar{q}g$) pair is merged into a $q$ ($\bar{q}$). If
the singular region is soft, the soft gluon is removed.
Observe that, for non singular limits (for example, in case two
quarks, or a quark and an antiquark of different flavour become collinear,
or in case a quark becomes soft) the flavour structure of the underlying
$n$-body process is undefined.

The factorization
remnants also have a flavour structure. We label it with
the index $\ctindpm$, and also for these configurations there is an
underlying $n$-body flavour structure.  We call $\{\ctindr|\fb\}$ and
$\{\ctindpm|\fb\}$ the set of all values of the indices $\ctindr$ and
$\ctindpm$ that have the flavour structure of the underlying
$n$-body-configuration equal to $\fb$.

We rewrite eq.~(\ref{eq:dsdOsub2}) according to the notation of this
section (making use of a straightforward extension of the context square
brackets introduced in eq.~(\ref{eq:context}))
\begin{eqnarray}
\label{eq:dsdOsub3}
 \langle O \rangle&=&\sum_\fb \left[
\langle O \rangle_B^\fb + \sum_{\frindsing \in \{\ctindr|\fb\}}
\langle O \rangle_R^\frindsing + \sum_{\ctindp \in \{\ctindp|\fb\}}
\langle O \rangle_{G_\splus}^\ctindp
 + \sum_{\ctindm \in \{\ctindm|\fb\}} \langle O \rangle_{G_\sminus }^\ctindm\right],
\\
\label{eq:OB} \label{eq:bbborn0}
\langle O \rangle_B^\fb&=&
\int d\Kinn\,
 O_n\!\(\Kinn\)\,\Big[B\(\Kinn\)+V\(\Kinn\)\Big]_\fb\;,
\\
\label{eq:OF}
 \langle O \rangle_{G_{\splusminus}}^\ctindpm &=&
\int d\Kinncpm\, O_n\!\(\BKinn\)\,
G_{\splusminus}^\ctindpm\!\(\Kinncpm\) \;,
\\
\label{eq:OR}
\langle O \rangle_R^\frindsing&=&  \int d\Kinnpo
 \,\Big[
O_{n+1}\!\(\Kinnpo\) R\!\(\Kinnpo\) -
 O_n\!\(\BKinn\)C \!\(\Kinnpo\)\Big]_\frindsing
\,.\phantom{aaaaa}
\end{eqnarray}
According to ref.~\cite{Nason:2004rx}, we now perform the following
manipulation
\begin{eqnarray} &&
\langle O \rangle_R^\frindsing = \langle O \rangle_{R,n}^\frindsing
+\langle O \rangle_{R,n+1}^\frindsing\;,
 \\
 \label{eq:ORn}
\langle O \rangle_{R,n}^\frindsing &=& \left[
\int d\Kinnpo\; O_n\!\(\BKinn\)
 \,\Big\{ R\(\Kinnpo\) -
 C\(\Kinnpo\)\Big\}\right]_\frindsing \;,
 \\ \label{eq:powhegradterm}
\langle O \rangle_{R,n+1}^\frindsing&=& \left[
\int d\Kinnpo\,
R\(\Kinnpo\) \,\Big\{
O_{n+1}\!\(\Kinnpo\) -
 O_n\!\(\BKinn\)\Big\}\right]_\frindsing \;.
\end{eqnarray}
The term of eq.~(\ref{eq:powhegradterm}) involves
real radiation. All other terms (i.e.~(\ref{eq:OB}),
(\ref{eq:OF}) and~(\ref{eq:ORn})), have $n$-body kinematics and, according to
ref.~\cite{Nason:2004rx}, should be all lumped together into an $n$-body
kinematics term, that was called $\bar{B}$.
However, we should now carefully distinguish the contributions to $\bar{B}$
according to their flavour structure.  We first rewrite eqs.~(\ref{eq:OF}),
(\ref{eq:ORn}) and~(\ref{eq:powhegradterm}) in the following
form
\begin{eqnarray}
 \label{eq:bbf0}
 \langle O \rangle_{G_{\splusminus}}^\ctindpm &=&
\int d\BKinn\,O_n\!\(\BKinn\)\, \frac{dz}{z}
\, G^\ctindpm_{\splusminus}\(\Kinncpm\)  \;,
 \\ \label{eq:bbr0}
\langle O \rangle_{R,n}^\frindsing &=& \lq
\int d\BKinn\,  O_n\!\(\BKinn\) \,
  d\Rad\,\lg R\(\Kinnpo\) - C\(\Kinnpo\) \rg\rq_\frindsing,
 \\ \label{eq:rad0}
\langle O \rangle_{R,n+1}^\frindsing&=& \lq
\int d\BKinn\, d\Rad\,
R\(\Kinnpo\) \lg
O_{n+1}\(\Kinnpo\) - O_n\!\(\BKinn\)\rg \rq_\frindsing.
\end{eqnarray}
According to ref.~\cite{Nason:2004rx}, we can write the $\bar{B}$ functions,
one for each flavour configuration, as
\begin{eqnarray}
\label{eq:bbdef}
\bar{B}^\fb(\Kinn) &=& \lq
B\(\Kinn\)+V\(\Kinn\)\rq_\fb
+ \sum_{\frindsing\in\{\frindsing|\fb\}} \int \Big[  d\Rad\,\lg R\(\Kinnpo\) -
 C\(\Kinnpo\)\rg\Big]_\frindsing^{\BKinn^\frindsing=\Kinn}
\nonumber \\
&&+ \sum_{\ctindp\in\{\ctindp|\fb\}} \int \frac{dz}{z}
\,G_\splus^\ctindp\(\Kinncp\)
+ \sum_{\ctindm\in\{\ctindm|\fb\}} \int \frac{dz}{z}
\,G_\sminus^\ctindm\(\Kinncm\) \;,
\end{eqnarray}
so that
\begin{equation}
\int d\Kinn\, 
 O_n\!\(\Kinn\)\,\bar{B}^\fb(\Kinn)=
\langle O \rangle_B^\fb + \!\!\!\!\! \sum_{\frindsing \in \{\ctindr|\fb\}}
\langle O \rangle_{R,n}^\frindsing +\!\!\!\!\! \sum_{\ctindp \in \{\ctindp|\fb\}}
\langle O \rangle_{G_\splus}^\ctindp
 +\!\!\!\!\!
 \sum_{\ctindm \in \{\ctindm|\fb\}} \langle O \rangle_{G_\sminus }^\ctindm\,,
\end{equation}
and
\begin{eqnarray}\label{eq:Owithbbar}
 \langle O \rangle
&=&\sum_\fb \int d\Kinn\,
 O_n\!\(\Kinn\)\,\bar{B}^\fb(\Kinn)
\nonumber \\
&+&
 \sum_\frindsing \lq
\int d\BKinn\ d\Rad\,
R\(\Kinnpo\) \lg
O_{n+1}\(\Kinnpo\) - O_n\!\(\BKinn\)\rg \rq_\frindsing\;.
\end{eqnarray}
We now define the Sudakov form factors
\begin{equation}
\label{eq:suddef}
\Delta^\fb(\Kinn,\pt)=\exp\lg
-\sum_{\frindsing\in\{\frindsing|\fb\}}
\int \frac{\Big[  d\Rad\, R\(\Kinnpo\)\, \stepf\(\kt(\Kinnpo)-\pt\)
\Big]_\frindsing^{\BKinn^\frindsing=\Kinn}}{ B^\fb\(\Kinn\)}
\rg\;.
\end{equation}
Notice that the identification $\BKinn^\ctindr=\Kinn$ is a sensible one only
if the underlying $n$-body-process flavour structure of $\ctindr$ is equal to
$\fb$.  In eq.~(\ref{eq:suddef}), $\kt^\ctindr$ is a function of the kinematics
variables that depends upon the particular singular region we are
considering (its $\ctindr$ index is omitted in eq.~(\ref{eq:suddef})
thanks to the context convention).
For initial-state collinear singularities, we require that
$\kt$ is proportional to the transverse momentum of the emitted parton
with respect to
the beam axis in the collinear limit.  For final-state collinear
singularities, assuming that the singular region corresponds to momenta $k_i$
and $k_j$ becoming collinear, we take as $\kt$ the (spatial) component of
$k_i$ (or equivalently $k_j$) orthogonal to the sum $\vec{k}_i+\vec{k}_j$.
In the following we assume that the transverse momentum is computed in the
CM frame of the colliding partons.

The factorization and renormalization scales adopted in the definition of
${\bar B}$, eq.~(\ref{eq:bbdef}), and in the definition of the Sudakov form
factors, eq.~(\ref{eq:suddef}), are different.  In the definition of ${\bar
B}$ one adopts a choice that is appropriate to the Born cross section.  In
the Sudakov exponents one must instead adopt a scale of the order of $\kt$.
In section~\ref{sec:sudak} we show that, with the above choice of scales, the
Sudakov form factor of eq.~(\ref{eq:suddef}) is equal, at least to the
leading logarithmic (LL) level, to the DDT~\cite{Dokshitzer:1978hw} Sudakov
form factor, and that, in some cases, with a simple prescription, one can
reach NLL accuracy.

The formula for the full \POWHEG{} cross section is
\begin{eqnarray}
\label{eq:POWHEGsigma}
d\sigma&=&\sum_\fb \bar{B}^\fb(\Kinn)\, d \Kinn
 \Bigg\{ \Delta^\fb\!\(\Kinn,\ptmin\)
\nonumber\\
&+&\!\!\!\!\!\!\sum_{\frindsing\in\{\frindsing|\fb\}}\!\!\!\!\!\!
 \frac{\Big[  d\Rad\;\stepf\(\kt-\ptmin\)
\Delta^\fb\!\(\Kinn,\kt\)\, R\(\Kinnpo\)
\Big]_\frindsing^{\BKinn^\frindsing=\Kinn}}{ B^\fb\!\(\Kinn\)}
\Bigg\},\phantom{aa}
\end{eqnarray}
where, for ease of notation, we have dropped the $\Kinnpo$ argument
in $\kt^\frindsing$.
The $\ptmin$ value introduced here is a lower cut-off on the transverse
momentum, that is needed in order to avoid to reach unphysical values of the
strong coupling constant and of the parton-density functions.

As discussed at the beginning of section~\ref{sec:SubtractionFormalism},
in case the $n$-body
cross section possesses singular regions, the observable $O_{n+1}$
should  vanish fast enough if $\Kinnpo$ approaches
two singular regions at the same time. Notice that, in the \POWHEG{}
cross section given in eq.~(\ref{eq:POWHEGsigma}),
the observable function $O$ has disappeared, so that this restriction
is no longer apparent in the formula.
However, thanks to the partition of the different singular regions,
it is sufficient to apply to the $\bar{B}$ function a damping factor
that suppresses the regions where the $n$-body configuration becomes
singular, in order to get a finite result. In this way, the \POWHEG{} approach
more closely resembles the standard Monte Carlo generators, where the hard
leading-order matrix element for jet production
is appropriately cut off in order to get a finite total cross section.

\subsection{Transverse-momentum ordering}\label{sec:trmomord}
A word of caution has to be said with regard to the separation of the
various singular contributions in \POWHEG{}, in cases where the underlying
$n$-body cross section also possesses singular regions. In standard NLO
calculations, these regions are avoided by simply requiring that the
physical observables one computes should be finite for the $n$-body term.
In \POWHEG{}, this requirement is in general not sufficient to guarantee
consistency. We should also require that the $\kt$ of the generated radiation
should not be harder than all the $\kt$'s associated with the underlying
Born kinematics. More precisely, we should require that radiation
with $\kt$ larger than the smallest $\kt$ of the underlying $n$-body
process should be suppressed. The separation of $R$ into contributions
from the various singular regions achieves to some extent this purpose:
$R^\frindsing$ is suppressed in any singular region different from
$\frindsing$. However, consistency with the treatment of soft singularities
requires that the suppression should be based upon $\kt$, rather than,
for example, virtuality. Thus, in the FKS case (for example), a most
appropriate choice for the $d_i^\splusminus$ is
\begin{eqnarray}\label{eq:ditmord}
d_i^\splusminus&=&\left(E_i\right)^{2b} 2^b (1\mp \cos \theta_i)^b\,,
\\ \label{eq:dijtmord}
d_{ij}&=&\(\frac{E_i E_j}{E_i+E_j}\)^{2b} 2^b \(1-\cos\theta_{ij}\)^b\,,
\end{eqnarray}
that correspond to the square of the transverse momentum to the power $b$, rather than
the form suggested in eqs~(\ref{eq:dispm}) and (\ref{eq:di2}).\newpage
\vspace*{-0.5cm}
\begin{minipage}{\textwidth}
\subsection{NLO accuracy of the \POWHEG{} formula}
In ref.~\cite{Nason:2004rx}, it was argued that the \POWHEG{} formula yields NLO
accuracy for infrared-finite observables, and that subsequent showering
by an SMC program does not spoil this conclusion.
The second point is a simple consequence of the fact that no radiation
with $\kt$ larger than that generated by \POWHEG{} is allowed
in the subsequent shower, and it requires no further discussion.
We wish instead to address the first point more rigorously.
In other words,
we want to show in detail that formula~(\ref{eq:POWHEGsigma}), when used
to compute an infrared-safe observable, yields the
correct NLO accuracy. The reader that finds this conclusion already obvious
can safely skip this section.

In order to ease the notation, we will drop the $\stepf\(\kt-\ptmin\)$
factor in the \POWHEG{} formula, always assuming that this factor
is present when there is real radiation.

If we apply formula~(\ref{eq:POWHEGsigma}) to an infrared-safe observable
$O$, we have
\begin{eqnarray}
\label{eq:POWHEGsigmaO}
\langle O \rangle &=&\sum_\fb \int d \Kinn \,\bar{B}^\fb(\Kinn)
 \Bigg\{ \Delta^\fb\!\(\Kinn,\ptmin\) O_n\!\(\Kinn\)
\nonumber\\
&&+\!\!\sum_{\frindsing\in\{\frindsing|\fb\}}\!\!\!\!\!\!
 \frac{\Big[  {\displaystyle \int} d\Rad\;
\Delta^\fb\!\(\Kinn,\kt\)\, R\(\Kinnpo\) O_{n+1}\!\(\Kinnpo\)
\Big]_\frindsing^{\BKinn^\frindsing=\Kinn}}{ B^\fb\!\(\Kinn\)}
\Bigg\}
\nonumber\\
&=&\sum_\fb \int d \Kinn \bar{B}^\fb\!\(\Kinn\)\,
\nonumber \\
 && \times\left\{\left[\Delta^\fb\!\(\Kinn,\ptmin\)
+\sum_{\frindsing\in\{\frindsing|\fb\}}\!\!\!\!
 \frac{\Big[{\displaystyle \int} d\Rad\;
\Delta^\fb\!\(\Kinn,\kt\)\, R\(\Kinnpo\)
\Big]_\frindsing^{\BKinn^\frindsing=\Kinn}}{ B^\fb\!\(\Kinn\)}\right]
 O_n (\Kinn)\right.
\nonumber\\ \label{eq:Opowheg}
&&+\!\!\!\!\!\! \left.\sum_{\frindsing\in\{\frindsing|\fb\}}\!\!\!\!\!\!
 \frac{\Big[  {\displaystyle \int} d\Rad\;
\Delta^\fb\!\(\Kinn,\kt\)\, R\(\Kinnpo\)
 \(O_{n+1}\!\(\Kinnpo\)-O_{n}\!\(\Kinn\)\) 
\Big]_\frindsing^{\BKinn^\frindsing=\Kinn}}{ B^\fb\!\(\Kinn\)}
\right\}\!,\phantom{aaaaa}
\end{eqnarray}
where, in the second equality, we have simply added and subtracted the
same term proportional to $R\(\Kinnpo\)O_n (\Kinn)$.
We now show that
the term in the large squared bracket in the third member of
eq.~(\ref{eq:Opowheg})
is equal to 1.
In fact
\begin{eqnarray}
\sum_{\frindsing\in\{\frindsing|\fb\}}\!\!\!\!\!\!
 \frac{\Big[{\displaystyle \int} d\Rad\;
\Delta^\fb\!\(\Kinn,\kt\)\, R\(\Kinnpo\)
\Big]_\frindsing^{\BKinn^\frindsing=\Kinn}}{ B^\fb\!\(\Kinn\)} &&
\nonumber \\
&&\hspace{-6.cm}=\int_\ptmin^\infty d\pt\;
\sum_{\frindsing\in\{\frindsing|\fb\}}\!\!\!\!\!\!
 \frac{\Big[{\displaystyle \int} d\Rad\;
\delta(\kt-\pt)\,\Delta^\fb\!\(\Kinn,\pt\) \, R\(\Kinnpo\)
\Big]_\frindsing^{\BKinn^\frindsing=\Kinn}}{ B^\fb\!\(\Kinn\)}
\nonumber \\
&&\hspace{-6.cm}=-\int_\ptmin^\infty d\pt\;\Delta^\fb\!\(\Kinn,\pt\)
\frac{d}{d\pt} 
\sum_{\frindsing\in\{\frindsing|\fb\}}\!\!\!\!\!\!
 \frac{\Big[{\displaystyle \int} d\Rad\;
\stepf\!\(\kt-\pt\) \,R\(\Kinnpo\)
\Big]_\frindsing^{\BKinn^\frindsing=\Kinn}}{ B^\fb\!\(\Kinn\)} 
\nonumber \\
&&\hspace{-6.cm}=
\int_\ptmin^\infty d\pt\; \frac{d}{d\pt} \Delta^\fb\!\(\Kinn,\pt\)
=1-\Delta^\fb\!\(\Kinn,\ptmin\)\;,
\end{eqnarray}
\end{minipage}
where we have used the fact that $\Delta^\fb\!\(\Kinn,\infty\)=1$.
Furthermore, in the last term
in the large curly bracket of eq.~(\ref{eq:POWHEGsigmaO}), small $\kt$
values in the integral are suppressed by the
$O_{n+1}(\Kinnpo)-O_{n}(\Kinn)$ factor, and therefore we can replace
$\Delta^\fb\! \to 1$ and $B \to \bar{B}$ up to higher orders in $\as$.
Equation~(\ref{eq:POWHEGsigmaO}) thus reduces to
\begin{eqnarray}\label{eq:PWGNLO}
\langle O \rangle &=& \sum_\fb \int d \Kinn \Bigg\{\bar{B}^\fb\!\(\Kinn\)\,
O_n(\Kinn) \nonumber \\
 &+&\sum_{\frindsing\in\{\frindsing|\fb\}}\!\!
 \lq {\displaystyle \int} d\Rad\;
R\(\Kinnpo\) \(O_{n+1}(\Kinnpo)-O_{n}(\Kinn)\)
\rq_\frindsing^{\BKinn^\frindsing=\Kinn}\Bigg\}
\;,
\end{eqnarray}
up to NNLO corrections. The restriction
$\stepf(\kt-\ptmin)$ can now be dropped
from the $d\Rad$ integration, its effect being
suppressed by powers of $\ptmin$, and formula eq.~(\ref{eq:PWGNLO})
is immediately found to agree with eq.~(\ref{eq:Owithbbar}),
thus concluding our proof.

\subsection{Practical implementation of the \POWHEG{} formulae}
The \POWHEG{} cross section (eq.~(\ref{eq:POWHEGsigma})) looks very complex,
but, in fact, from a numerical point of view, it is quite easy to implement
using few well-known Monte Carlo techniques.

\subsubsection{Generation of the Born variables}
\label{sec:gen_born_var}
To begin with, we must generate a Born-like configuration (a
point in the $\Kinn$ space) and a value of the index $\fb$, with a
probability given by $\bar{B}^\fb(\Kinn) \, d \Kinn$. The standard Monte
Carlo technique used in this cases is the hit-and-miss procedure: one finds
an upper bound to the cross section, generates randomly the phase-space
point, and then accepts it with a probability equal to the ratio of the value
of the cross section at the given point over the upper bound value. This
technique is inadequate for our case, since each evaluation of the $\bar{B}$
function requires an integration over the radiation variables. We thus
proceed as follows. For each singular region, we parametrize the radiation
variables $\Rad$ in terms of a set of three variables in the unit cube, that
we call $\Xrad=\left\{\Xrad^{(1)},\Xrad^{(2)},\Xrad^{(3)}\right\}$.
Similarly, the $z$ variable in the
collinear remnants is parametrized in terms of one of these three variables,
that we take to be $\Xrad^{(1)}$. We then introduce the function
\begin{eqnarray}
\label{eq:btdef}
\tilde{B}^\fb(\Kinn,\Xrad) &=& \lq
B\(\Kinn\)+V\(\Kinn\)\rq_\fb \nonumber \\
&+& \sum_{\frindsing\in\{\frindsing|\fb\}} \left[
\left|\frac{\partial\Rad}{\partial\Xrad}\right|\,\lg R\(\Kinnpo\) -
 C\(\Kinnpo\)\rg\right]_\frindsing^{\BKinn^\frindsing=\Kinn}
\nonumber \\
&+& \sum_{\ctindp\in\{\ctindp|\fb\}}  \frac{1}{z}
\left|\frac{\partial z}{\partial \Xrad^{(1)}}\right|
\,G_\splus^\ctindp\(\Kinncp\)
+ \sum_{\ctindm\in\{\ctindm|\fb\}} \frac{1}{z}
\left|\frac{\partial z}{\partial \Xrad^{(1)}}\right|
\,G_\sminus^\ctindm\(\Kinncm\) ,\nonumber\\
\end{eqnarray}
so that
\begin{equation}
\bar{B}^\fb(\Kinn)=
\int_0^1 d\Xrad^{(1)}\int_0^1d\Xrad^{(2)}\int_0^1d\Xrad^{(3)}\;
\tilde{B}^\fb(\Kinn,\Xrad)\,, 
\end{equation}
and we define
\begin{equation}
\tilde{B}(\Kinn,\Xrad)=\sum_\fb \tilde{B}^\fb(\Kinn,\Xrad)\,.
\end{equation}
There are computer programs that, after performing a single integration on a
given function, can efficiently generate points in the integration range,
distributed according to the integrand function. One such popular program is
the \BASES{}/\SPRING{} package~\cite{Kawabata:1995th}. The adaptive Monte
Carlo integration routine \BASES{} performs the integration of the
non-negative function, and stores the necessary intermediate results.  The
routine \SPRING{} uses these information to generate unweighted events.  In
our case, one integrates the $\tilde{B}(\Kinn,\Xrad)$ function in the full
$\(\Kinn,\,\Xrad\)$ space using \BASES{}. Then one generates
$\(\Kinn,\,\Xrad\)$ points using \SPRING{}. For each generated phase-space
point, one chooses an $\fb$ value with a probability equal to
$\tilde{B}^\fb(\Kinn,\Xrad)/\tilde{B}(\Kinn,\Xrad)$. At this point, the
$\Xrad$ values are discarded, and one has generated the $\(\Kinn,\,\fb\)$
values with probability proportional to $\bar{B}^\fb(\Kinn)$. In essence, by
doing a single $(n+1)$-body phase-space integration, one is able to
generate the Born configuration with reasonable efficiency.

As already pointed out, if the Born configuration possesses
singular regions, the $n$-body phase space should be suitably constrained,
in order to avoid them. This is also the case in standard SMC programs:
when one deals with cross sections with singular matrix elements,
typically cross sections that include the production of a light parton,
one specifies a transverse-momentum cut on its momentum, in order
to get a finite total cross section. Alternatively, a weight $W(\Kinn)$
should be attached to the $\tilde{B}$ function that suppresses the
$n$-body singular regions, so that the integral of $W\times \tilde{B}$
is finite. Events are then generated using $W\times \tilde{B}$
instead of $\tilde{B}$, and a weight $W^{-1}(\Kinn)$ should be attached
to each event. For example, in order to generate a sample of $Z+{\rm jet}$
events, knowing that we will select jets with transverse energy greater
than $E_T^{\rm cut}$, we can restrict the Born events so that
$E_T>E_T^{\rm cut}/2$, $E_T$ being the transverse energy of the
radiated parton. Alternatively, we weight $\tilde{B}$ with $E_T^2$, attach
the weight $1/E_T^2$ to the generated event, and unweight the events
after the event sample (including cuts) is generated. The cut on the
transverse energy of the associated jet should effectively cut off the events
with low parton $E_T$, so that the unweighting will be really possible. 

\subsubsection{Generation of the hardest-radiation variables}
\label{sec:gen_rad_var}
Given the Born kinematics $\(\Kinn,\,\fb\)$, we must now generate the
hardest-radiation configuration, characterized by $\(\frindsing,\,\Rad^\frindsing\)$,
with $\frindsing\in\{\frindsing|\fb\}$, with probability
\begin{equation} 
\label{eq:radprob}
\lq  \;
\frac{R\(\Kinnpo\)}{ B^\fb\!\(\Kinn\)}\;
\Delta^\fb\!\(\Kinn,\kt\(\Kinnpo\)\) \rq_\frindsing^{\BKinn^\frindsing=\Kinn}
d\Rad^\frindsing\,.
\end{equation}
The Sudakov form factor can be written as
\begin{equation}
\Delta^\fb(\Kinn,\pt)=
\prod_{\frindsing\in\{\frindsing|\fb\}}\Delta^\fb_\frindsing(\Kinn,\pt)\,,
\end{equation}
where
\begin{equation}
\label{eq:suddef_ar}
\Delta^\fb_\frindsing(\Kinn,\pt) = 
\exp\lg -\left[ \int d\Rad\,
\frac{R\(\Kinnpo\)}{B^\fb\(\Kinn\)} \, \stepf\(\kt(\Kinnpo)-\pt\) 
\right]_\frindsing^{\BKinn^\frindsing=\Kinn}\rg.
\end{equation}
Under these conditions, the problem of generating the radiation variables
according to eq.~(\ref{eq:radprob}) can be reduced to the problem of
generating them with probabilities
\begin{equation} 
\label{eq:radprobfb}
\lq \frac{R^\frindsing\(\Kinnpo\)}{ B^\fb\!\(\Kinn\)}\; \Delta^\fb_\frindsing
\!\(\Kinn,\kt\(\Kinnpo\)\) \rq^{\BKinn^\frindsing=\Kinn} d\Rad^\frindsing\,,
\end{equation}
by using the highest-bid method, illustrated in
appendix~\ref{sec:highest_pt}. We are thus left with the problem of
generating radiation variables according to eq.~(\ref{eq:radprobfb})
for a fixed value of $\frindsing$. This
problem can be dealt with using the veto technique, illustrated in
appendix~\ref{sec:veto_technique}.  In order to use this technique, we need a
sufficiently simple upper bounding function
\begin{equation}
\label{eq:uboundF}
\lq \frac{R^\frindsing\(\Kinnpo\)}{ B^\fb\!\(\Kinn\)
}\rq^{\BKinn^\frindsing=\Kinn} \le F(\Rad^\frindsing,\Kinn).
\end{equation}
This can be found by taking the singular limit of the left hand side of
eq.~(\ref{eq:uboundF}), that has, in general, a form suggested by the
factorization theorem, and by elementary properties of the parton densities
in the case of initial-state singular regions.  Once the functional form of
$F$ is guessed, its normalization is found by scanning the $\Kinnpo$ phase
space.

Within a given subtraction method, one has typically two kinds of upper
bounding functions: one for final-state radiation and one for initial-state
radiation. In section~\ref{sec:examples} we illustrate explicit forms for $F$
in the FKS and CS frameworks, for both initial- and final-state radiation.

Notice that, in general, it is not necessary to separate out all $\frindsing$
regions in order to apply the veto method. In many cases several regions can
be group together, thus simplifying the generation algorithm (see
section~\ref{sec:examples}).

\subsection{Sudakov form factors and NLL soft gluon resummation}%
\label{sec:sudak} 
The purpose of the \POWHEG{} method is to reach NLO accuracy for inclusive
quantities, and leading logarithmic (LL) accuracy for exclusive final states.
In this section we address the following question: to what extent we can do
better than LL accuracy in the \POWHEG{} framework? First of all, the
\POWHEG{} method deals with the hardest emission only. Subsequent emissions
are handled by the shower Monte Carlo to which \POWHEG{} is interfaced, and
will have, in general, only LL accuracy. However, exclusive observables that
are especially sensitive to the hardest emission will benefit from an
improved \POWHEG{} accuracy.

In this section, we show how to improve the logarithmic accuracy of
\POWHEG. We show that in many cases this improvement requires very minor
adjustments of the \POWHEG{} Sudakov form factor, and that in general, NLL
accuracy at least in the large $N_c$ limit (where $N_c$ is the number of
colours) should be easy to achieve. In all this section, for ease of notation,
$k$ will denote the momentum of the soft gluon.

The results of this section can be summarized as follows:
\begin{enumeratenumeric}
  \item \label{en:ktordmu} The \POWHEG{} Sudakov form factor is accurate at
  the LL level, provided that the strong coupling constant and the parton
  density functions in the Sudakov exponent are evaluated at a scale of order
  $\kt^2$.
  
  \item \label{en:3colpar} In case of processes involving no more than 3
  coloured partons (in the initial or final state), NLL accuracy is achieved
  by replacing the strong coupling constant in the Sudakov exponent with
  \begin{equation}
    \label{eq:nllcond} \as \rightarrow A \left( \as \left( \kt^2
    \right) \right), \qquad A (\as) = \as \lg 1 + \frac{\as}{2 \pi}
    \lq \left( \frac{67}{18} - \frac{\pi^2}{6} \right) \CA - \frac{5}{9}
    \NF \rq \rg,
  \end{equation}
  where the $\overline{\tmop{MS}}$, 1-loop expression of $\as$ should be
  used. Furthermore, the parton densities in the exponent must be evaluated
  at a scale of
  order $\kt^2$. The argument of $\as$ in eq.~(\ref{eq:nllcond}) can also be
  taken equal to a function of the radiation variable that is of order
  $\kt^2$ in the soft or collinear limit, but becomes exactly equal to
  $\kt^2$ in the soft {\em and} collinear region.
  
  \item \label{en:softNLL} In case of processes involving more than 3 coloured
  partons, the procedure of item \ref{en:3colpar} is not sufficient to
  guarantee NLL accuracy. There are in fact soft (NLL) contributions that do
  exponentiate only in a matrix sense, so that, in order to deal with them
  using standard Monte Carlo techniques suited for the evaluation of ordinary
  exponential (like, for example, the veto method), one should diagonalize
  their colour structure.
  
  \item We will show that, for processes involving more than 3 coloured
  partons, the correct exponentiation of the soft (NLL) contributions
  discussed in item \ref{en:softNLL} can be easily recovered for the dominant
  terms in the large-$N_c$ limit.
\end{enumeratenumeric}
In the rest of this section we will demonstrate the above points.  We assume,
in the following, that the reader is familiar with Sudakov resummation
techniques, and in particular with ref.~\cite{Bonciani:2003nt}. We also
reassure the reader that, if she/he is willing to accept the above points,
and is not interested in implementing the $4$th point of the above
list, the rest of this section can be safely skipped.

We begin by recalling the structure of Sudakov resummation of soft gluon
effects in QCD, taking the notation and the results given in ref.
{\cite{Bonciani:2003nt}}. One usually assumes that there are kinematic
constraints that limit the emission of large transverse-momentum partons,
called Sudakov weights $u$ in {\cite{Bonciani:2003nt}}. In the \POWHEG{}
Sudakov form factor, the constraint is given by the requirement that all
radiation processes have transverse momentum less than a given $\pt$, so that
$u = \stepf\!\(\pt^2 - \kt^2\)$ in our case. We now review the ingredients
that build up the Sudakov resummation factors at NLL level. Radiation from a
final-state massless parton $i$ carries a factor $J_i (\pt)$, given in
eq.~(10) of ref.~{\cite{Bonciani:2003nt}}
\begin{equation}
  \label{eq:Jsud} 
  \log J_i (\pt) = - 4 \pi \int \frac{d^4 k}{(2 \pi)^3}\,
  \delta^+\!\(k^2\) \,\stepf\!\(\kt^2 - \pt^2\) \,\frac{\stepf(z)}{k_i \cdot
  k}\, A\!\(\as\!\(\kt^2\) \) P_i (z),
\end{equation}
where the function $A$ is defined in eq.~(\ref{eq:nllcond}). Furthermore,
\begin{equation}
 z = 1 - \frac{k\cdot n}{ k_i\cdot n},
\end{equation}
where $n$ is a timelike vector, that we take to coincide
with the time direction in the partonic CM system,\footnote{In ref.~{\cite{Bonciani:2003nt}}
one $n$ is defined for each $k_i$,
Here, for simplicity, we identify all of them.}
and $\kt$ is the transverse momentum of $k$ relative to $k_i$,
defined as
\begin{equation}
\kt^2=2(1-z)k_i\cdot k\;
\end{equation}
in \cite{Bonciani:2003nt}. It corresponds in the soft-collinear limit
to the transverse momentum of $k$ with respect to $k_i$ in the CM frame
of the incoming partons.
$P_i$ are appropriate combination of the Altarelli-Parisi splitting
functions, defined in eq.~(11) of ref.~{\cite{Bonciani:2003nt}}. The soft
(i.e.\ $z \rightarrow 1$) and collinear singularities in formula
(\ref{eq:Jsud}) give rise to logarithmic terms, with a structure (at the NLL
level)
\begin{equation}
  \label{eq:Jsudexp} 
  \log J_i (\pt) = \sum^{\infty}_{k = 1} c^{J,\tmop{LL}}_k
  \as^k\!\(\mu^2\) \log^{k + 1} \frac{\pt}{\mu} + \sum^{\infty}_{k = 1}
  c^{J,\tmop{NLL}}_k \as^k\!\(\mu^2\) \log^k \frac{\pt}{\mu},
\end{equation}
where $\mu$ is a reference scale of the order of the renormalization scale.

The logarithms arise in the following way. The collinear region of
integration generates a $\log \pt$. The soft region (i.e.\ $z \rightarrow
1$) also generates a $\log \pt$, which arises from the $1 / (1 - z)$ terms in the
$P_i (z)$ functions. The NLL expansion of $\as (\kt^2)$ in powers of $\as$
evaluated at a reference scale $\mu$ has the structure
\begin{equation}
  \label{eq:asexp} 
  \as\!\(\kt^2\) = \as\!\(\mu^2\) \left[ \sum_{j = 0}^{\infty}
  c^{\alpha, \tmop{LL}}_j  \left( \as\!\(\mu^2\) \log \frac{\kt}{\mu} \right)^j +
  \sum_{j = 0}^{\infty} c^{\alpha,\tmop{NLL}}_j \as\!\(\mu^2\) \left( \as\!\(\mu^2\)
  \log \frac{\kt}{\mu^2} \right)^j + \ldots \right],
\end{equation}
and thus generates higher powers of logarithms. The coefficients
$c^{J,\tmop{LL}}_k$ in eq.~(\ref{eq:Jsudexp}) depend upon the coefficient of
the $1 / (1 - z)$ term in $P_i$, and upon the $c^{\alpha, \tmop{LL}}_j$ terms
in eq.~(\ref{eq:asexp}). The coefficients $c^{J,\tmop{NLL}}_k$ depend
also upon the full form of $P_i$, since non-soft, collinear terms can
generate single-log contributions, and upon the $c^{\alpha, \tmop{NLL}}_j$
terms in  eq.~(\ref{eq:asexp}). Furthermore,
the $\mathcal{O} (\as^2)$ term in $A$ also
contribute to the $c^{J,\tmop{NLL}}$ coefficients. It arises from the $z
\rightarrow 1$ singular part of the Altarelli Parisi splitting function, that
has the form~{\cite{Furmanski:1980cm}}
\begin{equation}
  P^{f f} (z) = \frac{c_f}{1 - z} \as  \lg 1 + \frac{\as}{2 \pi}
  \lq \left( \frac{67}{18} - \frac{\pi^2}{6} \right) \CA - \frac{5}{9} \NF
  \rq \rg, \quad c_g = 2\, \CA, \quad c_q = \CF .
\end{equation}
This relation holds for both the spacelike and timelike splitting function.
The corresponding $\as^2$ correction is associated with a collinear and soft singularity,
and thus yields two logarithms. When
combined with the $c^{\alpha, \tmop{LL}}_j$ terms of eq.~(\ref{eq:asexp})
it gives rise to NLL contributions of order $\as^2\log^2 \pt/\mu$,
 $\as^3\log^3 \pt/\mu$, and so on.
We notice that the expression in eq.~(\ref{eq:Jsud}) can be obtained from its own
${\cal O}(\as(\mu^2))$ expansion
\begin{equation}\label{eq:Jsudas}
  \log J_i (\pt) = - 4 \pi \int \frac{d^4 k}{(2 \pi)^3}\,
  \delta^+\!\(k^2\) \,\stepf\!\(\kt^2 - \pt^2\) \,
\frac{\stepf (z)}{k_i \cdot k} \, \as\!\(\mu^2\) \, P_i (z) +
{\cal O}\(\as^2\!\(\mu^2\)\)
\end{equation}
provided one replaces $\as (\mu^2)\rightarrow A(\as(\kt^2))$.

We now notice that the \POWHEG{}
Sudakov form factor has an exact expression of order $\as$ in the exponent. Since
formula~(\ref{eq:Jsudas}) generates the dominant $\log^2 \pt$ and $\log \pt$
terms, it must also be contained already in the \POWHEG{} Sudakov exponent.
It thus follows that, with the replacement of eq.~(\ref{eq:nllcond}) in the
\POWHEG{} Sudakov, we automatically include all LL and NLL terms of the $J
(\pt)$ factors. A similar reasoning holds for the resummation of initial
state emissions\footnote{In ref.~\cite{Nason:2006hf} this case is discussed
in details, in the framework of $Z$
pair production in hadronic collisions.}, called
$\Delta$ factors in~{\cite{Bonciani:2003nt}}, and leads to the conclusion
that, besides the replacement~(\ref{eq:nllcond}) one also needs to evaluate
the parton densities in the Sudakov exponent at a scale of order $\kt^2$, in
order to achieve NLL accuracy.

The exponentiation of genuine soft, non-collinear interference contributions
is more problematic. These contributions do not generally factorize in terms
of the Born cross section. The real emission cross section in the soft limit
has the well known structure
\begin{equation}
  \label{eq:softlim} \mathcal{R} \approx 4 \pi \as \left[ \sum_{i \neq j}
  \mathcal{B}_{i j} \frac{k_i \cdot k_j}{(k_i \cdot k) (k_j \cdot k)}
  +\mathcal{B} \sum_i \frac{k_i^2}{(k_i \cdot k)^2} \,C_i \right]
\end{equation}
where $\mathcal{B}_{i j}$ are the colour correlated Born amplitude, defined
in eq.~(\ref{eq:colourcorr}), and $C_i$ is the Casimir invariant of the
colour representation of parton $i$. If parton $i$ is massless, also
collinear singularities are present in formula (\ref{eq:softlim}). They can
be separated out by exploiting the techniques used to derive the results in
ref.~\cite{Bonciani:2003nt}. Since that technique is illustrated in
unpublished notes, in the following we sketch its main points.  Defining $k_i
+ k_j = k_{i j}$, we have
\begin{equation}
  \frac{k_i \cdot k_j}{(k_i \cdot k) (k_j \cdot k)} = \frac{k_i \cdot k_{i
  j}-k_i^2}{(k_i \cdot k) (k_{i j} \cdot k)} + \frac{k_j \cdot k_{i j}-k_j^2}{(k_j \cdot
  k) (k_{i j} \cdot k)}.
\end{equation}
In case of massless partons, for example, $k_i^2=0$, we can separate out the
collinear component
\begin{equation}
  \label{eq:onlysoft} \frac{k_i \cdot k_{i j}}{(k_i \cdot k) (k_{i j} \cdot
  k)} = \frac{1}{(k_i \cdot k)} \left[ \frac{k_i \cdot k_{i j}}{(k_{i j}
  \cdot k)} - \frac{k_i \cdot n}{k \cdot n} \right] + \frac{1}{(k_i \cdot k)}
  \frac{k_i \cdot n}{k \cdot n},
\end{equation}
where $n$ is an arbitrary timelike vector. In the first term of the right
hand side of eq.~(\ref{eq:onlysoft}) there are no collinear singularities,
since the factor in the square bracket vanishes when $k$ becomes collinear to
$k_i$.  The second term in (\ref{eq:onlysoft}) is collinear, but being
independent of $j$ gives a contribution of the form
\begin{equation}
  \sum_j \mathcal{B}_{i j} \frac{1}{(k_i \cdot k)} \frac{k_i \cdot n}{k \cdot
  n} \propto \mathcal{B}\, \frac{1}{(k_i \cdot k)} \frac{k_i \cdot n}{k \cdot
  n},
\end{equation}
factorized in terms of the Born cross section. Soft terms that factorize in
terms of the Born amplitude are automatically included in the \POWHEG{}
framework, since the \POWHEG{} exponent contains precisely the factor
$\mathcal{R}/\mathcal{B}$. Not so for the interference terms. We are thus
left with terms of the form
\begin{equation}
  \label{eq:softinterf} 
  \sum_{\stackrel{i \neq j}{\scriptscriptstyle k_i^2=0}} \mathcal{B}_{i j}
 \frac{2}{(k_i \cdot k)} \left[ \frac{k_i \cdot k_{i j}}{(k_{i j} \cdot k)} -
 \frac{k_i \cdot n}{k \cdot n} \right] + \sum_{\stackrel{i \neq
 j}{\scriptscriptstyle k_i^2\neq 0}} \mathcal{B}_{i j} \frac{2 (k_i \cdot
 k_{i j}-k_i^2)}{(k_i \cdot k) (k_{i j} \cdot k)}.
\end{equation}
We distinguish in the sum the massless and the massive case. The
$\stepf(\kt-\pt)$ function in the \POWHEG{} Sudakov exponent,
eq.~(\ref{eq:suddef}), could in principle be different for the various
collinear regions of the soft parton. However, since formula
(\ref{eq:softinterf}) does not carry collinear singularities, the dominant
integration region does not require any small angles. Under this condition,
$\kt$ is of the order of $k^0$, and at the single-logarithmic level one can
replace
\begin{equation}
  \stepf\!\(\kt^2 - \pt^2\) \Longrightarrow \stepf\!\(k_0^2 - \pt^2\).
\end{equation}
In other words, the integral in the radiation variables of the generic term
of formula (\ref{eq:softinterf}), with a theta function constraint
$\stepf(\kt-\pt)$, and $\kt$ defined relative to one generic $k_i$
yields a result of the form $a\log \pt +b$. If we replace the
$\stepf(\kt-\pt)$ with $\stepf(k_0-\pt)$ the result will become $a\log \pt
+b'$, i.e.\ the logarithmically enhanced term remains the same.

The angular integration of the $k$-dependent coefficients in
eq.~(\ref{eq:softinterf}) yields (see Appendix~\ref{app:Softints})
\begin{equation}\label{eq:softint1}
  I_{i j} = \int d \Omega \frac{1}{(k_i \cdot k)} \left[ \frac{k_i \cdot k_{i
  j}}{(k_{i j} \cdot k)} - \frac{k_i \cdot n}{n \cdot k} \right] = \frac{2
  \pi}{k_0^2} \log \frac{(k_{i j} \cdot k_i)^2 n^2}{k_{i j}^2 (n \cdot k_i)^2}
\end{equation}
if $k_i^2 = 0$, and
\begin{equation}\label{eq:softint2}
  I_{i j} = \int d \Omega \frac{k_i \cdot k_{i
  j}-k_i^2}{(k_i \cdot k)(k_{i j} \cdot k)}
 = \frac{2 \pi}{k_0^2} \frac{1-\frac{k_i^2}{k_i \cdot k_{i j}}}{\beta_{i j}}
  \log 
  \frac{1 + \beta_{i j}}{1 - \beta_{i j}},\qquad \beta_{i j} = \sqrt{1 -
  \frac{k_i^2 
  k_{i j}^2}{(k_i \cdot k_{i j})^2}}
\end{equation}
if $k_i^2 \neq 0$. The NLL resummed Sudakov form factor associated with these
soft emissions has the form
\begin{equation}
  \label{eq:deltaint} \Sigma^{\tmop{int}} = \frac{1}{|\mathcal{M}|^2}
\left| \exp \left[ - \int
  \sum_{i \neq j} \frac{d k^0}{k^0} \,\stepf\!\(k^0 - \pt\)
\frac{\as\!\(k_0^2\)}{4 \pi} \, \sum_a T^a_i \,T^a_j \,\left( I_{i j} + I_{j
  i} \right) \right]\! \mathcal{M}\, \right|^2 .
\end{equation}
where
$\Sigma^{\tmop{int}}$ corresponds to the expression for $\Sigma$
in eq.~(9) of~\cite{Bonciani:2003nt} (see also eqs.~(14) and~(15) there),
excluding the $J$ factors.\footnote{In
ref.~\cite{Bonciani:2003nt}, only the case of massless partons is considered.}
Here $\mathcal{M}$ is the Born matrix element, viewed as a complex vector in
all its colour components. More precisely, \tmtexttt{}$\mathcal{M}$ is a
tensor in colour space, carrying a colour index for each parton entering or
leaving the amplitude. It thus spans a linear space, equipped with a
sesquilinear product. The matrices $T^a_i$ can be seen as operators acting on
this linear space. In facts, they act on the colour index of the $i$th
particle in $\mathcal{M}$. Thus the exponential is to be seen as the
exponential of an operator in colour space.  In ref.~{\cite{Bonciani:2003nt}}
the exponential is written as an energy-ordered one, to remind that this kind
of exponentiation takes place because leading-log soft emission is dominated
by energy-ordered graphs, where softer emission take place later (i.e.\
closer to the final-state lines). In fact, at NLL accuracy the ordering has
no effect.

In the \POWHEG{} formalism, the exponentiation of the soft interference terms
has the form
\begin{equation}
  \label{eq:pwdeltaint} \Sigma^{\tmop{int}} = \exp \left[ - 2 \int \sum_{i
  \neq j} \frac{d k^0}{k^0}\, \stepf\!\(k^0 - \pt\)
  \frac{\as\!\(k_0^2\)}{4 \pi}\, 
  \frac{\mathcal{M}^{\dagger} \sum_a T^a_i \,T^a_j \,\left( I_{i j} + I_{j i}
  \right) \mathcal{M}}{|\mathcal{M}|^2} \right],
\end{equation}
since in \POWHEG{} the ratio $R / B$ appears in the Sudakov exponent. This
does not agree in general with eq.~(\ref{eq:deltaint}). There is, however,
one important case in which it agrees, that is to say, when $\mathcal{M}$ is
an eigenstate of all the operators $\sum_a T^a_i\, T^a_j$. In this case we
can replace $\sum_a T^a_i\, T^a_j$ in the exponentials in
eqs.~(\ref{eq:deltaint}) and~(\ref{eq:pwdeltaint}) with their eigenvalues,
and the two expressions become identical. It is also apparent that the
$\mathcal{B}_{i j}$ are in this case all proportional to $\mathcal{B}$, and
again we have complete factorization of the soft contribution in terms of the
Born cross section. As discussed in ref.~{\cite{Bonciani:2003nt}}, if there
are no more than 3 coloured partons entering or leaving the Born amplitude,
it turns out that $\mathcal{M}$ is always an eigenstate of the operator
$\sum_a T^a_i\, T^a_j$, for any $i, j$. We conclude that, in this case, the
prescription of eq.~(\ref{eq:nllcond}) is sufficient to guarantee NLL
accuracy.

If there are more than 3 coloured partons in the amplitude, the standard
\POWHEG{} formula does not allow one to simply obtain full NLL accuracy. It
should be very simple, however, to modify it in such a way that the
interference soft terms are correctly resummed at NLL accuracy, at least as
far as the dominant terms in the large-$N_c$ limit are concerned.
\begin{figure}[tbh]
\begin{center}
  \epsfig{file=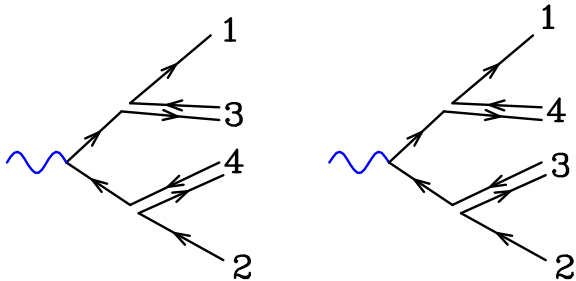,width=0.5\textwidth}
\end{center}
  \caption{\label{fig:epemqqggpl}The two inequivalent planar colour
  configurations in $e^+ e^- \rightarrow q \bar{q} g_3 g_4$. Notice that the
  square of each amplitude is not symmetric in the exchange of the final
  state gluons.}
\end{figure}
In the large-$N_c$ limit, the Born amplitude can be written as the sum of
independent planar colour structures, that differ only by permutations of the
external lines, as illustrated in fig.~\ref{fig:epemqqggpl}. We write
\begin{equation}
  \mathcal{M}_{\tmop{pl}} = \sum_{\rho} \mathcal{M}_{\tmop{pl}}^{\rho}, \quad\quad
  |\mathcal{M}_{\tmop{pl}} |^2 = \sum_{\rho} |\mathcal{M}^{\rho}_{\tmop{pl}}
  |^2,
\end{equation}
where the subscript $\tmop{pl}$ stands for planar, and $\rho$ labels the
different planar colour components. Observe that interference terms in the
square of $\mathcal{M}_{\tmop{pl}}$ do not appear, since they give rise to
non-planar structures, and are suppressed in the large $N_c$ limit. Soft
emission factors give rise to leading $N_c$ contributions only if they act
upon adjacent colour lines in the planar structures
\begin{figure}[tbh]
\begin{center}
  \epsfig{file=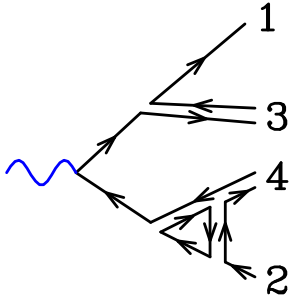,width=0.25\textwidth}
\end{center}
  \caption{\label{fig:planarTTaction}Soft emission colour factor $T^a_3
  T^a_4$ does not alter the colour structure of the Born planar amplitude,
  and it acts as a multiplicative constant. The close colour loop provides an
  extra factor of $N_c$.}
\end{figure}. 
In this case their effect amounts to a factor of $N_c / 2$, and they do not
alter the colour connections of the Born planar amplitudes. It follows that
\begin{eqnarray}
  \Sigma^{\tmop{int}} \!& = & \frac{1}{|\mathcal{M}_{\tmop{pl}} |^2}
\left| \exp \left[ - \int \sum_{i \neq j}
  \frac{d k^0}{k^0}\, \stepf\!\(k^0 - \pt\) \frac{\as\!\(k_0^2\)}{4 \pi}
  \sum_a  T^a_i \, T^a_j \left( I_{i j} + I_{j i} \right) \right]\!
\mathcal{M}_{\tmop{pl}} \,  \right|^2 \nonumber\\
  & = & \frac{1}{\sum_{\rho}
  |\mathcal{M}_{\tmop{pl}}^{\rho} |^2}
  \sum_{\rho} |\mathcal{M}_{\rho} |^2 \exp \left[ - 2 \int
  \sum_{i = j \pm 1} \frac{d k^0}{k^0} \,\stepf\!\(k^0 - \pt\) 
  \frac{\as\!\(k_0^2\)}{4 \pi} \, \frac{N_c}{2} \(I_{i j} + I_{j i}\) \right]
  \nonumber\\ 
  & = & \frac{1}{\sum_{\rho} |\mathcal{M}_{\tmop{pl}}^{\rho} |^2}
\nonumber\\
&&\times
\sum_{\rho} |\mathcal{M}_{\rho} |^2 \exp \left[ - 2\! \int
  \sum_{i \neq j} \frac{d k^0}{k^0} \,\stepf\!\(k^0 - \pt\) 
  \frac{\as\!\(k_0^2\)}{4 \pi}\, \frac{\mathcal{M}^{\rho \dagger}_{\tmop{pl}}
   \, \sum_a T^a_i \,  T^a_j \(I_{i j} +
   I_{ji}\)\mathcal{M}_{\tmop{pl}}^{\rho}}{|\mathcal{M}_{\tmop{pl}}^{\rho}
    |^2} \right]\!,
\nonumber\\
 \label{eq:intpl}
\end{eqnarray}
where with the notation $i = j \pm 1$ we (somewhat imprecisely) indicate that
we only consider adjacent colour lines for a given planar ordered amplitude.
The last line of eq.~(\ref{eq:intpl}) is similar to the form that appears in
eq.~(\ref{eq:pwdeltaint}), except that there is one term for each planar
amplitude. This suggests how to implement it in the \POWHEG{} framework. In
section~\ref{sec:powhegmth} we have introduced a classification of the Born
contributions in terms of their flavour structure $f_b$, and a classification
of the real amplitude contributions in terms of their
singularity regions and flavour
structures $\alpha_r$. We need to extend this classification to include also
the colour structure in the large $N_c$ limit. One simple way to do so is to
compute $B_{\tmop{pl}}$ and $R_{\tmop{pl}}$, the Born and real terms in the
planar limit, for each of their given flavour structure. Then we define
\begin{equation}
  \label{eq:BRpldecomp}
  B^{f_b, \rho} =
  B^{f_b} \frac{B^{f_b, \rho}_{\tmop{pl}}}{\sum_{\rho} B^{f_{b,
  \rho}}_{\tmop{pl}}},
  \quad\quad  R^{\alpha_r,
  \rho_r} = R^{\alpha_r} \frac{R^{\alpha_r, \rho_r}_{\tmop{pl}}}{\sum_{\rho_r}
  R_{\tmop{pl}}^{\alpha_r, \rho_r}},
\end{equation}
so that
\begin{equation}
\label{eq:BRpldecomp1}
 B^{f_b} = \sum_{\rho} B^{f_b, \rho},\quad\quad
 R^{\alpha_r} = \sum_{\rho_r} R^{\alpha_r, \rho_r}.
\end{equation}
The labels $\rho$ and $\rho_r$ refer to the contribution of a given planar
structure in the Born and Real contributions respectively.
Next we must associate with each $(\alpha_r, \rho_r)$ pair an
underlying Born $(f_b, \rho)$ pair. As far as $f_b$ is concerned, the
association is performed along the lines explained in
section~\ref{sec:powhegmth}. For $\rho$ and $\rho_r$, one proceeds as
follows. If the singular region is soft, the Born planar colour structure
is obtained by simply removing the soft gluon (and joining its colour lines)
from the real emission colour structure. If the singular region is collinear,
we need to show first that each $R^{\alpha_r, \rho_r}$ has collinear
singularities only for collinear particles that are nearby in its planar
structure. This property follows from the fact that the ratio
\begin{equation}
\frac{R^{\alpha_r}}{\sum_{\rho_r} R_{\tmop{pl}}^{\alpha_r, \rho_r}} 
\end{equation}
has no collinear singularities, since the denominator is the large $N_c$
limit of the numerator, and therefore it has the same singular structure.  It
follows then, from the second equality in~(\ref{eq:BRpldecomp}), that
$R^{\alpha_r, \rho_r}$ has the same collinear singularities of $R^{\alpha_r,
\rho_r}_{\tmop{pl}}$.
Thus, since planar cross sections have collinear singularities only when
nearby partons are collinear, $R^{\alpha_r, \rho_r}$ has the same property.
Joining
the planar colour of nearby partons yields a valid Born planar colour
configuration. We now introduce
\begin{equation}
  \label{eq:BBpl} \bar{B}^{f_b, \rho} = \bar{B}^{f_b} \frac{B^{f_b,
  \rho}_{\tmop{pl}}}{\sum_{\rho} B^{f_{b, \rho}}_{\tmop{pl}}},
\end{equation}
and the corresponding \POWHEG{} Sudakov form factor
\begin{equation}
  \label{eq:pwsudpl} \Delta^{f_b, \rho} ( \tmmathbf{\Phi}_n, \pt) = \exp
  \left\{ - \sum_{\alpha_r, \rho_r \in \{\alpha_r, \rho_r |f_b, \rho\}} \int
  \frac{\Big[ d\Rad \, R ( \tmmathbf{\Phi}_{n + 1})\, \stepf(\kt -
  \pt) \Big]_{\alpha_r, \rho_r}^{\bar{\Phi}_n^{\alpha_r} =
  \Phi_n}}{B^{f_b, \rho} ( \tmmathbf{\Phi}_n)} \right\} ,
\end{equation}
that should correctly generate large angle soft radiation in the
large $N_c$ limit.
Equations~(\ref{eq:BBpl}) and~(\ref{eq:pwsudpl}) certainly satisfy the
formula~(\ref{eq:intpl}) for the soft interference terms in the large $N_c$
limit. We must still show that collinear singularities are treated
correctly for all $N_c$, i.e.\ that formula~(\ref{eq:pwsudpl}) is
equivalent to formula~(\ref{eq:suddef}) in the collinear regions.
This is the case if
\begin{equation}
  \sum_{\rho_r \in \{\alpha_r, \rho_r |f_b, \rho\}} \frac{R^{\alpha_r,
  \rho_r}}{B^{f_b, \rho}} = \frac{R^{\alpha_r}}{B^{f_b}}
\end{equation}
in the $\alpha_r$ collinear region. Using eqs.~(\ref{eq:BRpldecomp})
we write
\begin{equation}
  \label{eq:showcollpl} \sum_{\rho_r \in \{\alpha_r, \rho_r |f_b, \rho\}}
  \frac{R^{\alpha_r, \rho_r}}{B^{f_b, \rho}} = \frac{R^{\alpha_r}}{B^{f_b}} 
  \frac{\left[ \sum_{\rho_r \in \{\alpha_r, \rho_r |f_b, \rho\}} R^{\alpha_r,
  \rho_r}_{\tmop{pl}} \right] / B^{f_b, \rho}_{\tmop{pl}}}{\left[
  \sum_{\rho_r} R_{\tmop{pl}}^{\alpha_r, \rho_r} \right] / \sum_{\rho}
  B^{f_{b, \rho}}_{\tmop{pl}}} .
\end{equation}
The second term on the r.h.s. of eq.~(\ref{eq:showcollpl}) yields 1
in the collinear limit. In fact, it is easy to convince oneself that, in this
limit, both its numerator and denominator yield the (planar) Altarelli Parisi
splitting kernel, that simplifies in the ratio. Thus $\Delta^{f_b, \rho} (
\tmmathbf{\Phi}_n, \pt)$ becomes $\rho$ independent in this limit, and the
overall $\bar{B}^{f_b, \rho}$ factor can be summed in $\rho$, to yield back
$\bar{B}^{f_b}$.

We conclude this section by reminding the reader that the importance of the
inclusion of soft-interference terms should not be overemphasized at this
stage. After all, the \POWHEG{} formula in eq.~(\ref{eq:pwdeltaint}) differs
from formula~(\ref{eq:deltaint}) only starting at order $\as^2$. It is thus
unlikely that these terms will have important effects in a full \POWHEG{}
implementation. Nevertheless, if one wants to assess their importance, one
can, to begin with, implement their large $N_c$ resummation and study its
impact.

\subsection{Interfacing \POWHEG{} to an SMC}
\label{sec:colour}
The \POWHEG{} algorithm generates the kinematics and flavour configuration
of the hardest-emission event. The event should be fed into a SMC
using the {\em Les Houches Interface
for User Processes}~\cite{Boos:2001cv} (LHIUP from now on). In particular,
one should require that no events harder than the one generated by \POWHEG{}
should be generated by the SMC. This is achieved by setting the variable
{\tt SCALUP} of the LHIUP equal to the $\kt$ of the  \POWHEG{} event.
The LHIUP specifies how to pass the kinematics and flavour structure
of the hard event to the SMC.
\subsubsection{Colour assignment}
The LHIUP also requires that the colour connections of the hard event
(in the large $N_c$ limit) should also be specified. \POWHEG{} does not, in general,
generate these large-$N_c$ colour structures. They are needed (and can be generated)
only if one wishes to reach large-$N_c$ NLL accuracy of the Sudakov
form factor, in events with more than 3 coloured partons at the Born level,
as discussed in section~\ref{sec:sudak}.
If this is not the case, the generation of the colour configuration should be
performed after the \POWHEG{} event has been generated. Here we illustrate
two acceptable ways to perform colour assignment. The first (and simpler)
approach is the following:
\begin{itemize}
\item
Generate the \POWHEG{} event in the standard way.
\item
Compute the different (planar) colour contributions to the Born cross section,
at the kinematics of the generated underlying Born configuration.
\item
Pick an underlying Born colour configuration, with a probability proportional
to the corresponding (planar) colour Born contribution.
\item
If no radiation has been generated, this is the colour structure of the event.
\item
If radiation has been generated, \POWHEG{} has also generated a $\frindsing$
index, specifying a singular region. In this case we always
assume that the emitted parton is (planar) colour-connected to the emitter.
This fully specifies the planar colour structure of the event.
\end{itemize}
This method only requires the calculation of the planar colour-structures of
the Born term.
In the second method (which is usually implemented in \MCatNLO{}) one computes
all the planar colour contributions to $R$, and chooses the colour configuration
with a probability proportional to the corresponding contribution.

\section{\POWHEG{} in the \FKS{} framework}
\label{sec:powhegfks}
In this section we construct a mapping from the $\Kinnpo$ phase space
to the barred and radiation variables
of eq.~(\ref{eq:phspfact}) (and its corresponding inverse mapping)
for the two kinds of singular regions $i$ (initial state) and $ij$ (final state),
that is compatible with the FKS subtraction method.
These is the only missing ingredients one needs to construct
a \POWHEG{} generator in the FKS framework.

We stress that the mapping that we propose is not the only possible one. In practical
examples one may find convenient to depart from this approach. It is however fully
general, and as such it may be implemented once and for all in a computer code
to be used for all processes.

\subsection{Initial-state singularity}
\label{sec:powhegfksiss}

\subsubsection{Radiation and barred variables}
\label{sec:FKS_IS}
We first show how to construct the underlying Born (i.e.\ the barred variables)
and the three radiation variables from the $\Kinnpo$ kinematics,
for the case of the initial-state singularity region.

Without loss of generality, we assume that the FKS parton is the $(n+1)$th
one, so that the mapping from $k_i$ to $\bar{k}_i$ ($i=1,\ldots,n$)
does not requires a relabeling of the momenta.
In the FKS formulation, $k_{n+1}$ is a function of $\xi$,
$y=\cos\theta$ and $\phi$ (see eq.~(\ref{eq:FKSvar})), so that, in the centre
of mass of the colliding partons, i.e.\ the CM of the $\(\xplus \,
\Kplus+\xminus \, \Kminus\)$ system, we have
\begin{equation}
\label{eq:FKS_IS_knpo}
k_{n+1}^0=
\frac{\sqrt{s}}{2}\xi\,,\qquad\quad
k_{n+1}=k_{n+1}^0\(1,\sin\theta\,\sin\phi,\,\sin\theta\,\cos\phi,\,
\cos\theta\)\,. 
\end{equation}
We take $\xi$, $y$ and $\phi$ as the radiation
variables for the singular region, and obtain
\begin{equation}
\frac{d^3 k_{n+1}}{2k^0_{n+1} (2\pi)^3}
=\frac{s}{(4\pi)^3} \,\xi\, d\xi\, dy\,d\phi\,.
\end{equation}
We introduce the momentum
\begin{equation}
k_{\rm tot}=\sum_{i=1}^n k_i=\xplus \Kplus+\xminus \Kminus -k_{n+1}  \,,
\end{equation}
and construct a longitudinal boost $\boost_L$ (longitudinal
with respect to the incoming beams) such that $\boost_L k_{\rm tot}$
has zero longitudinal component. Notice that $\boost_L$ is unique, being given
by a longitudinal boost with boost angle equal to minus the rapidity of $k_{\rm tot}$.
Then we construct a transverse boost $\boost_T$ such that $\boost_T \boost_L  k_{\rm tot}$
has zero transverse momentum. We then define the barred momenta as
\begin{equation}
\label{eq:kbardef}
\bar{k}_i=\boost_L^{-1} \boost_T \boost_L \;k_i\,,\qquad i=1,\ldots ,n\;,
\end{equation}
and define
\begin{equation}\label{eq:ktot}
\bar{k}_{\rm tot}=\sum_{i=1}^n \bar{k}_i=\boost_L^{-1} \boost_T
\boost_L k_{\rm tot}\,.
\end{equation}
We notice that, by construction, $k_{\rm tot}$ and $\bar{k}_{\rm tot}$ have the same
invariant mass and rapidity.
We now define $\bar{x}_\splus$ and $\bar{x}_\sminus $ in such a way that
\begin{equation}
\bar{x}_\splus \Kplus+\bar{x}_\sminus  \Kminus =\bar{k}_{\rm tot}\,.
\end{equation}
An explicit expression for $\bar{x}_\splus$ and $\bar{x}_\sminus $ is easily
obtained,
by using the fact that $\bar{k}_{\rm tot}$ and $k_{\rm tot}$ have the
same invariant mass and rapidity. We get
\begin{equation}
\label{eq:x1x2}
\bar{x}_\splus = \xplus \sqrt{1-\xi} \, 
\sqrt{\frac{2-\xi(1+y)}{2-\xi(1-y)}}\,,  \qquad\quad
\bar{x}_\sminus =\xminus\sqrt{1-\xi} \,
\sqrt{\frac{2-\xi(1-y)}{2-\xi(1+y)}}\,,
\end{equation}
and
\begin{equation}
d\xplus \,d\xminus  = \frac{d\bar{x}_\splus d\bar{x}_\sminus }{1-\xi}\;.
\end{equation}
Observe that we always have $\bar{x}_\splus \le \xplus$ and
$\bar{x}_\sminus \le \xminus$.
The phase space $d\Kinnpo$ can be
rewritten as follows
\begin{eqnarray}
\label{eq:dPhinpo_IS_FKS}
d\Kinnpo &=& d\xplus \, d\xminus  \,
(2\pi)^4\delta^4\!\left(\xplus \Kplus+\xminus \Kminus
-\sum_{i=1}^{n+1}k_i\right)\, 
\prod_{i=1}^{n+1} \frac{d^3k_i}{2k_i^0(2\pi)^3}
\nonumber \\
&=& d\bar{x}_\splus\, d\bar{x}_\sminus \,\frac{s}{(4\pi)^3}\,
\frac{\xi}{1-\xi}\,d\xi\,dy\, d\phi 
\nonumber\\
&&\times\, (2\pi)^4\delta^4\!\left(\xplus \Kplus+\xminus \Kminus
-k_{n+1}-\sum_{i=1}^{n}k_i\right)\, 
\prod_{i=1}^{n} \frac{d^3k_i}{2k_i^0(2\pi)^3}
\nonumber \\
&=& d\bar{x}_\splus\, d\bar{x}_\sminus
\,\frac{s}{(4\pi)^3}\,\frac{\xi}{1-\xi}\, d\xi \,dy\, d\phi
\,(2\pi)^4
\delta^4\!\left(\bar{x}_\splus \Kplus+\bar{x}_\sminus \Kminus
-\sum_{i=1}^{n}\bar{k}_i\right)\, 
\prod_{i=1}^{n} \frac{d^3\bar{k}_i}{2\bar{k}_i^0(2\pi)^3}
\nonumber \\ \label{eq:fksphspiss}
&=& \frac{s}{(4\pi)^3}\,\frac{\xi}{1-\xi}\,d\xi\,dy\, d\phi\,
d\BKinn\;,
\end{eqnarray}
where we have used the boost invariance of the $n$-body phase space.
Thus, from eq.~(\ref{eq:emfacphspe0}) we obtain
\begin{equation}
\label{eq:dRad_IS_FKS}
d\Rad=\frac{s}{(4\pi)^3}\,\frac{\xi}{1-\xi}\,d\xi\,dy\, d\phi\,.
\end{equation}

\subsubsection{Inverse construction}
\label{sec:inv_construction_FKS}
We describe now the construction of the full $(n+1)$-particle
phase space, given the barred variables $\BKinn$ and the radiation variables
$\xi$, $y=\cos\th$ and $\phi$.  Using
eq.~(\ref{eq:FKS_IS_knpo}) we fully construct $k_{n+1}$.  Inverting
eq.~(\ref{eq:x1x2}), we have
\begin{equation}
\label{eq:xp_xm_IS_FKS}
\xplus = \frac{\bar{x}_\splus}{\sqrt{1-\xi}} \, 
\sqrt{\frac{2-\xi(1-y)}{2-\xi(1+y)}},  \qquad
\xminus =\frac{\bar{x}_\sminus }{\sqrt{1-\xi}} \,
\sqrt{\frac{2-\xi(1+y)}{2-\xi(1-y)}}\;.
\end{equation}
The range for the variable $\xi$ is now restricted by the requirement that both
$\xplus $ and $\xminus $ be less than 1. This gives
\begin{equation}
\label{eq:xi_limits}
0 \leq \xi \leq \xi_{\rm max}
\end{equation}
with
\begin{eqnarray}
\label{eq:xi_max}
\xi_{\rm max}= 1- {\rm max}\!\!\!&&\lg
\frac{2(1+y)\,\bar{x}_\splus^2}{\sqrt{(1+\bar{x}_\splus^2)^2(1-y)^2 +
 16\,y\,\bar{x}_\splus^2}+(1-y)(1-\bar{x}_\splus^2)},\right.
\nonumber\\
&&\phantom{\Bigg\{}\left.
\frac{2(1-y)\,\bar{x}_\sminus ^2}{\sqrt{(1+\bar{x}_\sminus ^2)^2(1+y)^2 -
    16\,y\,\bar{x}_\sminus ^2} +(1+y)(1-\bar{x}_\sminus ^2)}\rg .
\end{eqnarray}
From $k_{n+1}$ and $x_\splusminus$ we
can construct $k_{\rm tot}=\xplus \Kplus+\xminus \Kminus -k_{n+1}$, and
summing the $n$ barred momenta, according to eq.~(\ref{eq:ktot}), we can
compute 
\begin{equation}
\bar{k}_{\rm tot} = \sum_{i=1}^n \bar{k}_i\,.
\end{equation}
The four vectors $\bar{k}_{\rm tot}$ and $k_{\rm tot}$ have
the same invariant mass and rapidity by construction, since
the relation between the $x_\splusminus$ and the $\bar{x}_\splusminus$
was obtained precisely from these conditions.
We then construct the boost $\boost_L$ such that
$\boost_L\bar{k}_{\rm tot}$ has zero rapidity.
We will also have that $\boost_L k_{\rm tot}$ has zero rapidity.
Then we compute the transverse boost $\boost_T$ such that
\begin{equation}
 \boost_T \,\boost_L \, k_{\rm tot} =  \boost_L \, \bar{k}_{\rm tot}\,.
\end{equation}
Finally, the momenta $k_i$, $i=1,\ldots ,n$ are obtained as
\begin{equation}
k_i =  \boost_L^{-1}\, \boost_T^{-1} \, \boost_L \,
\bar{k}_i\,, \qquad  i=1,\ldots ,n\,.
\end{equation}
This completes the construction of the $(n+1)$-body phase space, starting
from an underlying Born configuration and the three radiation variables.

\subsection{Final-state singularity}

\subsubsection{Radiation and barred variables}
\label{subsub:fksbar}
In this section, we show how to construct the underlying Born and the three
radiation variables, given the $\Kinnpo$ variables, in the case of
final-state singularity.  We assume, without loss of generality, that the
singular region is associated with partons $(n+1)$ and $n$, that is, the
$(n+1)$th parton becoming collinear to the $n$th parton, or becoming soft.

Our mapping is constructed in such a way that
\begin{equation}
\bar{x}_\splus=\xplus\,, \qquad\quad \bar{x}_\sminus =\xminus\,.
\end{equation}
Thus, the partonic CM frame of the final-state $(n+1)$-particle system
coincides with the CM frame of the $n$ barred momenta $\bar{k}_i$,
and we work in this frame from now on.
In this section (and only here), we  need to introduce a short-hand
notation for the modulus of the space component of a four-vector in the CM
frame
\begin{equation}
\mmod{p}\equiv \abs{\vec{p}}.
\end{equation}
We then define
\begin{equation}
q=\Kplus\,\xplus +\Kminus \,\xminus  = \sum_{i=1}^{n+1} k_i\,,
\end{equation}
and, in the CM frame, we have
\begin{equation}
\vec{q}=0\,,\quad\quad  q^2=\(q^0\)^2\,.
\end{equation}
Introducing the four momentum
\begin{equation}
\label{eq:k_def}
k=k_{n}+k_{n+1}\,,
\end{equation}
we define the radiation variables as
\begin{equation}\label{eq:FKSfsr}
\xi=\frac{2 k_{n+1}^0}{q^0}\,,\qquad \quad
y=\frac{\vec{k}_{n+1}\cdot \vec{k}_{n}}{\mmod{k}_{n+1} \mmod{k}_n}\,,\qquad
\quad 
\phi=\phi\(\vec{\eta} \times \vec{k},\; \vec{k}_{n+1} \times \vec{k}\),
\end{equation}
where $\vec\eta\,$ is an arbitrary direction that serves as origin for the
azimuthal angle of $\vec{k}_{n+1}$ around $\vec{k}$, and ``$\times$'' is the
cross vector product.  The notation $\phi(\vec{v}_1,\vec{v}_2)$ denotes the
angle between $\vec{v}_1$ and $\vec{v}_2$. Thus, $\phi$ is the azimuth of the
vector $\vec{k}_{n+1}$ around the direction $\vec{k}$.  Notice that only
$\xi$ and $y$ correspond exactly to FKS variables, since the FKS azimuth is
usually defined with respect to $\vec{k}_n$ rather than $\vec{k}$. We prefer
the choice (\ref{eq:FKSfsr}) because we want to defined the mapping in such a
way that the $\vec{k}$ direction (rather than the $\vec{k}_n$ one) is
preserved.  Our choice, however, makes only irrelevant differences in the FKS
formalism, since the real-emission cross section has singular distributions
only in the variables $\xi$ and $y$.

We introduce the recoil four-momentum and mass
\begin{equation}
\label{eq:Mrec}
k_{\rm rec}=\sum_{i=1}^{n-1} k_i\,,\quad\quad M_{\rm rec}^2=k_{\rm rec}^2\;.
\end{equation}
We have
\begin{equation}
k_{\rm rec}=q-k\,,\qquad\vec{k}_{\rm rec}=-\vec{k}\;.
\end{equation}
We construct a boost $\boost$ along the $\vec{k}_{\rm rec}$
direction, such that the 4-momentum $(q-\boost\,k_{\rm rec})$ is
light-like, that is to say
\begin{equation}
(q-\boost\,k_{\rm rec})^2=0\,.
\end{equation}
The boost velocity is easily computed
\begin{equation}
\label{eq:betaboost}
\beta=
\frac{q^2-(k^0_{\rm rec}+\mmod{k}_{\rm rec})^2}
{q^2+(k^0_{\rm rec}+\mmod{k}_{\rm rec})^2}\,.
\end{equation}
Since $q^0=k^0+k^0_{\rm rec}$, and $k^0\ge \mmod{k}=\mmod{k}_{\rm rec}$,
$\beta$ is positive and smaller than one. Thus, the
boost $\boost$ always exists.
We define the barred momenta as
\begin{equation}
\label{eq:q_minus_krec}
\bar{k}_i=\boost\,k_i,\qquad i=1,\ldots,n-1\;,
\qquad\qquad \bar{k}_n=q-\boost\,k_{\rm rec}\,,
\end{equation}
so that they clearly satisfy momentum conservation
\begin{equation}
\label{eq:sum_kbar_q}
\sum_{i=1}^n \bar{k}_i=q\,.
\end{equation}
In order to obtain $d\Rad$, we write the $(n+1)$-body phase space as
\begin{eqnarray}
d\Phi_{n+1}&=&\prod_{i=1}^{n+1}
\frac{d^3 k_i}{2k_i^0\,(2\pi)^3}\,(2\pi)^4\,
\delta^4\!\left(q-\sum_{i=1}^{n+1}k_i\right)
\nonumber \\ \label{eq:fsrphspnp1}
&=&\frac{d^3 k_{n+1}}{2k_{n+1}^0\,(2\pi)^3}
\,\frac{d^3 k}{2k_n^0\,(2\pi)^3}\,\prod_{i=1}^{n-1}
\frac{d^3 k_i}{2k_i^0\,(2\pi)^3}\,(2\pi)^4\,
\delta^4\!\left(q-k-\sum_{i=1}^{n-1} k_i\right),
\end{eqnarray}
where, in the last equality, using eq.~(\ref{eq:k_def}), we have traded
$\vec{k}_n$ for $\vec{k}$ 
as independent variable. We thus have now
\begin{equation}
k^0_n=\abs{\vec{k}-\vec{k}_{n+1}}\,,\qquad\qquad k^0=k^0_{n+1}+k^0_n\,.
\end{equation}
We identify the phase space in eq.~(\ref{eq:fsrphspnp1}) with the phase space
written in terms of the barred and radiation variables, multiplied by a
Jacobian factor $J$
\begin{eqnarray}
\label{eq:FKS_fac_phinpo}
d\Phi_{n+1}
&=& \frac{d^3 k_{n+1}}{2k_{n+1}^0\,(2\pi)^3}\,
\frac{d^3 k}{2k_n^0\,(2\pi)^3}\,\prod_{i=1}^{n-1}
\frac{d^3 k_i}{2k_i^0\,(2\pi)^3}\,(2\pi)^4\,
\delta^4\!\left(q-k-\sum_{i=1}^{n-1} k_i\right)
\nonumber \\
\label{eq:phasespaceident}
&=&  d\Rad \, d\bar{\Phi}_n\, = \( J\,d\xi \,dy\, d\phi \)
\left[\prod_{i=1}^{n} 
\frac{d^3 \bar{k}_i}{2\bar{k}_i^0\,(2\pi)^3}\,(2\pi)^4\,
\delta^4\left(q-\sum_{i=1}^{n} \
\bar{k}_i\right)\right].\;\;\;\;\;
\end{eqnarray}
We work out this equality simplifying common factors, until we obtain an
expression for $J$.  We consider the barred and radiation variables to be
functions of the unbarred variables.  First of all, we remind that $\vec{k}$
and 
$\vec{\bar{k}}_n$ have the same direction. We thus make the replacements
\begin{equation}
d^3 k = d\Omega^2\, \mmod{k}^2\, d\mmod{k}\,,\qquad d^3 \bar{k}_n=  d\Omega^2\,
 \mmod{\bar{k}}_n^2 \, d\mmod{\bar{k}}_n\,,
\end{equation}
and cancel the common factor $d\Omega^2$ on both sides of
eq.~(\ref{eq:phasespaceident}).  Boost invariance of the phase-space elements
and of the four-dimensional delta function also guarantees that
\begin{equation}
\label{eq:boostedid}
\prod_{i=1}^{n-1}
\frac{d^3 k_i}{2k_i^0\,(2\pi)^3}\,(2\pi)^4\,
\delta^4\!\left(q-k-\sum_{i=1}^{n-1} k_i\right)=
\prod_{i=1}^{n-1}
\frac{d^3 \bar{k}_i}{2\bar{k}_i^0\,(2\pi)^3}\,(2\pi)^4\,
\delta^4\!\left(q-\bar{k}_n-\sum_{i=1}^{n-1}\bar{k}_i\right).
\end{equation}
In fact, $\vec{k}_n$, $\xi$, $y$ and $\phi$ are functions of $\vec{k}_{n+1}$
and $\vec{k}$ only, while $\bar{k}_1,\ldots,\bar{k}_{n-1}$ depend upon
$k_1,\ldots,k_{n-1}$ via the boost $\boost$, that depends only upon
$\vec{k}_{n+1}$ and $\vec{k}$.  Furthermore, according
to~(\ref{eq:q_minus_krec}),
\begin{equation}
\boost\, (q-k) = q-\bar{k}_n\;,
\end{equation}
so that eq.~(\ref{eq:boostedid}) is implied by boost invariance.  The members
of eq.~(\ref{eq:boostedid}) can thus be divided out on both sides of
eq.~(\ref{eq:phasespaceident}). Performing also the replacement
\begin{equation}
\frac{d^3 k_{n+1}}{2k_{n+1}^0\,(2\pi)^3}
=\frac{q^2}{(4\pi)^3} \,\xi \,d\xi\,d\cos\psi \,d\phi\,,
\end{equation}
where $\psi$ is the angle between $\vec{k}_{n+1}$ and $\vec{k}$,
on the left hand side of eq.~(\ref{eq:phasespaceident}), we can also divide out
$d\xi\,d\phi$.  We are left with the identity
\begin{equation}
\frac{q^2}{(4\pi)^3}\, \xi \, d\cos\psi \,\frac{\mmod{k}^2\,d\mmod{k}}{k_n^0}
=J dy \,\mmod{\bar{k}}_n\,d\mmod{\bar{k}}_n\,,
\end{equation}
and, thus, we just need to express $y$ and $\mmod{\bar{k}}_n$ in terms of
$\cos\psi$ and $\mmod{k}$, at fixed $\xi$, and compute the Jacobian of this
two-variable transformation. The following sequence of relations give $y$ and
$\mmod{\bar{k}}_n$ as functions of $\cos\psi$ and $\mmod{k}$ (at fixed $\xi$)
\begin{equation}
\renewcommand{\arraystretch}{2}
\begin{array}{rclrcl}
\mmod{k}_n &=&\sqrt{\mmod{k}^2+\mmod{k}_{n+1}^2
-2\,\mmod{k}\,\mmod{k}_{n+1}\,\cos\psi}\,,\qquad\quad
& \Mrec^2 &=&\(q^0-\mmod{k}_{n+1}-\mmod{k}_n\)^2-\mmod{k}^2\,,
\\
\mmod{\bar{k}}_n&=&\displaystyle{\frac{q^2-\Mrec^2}{2\,q^0}}\,,\qquad
&y &=&\displaystyle{\frac{\mmod{k}^2-\mmod{k}_n^2-\mmod{k}_{n+1}^2}
{2\,\mmod{k}_n\,\mmod{k}_{n+1}}}\,.
\end{array}
\end{equation}
Thus
\begin{equation}
\left| \begin{array}{cc}
\displaystyle
\frac{\partial \mmod{\bar{k}}_n }{\partial \mmod{k}}
&
\phantom{aa}\displaystyle
\frac{\partial y}{\partial \mmod{k}} \\[3mm]
\displaystyle
\frac{\partial \mmod{\bar{k}}_n }{\partial \cos\psi}
&\phantom{aa}
\displaystyle
\frac{\partial y}{\partial \cos\psi}
\end{array}\right|
=\frac{\mmod{k}^2}{\mmod{k}_n^3}\(\mmod{k}_n-\frac{k^2}{2q^0}\)\,,
\end{equation}
where 
\begin{equation}
k^2 = 2\, \mmod{k}_n\, \mmod{k}_{n+1} \, (1-y)\,,
\end{equation}
and we get
\begin{equation}
\label{eq:jfsrfks}
J=\frac{q^2\,\xi}{(4\pi)^3}\,\frac{\mmod{k}_n^2}{\mmod{\bar{k}}_n}
\( \mmod{k}_n-\frac{k^2}{2q^0}\)^{-1},
\end{equation}
so that
\begin{equation}
\label{eq:FKS_dRad}
d\Rad = \frac{q^2\,\xi}{(4\pi)^3}\,\frac{\mmod{k}_n^2}{\mmod{\bar{k}}_n}
\( \mmod{k}_n-\frac{k^2}{2q^0}\)^{-1} d\xi \,dy\, d\phi\,.
\end{equation}

\subsubsection{Inverse construction}
In this section, we describe the construction of the full $(n+1)$-particle
phase space, given the barred variables $\BKinn$ and the radiation variables
$\xi$, $y$ and $\phi$. We immediately have
\begin{equation}
k_{n+1}^0=\mmod{k}_{n+1}=\xi \,\frac{q^0}{2}\,.
\end{equation}
The absolute value of the three-momentum of $k_n$ must instead be obtained by
solving the equation for the energy conservation
\begin{equation}
\label{eq:solveform1} 
\mmod{k}_n+\mmod{k}_{n+1}+\sqrt{\mmod{k}^2 + \Mrec^2}=q^0\,,
\end{equation}
where
\begin{equation}
  \mmod{k}=\sqrt{\mmod{k}_n^2+\mmod{k}_{n+1}^2+2\,\mmod{k}_n\,\mmod{k}_{n+1}\,
  y}\,,  
\end{equation}
and $\Mrec^2$, according to eqs.~(\ref{eq:Mrec}),
(\ref{eq:q_minus_krec}) and~(\ref{eq:sum_kbar_q}), is given by
\begin{equation}
\Mrec^2 = (q-\bar{k}_n)^2\,.
\end{equation}
Under the obvious condition
\begin{equation}
k_{n+1}^0 < \frac{q^2-\Mrec^2}{2q^0}\,,
\end{equation}
there is one and only one (positive) solution of eq.~(\ref{eq:solveform1})
\begin{equation}\label{eq:inversefkskn}
\mmod{k}_{n}=\frac{q^2-\Mrec^2-2q^0\mmod{k}_{n+1}}
{2\lq q^0-\mmod{k}_{n+1}\,(1-y)\rq}\;.
\end{equation}
Having specified the size of $k_{n}$, we construct the vectors $\vec{k}_{n}$
and $\vec{k}_{n+1}$ in such a way that their vector sum
$\vec{k}=\vec{k}_{n}+\vec{k}_{n+1}$ is parallel to $\vec{\bar{k}}_n$, and
that the azimuth of $\vec{k}_{n+1}$ relative to $\vec{k}$ (the given
reference direction) is $\phi$.  Having constructed $k_n$ and $k_{n+1}$, we
can define $k=k_n+k_{n+1}$ and find $k_{\rm rec}=q-k$. We can now compute
$\beta$ according to eq.~(\ref{eq:betaboost}), and obtain
\begin{equation}
k_i=\boost^{-1} \,\bar{k}_i,\qquad\qquad i=1,\ldots,n-1\,.
\end{equation}
Thus, the inverse mapping exists provided
\begin{equation}\label{eq:xiboundFSR}
\xi < \frac{q^2-\Mrec^2}{q^2}\,,
\end{equation}
and it is always unique.

\section{\POWHEG{} in the \CS{} framework}
\label{sec:powhegcs}
The separation of $R$ into singular components may require particular attention.
First of all, if one uses the formula
\begin{equation}\label{eq:RpartCS}
R^\ctindr = \frac{{\cal D}_\ctindr}
{\sum_{\alpha_r^\prime} {\cal D}_{\alpha_r^\prime}}\,R\,,
\end{equation}
a problem may arise due to zeros in the denominator (in fact, the CS
counterterms are not necessarily positive). We have not tried to prove
that this can really happen. However, the problem is easily
solved by using instead (for example)
\begin{equation}
R^\ctindr = \frac{{\cal D}_\ctindr^2}
{\sum_{\alpha_r^\prime} {\cal D}_{\alpha_r^\prime}^2}\,R\,.
\end{equation}
In case  the $n$-body
cross section possesses singular regions, as discussed at the beginning of
section~\ref{sec:SubtractionFormalism}, a further problem may arise.
Formula~(\ref{eq:RpartCS})
may not be adequate to partition the different singular components
of $R$. This is better seen with an example.
Consider the $Z+{\rm jet}$ production
process. The $n$-body process corresponds to a $Z+l$ final state,
where $l$ is a light parton, and the $(n+1)$-body process corresponds
to a $Z+l_1+l_2$ final state, where $l_1$ and $l_2$ are light partons.
Consider now the counterterm corresponding to $l_2$ becoming collinear to
an initial-state parton. It is proportional
to the underlying $Z+l_1$~parton cross
section. It is therefore also singular when $l_1$ becomes collinear to
an initial-state parton, since the $Z+l_1$ cross section is singular
in this limit.
In standard NLO calculation, an infrared-safe observable $O$,
that vanishes when two singular regions are approached at
the same time, suppresses the singular regions
of the underlying Born process in the counterterm
(see eq.~(\ref{eq:sub0})).
This problem is easily solved by writing
\begin{equation}\label{eq:rpartCS}
R^\ctindr = \frac{H\(\BKinn^{(\ctindr)}\){\cal D}_\ctindr}
{\sum_{\alpha_r^\prime} H\(\BKinn^{(\alpha_r^\prime)}\)
{\cal D}_{\alpha_r^\prime}}\,R\,,
\end{equation}
where $H$ is a positive function that vanishes when its argument approaches
an $n$-body singular configuration. We again have
\begin{equation}
\sum_{\ctindr} R^\ctindr = R\,,
\end{equation}
and now $R^\ctindr$ is singular only in the $\ctindr$ region.

If one wishes to enforce transverse-momentum ordering (see sec.~\ref{sec:trmomord}),
one can first separate $R$ as in eqs.~(\ref{FKSpart}) and~(\ref{FKSpart1}),
using the $\Sfun$ functions defined in eqs.~(\ref{Soijdef}) and~(\ref{eq:szipmdef}),
with the $d$ functions defined as in eqs.~(\ref{eq:ditmord}) and~(\ref{eq:dijtmord}).
The contributions to $R$
corresponding to a given singular region can then be separated according
to each possible spectator using a formula similar to eq.~(\ref{eq:rpartCS}).

We now discuss the various dipole configurations of the CS
approach. We use the notation introduced in ref.~\cite{Catani:1996vz},
slightly modified in order to be consistent with the notation of the
rest of this paper.

\subsection{Final-state singularity with final-state spectator}
\subsubsection{Radiation and barred variables}
The labels $i,j,k$ denote the radiator, the emitted, and the
spectator partons respectively.
Without loss of generality, we assume that the radiated parton $j$ is
the $(n+1)$th parton.
We introduce the variable $y_{ij,k}$, that is zero in the
soft and collinear limit,
\begin{equation}
\label{eq:y_ijk}
y_{ij,k} = \frac{k_i\cdot k_j}{k_i \cdot k_j+(k_i+k_j)\cdot k_k} =
\frac{2\,k_i\cdot  k_j}{(k_i+k_j+k_k)^2} \,.
\end{equation}
The $n$ barred momenta, are defined as follows
\begin{eqnarray}
\label{eq:k_k_FS_FS}
\bar{k}_k &=& \frac{1}{1-y_{ij,k}} \,k_k \;,\\
\label{eq:k_ij_FS_FS}
\bar{k}_i &=& \bar{k}_{ij} = k_i  + k_j  - \frac{y_{ij,k}}{1-y_{ij,k}}
\,k_k\;, \\
\label{eq:k_r_FS_FS}
\bar{k}_r &=& k_r\,,  \qquad\qquad r=1,\ldots, n+1, \qquad\quad  r\ne i,j,k\,.
\end{eqnarray}
Notice that we have introduced the variable $\bar{k}_{ij}=\bar{k}_i$, so that
$\bar{k}_k$ and $\bar{k}_{ij}$ correspond to $\tilde{p}_k$ and $\tilde{p}_{ij}$
of ref.~\cite{Catani:1996vz}.
The expression~(\ref{eq:y_ijk}) for $y_{ij,k}$ is determined by imposing
$\bar{k}^2_{ij}=0$, and since
\begin{equation}
\label{eq:mom_cons_FS_FS}
k_i+k_j+k_k=\bar{k}_k +\bar{k}_{ij}\,,
\end{equation}
momentum conservation is enforced.

Observe that, as required, the barred momenta satisfy the momentum
conservation relation
\begin{equation}
\bar{k}_\splus+\bar{k}_\sminus   = \sum_{i=1}^{n} \bar{k}_i\,,
\end{equation}
where the initial-state barred momenta are defined as
\begin{eqnarray}
\bar{k}_\splus &=&   \kplus =\bxplus \Kplus\\
\bar{k}_\sminus &=& \kminus  = \bxminus \Kminus 
\end{eqnarray}
so that
\begin{equation}
\bxplusminus = \xplusminus\,.
\end{equation}
In addition, we define
\begin{equation}
\label{eq:tz_i}
{\tilde z}_i = \frac{k_i\cdot  k_k}{(k_i+k_j)\cdot k_k} =
\frac{k_i \cdot \bar{k}_k}{\bar{k}_{ij} \cdot \bar{k}_k}\,,
\end{equation}
that, in the collinear limit, is equal to the fraction of the momentum
carried by the $i$th particle.  We write eq.~(5.20) of ref.~\cite{Catani:1996vz}
(in $d=4$) in our notation as
%
%
\begin{equation}
\label{eq:dki_fin_state}
\lq dk_i \( \bar{k}_{ij}, \bar{k}_k \) \rq
= \frac{(2{\bar{k}}_{ij}\cdot {\bar{k}}_k)}{16\pi^2}
\;\frac{d\phi}{(2\pi)}
\;d{\tilde z}_i \;dy_{ij,k}
\;\stepf\big({\tilde z}_i(1-{\tilde z}_i)\big)
\,\stepf\big( y_{ij,k}(1-y_{ij,k})\big) \,\left( 1-y_{ij,k} \right)  \,,
\end{equation}
where $\phi$ is the azimuthal angle of $k_i$ in the centre-of-mass system of
$({\bar{k}}_{ij}+{\bar{k}}_k)$ and
\begin{equation}
\int d\phi = 2\pi \,.
\end{equation}
The integral of a generic finite quantity $F$ over the $(n+1)$-parton phase
space can then be written as
\begin{eqnarray}
 && 
\int d\xplus \;d\xminus \;d\PSnpo\(\xplus\Kplus+\xminus\Kminus;
k_1,\ldots,k_{n+1}\)\, 
\Lum(\xplus,\xminus)\nonumber\\
&&
 \hspace{3cm}  \times \;
F\(\xplus, \xminus;k_1,\ldots,k_{n+1}\) \nonumber\\
&=& \int d\bxplus \;d\bxminus \;d\PSn\(\bxplus\Kplus+\bxminus\Kminus;
\bar{k}_1,\ldots,\bar{k}_{n}\)\,  \lq dk_i \( \bar{k}_{ij}, \bar{k}_k \)\rq\, 
\Lum(\bxplus,\bxminus)\nonumber\\
&&
 \hspace{3cm}  \times \;
 F\(\xplus, \xminus;k_1,\ldots,k_{n+1}\)\nonumber\\
 &=& \int d\BKinn \; d\Rad \;
\Lum(\bxplus,\bxminus) \;  F\(\xplus, \xminus;k_1,\ldots,k_{n+1}\)\,,
\end{eqnarray}
where
\begin{equation}
\label{eq:Rad_FS_FS}
d \Rad = \frac{(2{\bar{k}}_{ij}\cdot {\bar{k}}_k)}{16\pi^2}
\;\frac{d\phi}{(2\pi)}
\;d{\tilde z}_i \;dy_{ij,k}
\;\stepf\big({\tilde z}_i(1-{\tilde z}_i)\big)
\,\stepf\big( y_{ij,k}(1-y_{ij,k})\big) \,\left( 1-y_{ij,k} \right).
\end{equation}

\subsubsection{Inverse construction}
\label{sec:CS_inverse_construction_FS_FS}
In this section, we describe the construction of the full $(n+1)$-particle
phase space, given the barred variables $\BKinn$ and the radiation variables.
The goal is to build the
momenta of the emitted particle, $k_j$, of the emitting particle, $k_i$, and
of the spectator particle, $k_k$, given $\bar{k}_{ij}$, $\bar{k}_k$ and the
three radiation variable: $y_{ij,k}$, ${\tilde z}_i$ and $\phi$. All the
other momenta remain unchanged, according to eq.~(\ref{eq:k_r_FS_FS}).

From eqs.~(\ref{eq:y_ijk}), (\ref{eq:tz_i}) and from momentum
conservation~(\ref{eq:mom_cons_FS_FS}) we have
\begin{eqnarray}
\label{eq:dotp_FS_FS_1}
k_i\cdot  k_j &=& y_{ij,k} \, \bar{k}_k \cdot \bar{k}_{ij}\\
\label{eq:dotp_FS_FS_2}
k_i \cdot \bar{k}_k &=& {\tilde z}_i \, \bar{k}_k \cdot \bar{k}_{ij}\\
\label{eq:dotp_FS_FS_3}
k_i \cdot \bar{k}_{ij} &=& y_{ij,k} \(1- {\tilde z}_i\) \, \bar{k}_k \cdot
\bar{k}_{ij}\,. 
\end{eqnarray}
We make a Lorentz boost $\boost$ to centre-of-mass frame of $(\bar{k}_{ij} +
\bar{k}_k)$, and we denote with a ``prime'' the momenta in this frame. We fix
the $z'$ axis parallel to $\bar{k}'_k$.  In this reference frame, the boosted
momenta have the following Lorentz components
\begin{eqnarray}
\bar{k}'_k &=& E(1,0,0,1)\\
\bar{k}'_{ij} &=& E(1,0,0,-1)\\
\label{eq:ki_prime_FS_FS}
k'_i &=& E'_i (1,\sin\th'_i\cos\phi,\sin\th'_i\sin\phi, \cos\th'_i)
\end{eqnarray}
where $\bar{k}_k \cdot \bar{k}_{ij} = 2E^2$, so that
\begin{equation}
E = \sqrt{\frac{\bar{k}_k \cdot \bar{k}_{ij}}{2}}\,,
\end{equation}
and eqs.~(\ref{eq:dotp_FS_FS_2}) and~(\ref{eq:dotp_FS_FS_3}), evaluated in
this frame, give
\begin{eqnarray}
2E^2 {\tilde z}_i &=& E'_i \, E\(1-\cos\th'_i\)\\
2E^2 \,y_{ij,k}(1-{\tilde z}_i) &=& E'_i \, E\(1+\cos\th'_i\).
\end{eqnarray}
Solving these equations, we derive
\begin{eqnarray}
\label{eq:E'_i}
 E'_i  &=& \sqrt{\frac{\bar{k}_k \cdot \bar{k}_{ij}}{2}} \lq y_{ij,k}(1-{\tilde
   z}_i)+{\tilde z}_i \rq\\ 
\label{eq:cos_th_i}
\cos\th'_i &=& = \frac{y_{ij,k}(1-{\tilde z}_i)-{\tilde
    z}_i}{y_{ij,k}(1-{\tilde z}_i)+{\tilde z}_i}\,, 
\end{eqnarray}
and we have so determined all the four components of $k'_i$.  We boost back
to the original frame and we obtain
\begin{equation}
k_i =\boost^{-1} k'_i\,.
\end{equation}
From eq.~(\ref{eq:k_k_FS_FS}) and from momentum conservation we can then
write
\begin{eqnarray}
k_k &=& (1-y_{ij,k}) \bar{k}_k\\
k_j &=& y_{ij,k} \bar{k}_k + \bar{k}_{ij} - k_i\,,
\end{eqnarray}
and this completes the task of building the $(n+1)$ final-state momenta.

\subsection{Final-state singularity with initial-state spectator}

\subsubsection{Radiation and barred variables}
In this case the radiator $i$ is a final-state parton and the spectator
$k$ is an initial-state one, that we assume for definiteness to be the
$\nplus$ parton. As before, without loss of generality, we assume that the
radiated parton $j$ is the $(n+1)$th parton. We introduce the variable
\begin{equation}
\label{eq:x_FS_IS}
x_{ij,\splus} = 1 - \frac{k_i\cdot  k_j}{(k_i+k_j)\cdot \kplus}\,,
%
\end{equation}
that approaches 1 in the soft and collinear limit, and the following $n$
barred momenta
\begin{eqnarray}
\bar{k}_i  &=& \bar{k}_{ij}  = k_i  + k_j  - (1 - x_{ij,\splus} ) \,\kplus  \;,\\
\bar{k}_r &=& k_r\,,  \qquad\qquad r=1,\ldots, n+1, \qquad\quad  r\ne i,j\,.
\end{eqnarray}
Momentum conservation reads
\begin{equation}
\bar{k}_\splus + \bar{k}_\sminus   = \sum_{i=1}^{n} \bar{k}_i\,,
\end{equation}
where
\begin{eqnarray}
\label{eq:bar_kp_FS_IS}
\bar{k}_\splus &=& x_{ij,\splus} \kplus = \bxplus \Kplus\,,\\
\bar{k}_\sminus &=& \kminus = \bxminus \Kminus\,,
\end{eqnarray}
and
\begin{eqnarray}
\label{eq:xp_FS_Is}
\bxplus &=& x_{ij,\splus} \; \xplus\,,\\
\label{eq:xm_FS_Is}
\bxminus &=& \xminus\,.
\end{eqnarray}
Introducing the variable
\begin{equation}
\label{eq:z_FS_IS}
{\tilde z}_i = \frac{k_i \cdot \kplus}{(k_i+k_j)\cdot \kplus}\,,
\end{equation}
that, in the collinear limit, is equal to the fraction of the momentum
carried by the $i$th particle, we can write the integral of a generic finite
quantity $F$ over the $(n+1)$-parton phase space as
\begin{eqnarray}
&& \int d\xplus \;d\xminus \;d\PSnpo\(\xplus\Kplus+\xminus\Kminus;
k_1,\ldots,k_{n+1}\)\, 
\Lum(\xplus,\xminus)\nonumber\\
&& \hspace{3cm}  \times \;
F\(\xplus, \xminus;k_1,\ldots,k_{n+1}\) \nonumber\\
&=& \int d\xplus \;d\xminus \; dx \; 
d\PSn\(x\, \xplus  \Kplus +\xminus\Kminus; \bar{k}_1\ldots \bar{k}_n\) 
\lq dk_i\(\bar{k}_{ij}; \kplus, x\) \rq \, 
\Lum(\xplus,\xminus) \nonumber\\
&& \hspace{3cm}  \times \;
F\(\xplus, \xminus;k_1,\ldots,k_{n+1}\) \,,
\end{eqnarray}
where $\left[ dk_i(\bar{k}_{ij};\kplus, x) \right]$ is given by eq.~(5.48) of
ref.~\cite{Catani:1996vz} in $d=4$ dimensions
%
%
\begin{equation}
\left[ dk_i(\bar{k}_{ij};\kplus, x) \right]
= \frac{(2\bar{k}_{ij} \cdot \kplus)}{16\pi^2}
\;\frac{d\phi}{2\pi}
\;d{\tilde z}_i \;dx_{ij,\splus}
\;\stepf\big({\tilde z}_i(1-{\tilde z}_i)\big)
\;\stepf\big(x(1-x)\big)\;\delta(x-x_{ij,\splus})\,,
\end{equation}
and $\phi$ is the azimuthal angle of $k_i$ in the centre-of-mass system of
$(\bar{k}_{ij}+\kplus)$.

Performing the integration in $x$, that fixes $x=x_{ij,\splus}$, with the
change of variable of eqs.~(\ref{eq:xp_FS_Is}) and~(\ref{eq:xm_FS_Is}), we
obtain
\begin{eqnarray}
\label{eq:O_FS_Is}
&&\int d\bxplus \;d\bxminus \;
\frac{dx_{ij,\splus}}{x_{ij,\splus}}\; 
d\PSn\( \bxplus  \Kplus +\bxminus\Kminus; \bar{k}_1\ldots \bar{k}_n\) 
\Lum\(\frac{\bxplus}{x_{ij,\splus}},\bxminus\) \nonumber\\
&& \hspace{2cm}  \times  \;
\frac{(2\bar{k}_{ij} \cdot \kplus)}{16\pi^2}
\;\frac{d\phi}{2\pi}
\;d{\tilde z}_i 
\;\stepf\big({\tilde z}_i(1-{\tilde z}_i)\big)
\;\stepf\big(x_{ij,\splus}(1-x_{ij,\splus})\big)\nonumber\\
&& \hspace{2cm}  \times \;
\stepf\(x_{ij,\splus} - \bxplus\) 
F\(\xplus, \xminus;k_1,\ldots,k_{n+1}\) \nonumber\\
&=& 
\int d\BKinn \; d\Rad \;
\Lum\(\frac{\bxplus}{x_{ij,\splus}},\bxminus\) \;  
F\(\xplus, \xminus;k_1,\ldots,k_{n+1}\)\,,
\end{eqnarray}
where
\begin{eqnarray}
d \Rad &=&
\frac{(2\bar{k}_{ij}\cdot  \kplus)}{16\pi^2}
\;\frac{d\phi}{2\pi}
\;d{\tilde z}_i  \; \frac{dx_{ij,\splus}}{x_{ij,\splus}}
\;\stepf\big({\tilde z}_i(1-{\tilde z}_i)\big)
\;\stepf\big(x_{ij,\splus}(1-x_{ij,\splus})\big)
\, \stepf\(z - \bxplus\) .
\end{eqnarray}

\subsubsection{Inverse construction}
\label{sec:CS_inverse_construction_FS_IS}
In order to reconstruct the $(n+1)$ momenta from $\BKinn$ and $\Rad$, we
follow closely the procedure used in
section~\ref{sec:CS_inverse_construction_FS_FS}.  More precisely, we have to
reconstruct $k_i$, $k_j$ and $\kplus$ given $\bar{k}_{ij}$, $\bar{k}_\splus$
and the radiation variables $x_{ij,\splus}$, ${\tilde z}_i$ and $\phi$.

From eq.~(\ref{eq:bar_kp_FS_IS}) we immediately have
\begin{equation}
\kplus = \frac{\bar{k}_\splus}{ x_{ij,\splus}}\,,
\end{equation}
and using eqs.~(\ref{eq:x_FS_IS}) and~(\ref{eq:z_FS_IS}) we can write
\begin{eqnarray}
k_i\cdot  k_j &=& (1-x_{ij,\splus}) \, \kplus \cdot \bar{k}_{ij}\\
k_i \cdot \kplus &=& {\tilde z}_i \, \kplus \cdot \bar{k}_{ij}\\
k_i\cdot  \bar{k}_{ij} &=& (1-x_{ij,\splus}) \(1- {\tilde z}_i\) \, \bar{k}_k
\cdot \bar{k}_{ij}\,.
\end{eqnarray}
These equations are similar to
eqs.~(\ref{eq:dotp_FS_FS_1})--(\ref{eq:dotp_FS_FS_3}) with the substitutions
\begin{eqnarray}
\bar{k}_k &\leftrightarrow& \kplus\\
 y_{ij,k} &\leftrightarrow& 1-x_{ij,\splus}
\end{eqnarray}
so that, in the centre-of-mass of the $(\bar{k}_{ij}+\kplus)$, the energy and
the angle $\th'_i$ of $k_i$ with the $\kplus'$ direction are given by (see
eqs.~(\ref{eq:E'_i}) and~(\ref{eq:cos_th_i}))
\begin{eqnarray}
 E'_i  &=& \sqrt{\frac{\kplus \cdot \bar{k}_{ij}}{2}} \lq (1-x_{ij,\splus})
 (1-{\tilde z}_i)+{\tilde z}_i \rq\\ 
\cos\th'_i &=& = \frac{(1-x_{ij,\splus})(1-{\tilde z}_i)-{\tilde
    z}_i}{(1-x_{ij,\splus})(1-{\tilde z}_i)+{\tilde z}_i}\,. 
\end{eqnarray}
We can boost back to the original frame and obtain $k_i$.  The last momentum
$k_j$ is constrained by momentum conservation
\begin{equation}
k_j = \bar{k}_{ij} - k_i + (1-x_{ij,\splus})\, \kplus\,.
\end{equation}

\subsection{Initial-state singularity with final-state spectator}

\subsubsection{Radiation and barred variables}
In the case where the parton $i$ (assumed here to be the $(n+1)$th parton)
is emitted from an initial-state parton,
that we take for definiteness to be the $\nplus$ one, and in the presence of
a final-state spectator $k$, we define the following $n$ barred final-state momenta
\begin{eqnarray}
\bar{k}_{k}  &=& k_i  + k_k  - (1 - x_{ik,\splus} ) \,\kplus \\
\bar{k}_{r}  &=& k_{r}   \qquad\qquad r=1,\ldots, n+1, \qquad\quad
r\ne i,k\,,
\end{eqnarray}
where the variable
\begin{equation}
x_{ik,\splus} = 1 - \frac{k_i\cdot  k_k}{(k_i+k_k)\cdot \kplus}
\end{equation}
approaches 1 when $k_i$ becomes soft, and approaches $z$ in the collinear
limit, when $k_i = (1-z)\, \kplus$. 
Momentum conservation reads
\begin{equation}
\bar{k}_\splus + \bar{k}_\sminus   = \sum_{i=1}^{n} \bar{k}_i\,,
\end{equation}
where
\begin{eqnarray}
\bar{k}_\splus &=& x_{ik,\splus}\, \kplus = \bxplus \Kplus\\
\bar{k}_\sminus &=& \kminus = \bxminus \Kminus
\end{eqnarray}
so that
\begin{eqnarray}
\label{eq:xp_IS_Fs}
\bxplus &=& x_{ik,\splus}  \; \xplus  \\
\label{eq:xm_IS_Fs}
\bxminus &=& \xminus\,.
\end{eqnarray}
Introducing the variable 
\begin{equation}
u_i = \frac{k_i\cdot \kplus}{(k_i+k_k)\cdot \kplus}\,,
\end{equation}
we can write the integral of a generic finite quantity $F$ over the
$(n+1)$-parton phase space as
\begin{eqnarray}
&& \int d\xplus \;d\xminus \;d\PSnpo\(\xplus\Kplus+\xminus\Kminus;
k_1,\ldots,k_{n+1}\)\, 
\Lum(\xplus,\xminus)\nonumber\\
&& \hspace{3cm}  \times \;
F\(\xplus, \xminus;k_1,\ldots,k_{n+1}\) \nonumber\\
&=& \int d\xplus \;d\xminus \; dx \; 
d\PSn\(x\, \xplus  \Kplus +\xminus\Kminus; \bar{k}_1\ldots \bar{k}_n\) 
\lq dk_i\(\bar{k}_{k}; \kplus, x\) \rq \, 
\Lum(\xplus,\xminus) \nonumber\\
&& \hspace{3cm}  \times \;
F\(\xplus, \xminus;k_1,\ldots,k_{n+1}\) \,,
\end{eqnarray}
where $\left[ dk_i(\bar{k}_{k};\kplus, x) \right]$ is given by eq.~(5.72)
of ref.~\cite{Catani:1996vz} in $d=4$ dimensions
%
%
\begin{equation}
\left[ dk_i(\bar{k}_{k};\kplus, x) \right]
= \frac{(2\bar{k}_{k} \cdot \kplus)}{16\pi^2}
\;\frac{d\phi}{2\pi}
\;du_i \;dx_{ik,\splus}
\;\stepf\big(u_i(1-u_i)\big)
\;\stepf\big(x(1-x)\big)\;\delta(x-x_{ik,\splus})\,,
\end{equation}
and $\phi$ is the azimuthal angle of $k_i$ in the centre-of-mass system of
$(\bar{k}_{k}+\kplus)$.

Performing the integration in $x$, that fixes $ x=x_{ik,\splus}$, with the
change of variable of eqs.~(\ref{eq:xp_IS_Fs}) and~(\ref{eq:xm_IS_Fs}), we
obtain
\begin{eqnarray}
\label{eq:O_IS_Fs}
&& \int d\bxplus \;d\bxminus \; \frac{dx_{ik,\splus}}{x_{ik,\splus}}\;
d\PSn\( \bxplus  \Kplus +\bxminus\Kminus; \bar{k}_1\ldots \bar{k}_n\) 
\Lum\(\frac{\bxplus}{x_{ik,\splus}},\bxminus\) \nonumber\\
&& \hspace{2cm}  \times  \;
\frac{(2\bar{k}_{k} \cdot \kplus)}{16\pi^2}
\;\frac{d\phi}{2\pi}
\;du_i 
\;\stepf\big(u_i(1- u_i)\big)
\;\stepf\big(x_{ik,\splus}(1-x_{ik,\splus})\big)\nonumber\\
&& \hspace{2cm}  \times \;
\stepf\(x_{ik,\splus} - \bxplus\) 
F\(\xplus, \xminus;k_1,\ldots,k_{n+1}\) \nonumber\\
&=& 
\int d\BKinn \; d\Rad \;
\Lum\(\frac{\bxplus}{x_{ik,\splus}},\bxminus\) \;  
F\(\xplus, \xminus;k_1,\ldots,k_{n+1}\)\,,
\end{eqnarray}
where
\begin{equation}
d \Rad =
\frac{(2\bar{k}_{k} \cdot \kplus)}{16\pi^2}
\;\frac{d\phi}{2\pi}
\;du_i \; \frac{dx_{ik,\splus}}{x_{ik,\splus}}
\;\stepf\big(u_i(1-u_i)\big)
\;\stepf\big(x_{ik,\splus}(1-x_{ik,\splus})\big)
\; \stepf\(x_{ik,\splus} - \bxplus\) .
\end{equation}

\subsubsection{Inverse construction}
\label{sec:CS_inverse_construction_IS_FS}
This case is completely analogous to the one in
section~\ref{sec:CS_inverse_construction_FS_IS} with the substitutions
\begin{eqnarray}
k_j &\leftrightarrow& k_k \\
\bar{k}_{ij}  &\leftrightarrow& \bar{k}_{k} \\
x_{ij,\splus} &\leftrightarrow& x_{ik,\splus}\\
{\tilde z}_i &\leftrightarrow& u_i
\end{eqnarray}
so that
\begin{eqnarray}
\kplus &=& \frac{\bar{k}_\splus}{ x_{ik,\splus}}\\
 E'_i  &=& \sqrt{\frac{\kplus \cdot \bar{k}_{k}}{2}} \lq (1-x_{ik,\splus})
 (1-u_i)+u_i \rq\\  
\cos\th'_i &=& = \frac{(1-x_{ik,\splus})(1-u_i)-u_i}{(1-x_{ik,\splus})
  (1-u_i)+u_i}\,.   
\end{eqnarray}
where $E'_i$ and $\th'_i$ are the energy and the angle that the vector $k'_i$
forms with the direction of $\kplus'$, in the centre-of-mass of the
$(\bar{k}_{k}+\kplus)$.  The four-vector $k_i$ is obtained from $k'_i$ with a
boost back in the original frame, while $k_k$ is constrained by momentum
conservation
\begin{equation}
k_k = \bar{k}_{k} - k_i + (1-x_{ik,\splus})\, \kplus\,.
\end{equation}

\subsection{Initial-state singularity with initial-state spectator}
\label{sec:CS_IS_IS}
\subsubsection{Radiation and barred variables}
In the case where the parton $i$ (that we take to be the $(n+1)$th parton)
is emitted from an initial-state parton,
that we take for definiteness to be the $\nplus$ one, and in the presence of
an initial-state spectator, the $\nminus$ parton, we define the following $n$
barred final-state momenta
\begin{equation}
\bar{k}^{\mu}_r = \Lambda^{\mu}_{\;\; \nu}(K,{\bar K})
\;k^{\nu}_r  \qquad\qquad r=1,\ldots, n+1, \qquad\quad r\ne i\,,
\end{equation}
where the boost tensor is given by
\begin{eqnarray}
\Lambda^{\mu}_{\;\; \nu}(K,{\bar K}) &=& g^{\mu}_{\;\; \nu} -
\frac{2 (K+{\bar K})^\mu (K+{\bar K})_\nu}{(K+{\bar K})^2}
+\frac{2 {\bar K}^\mu K_\nu}{K^2}\;,\\
K &=& \kplus + \kminus - k_i = \sum_{\stackrel{r=1}{r\ne i}}^{n+1}k_r\;, \\
{\bar K} &=& x_{i,\splus\sminus} \, \kplus + \kminus = 
\sum_{\stackrel{r=1}{r\ne i}}^{n+1} \bar{k}_r\;,
\end{eqnarray}
and
\begin{equation}
x_{i,\splus\sminus} = 1 - \frac{\(\kplus + \kminus\) \cdot k_i}{\kplus\cdot
  \kminus} \,. 
\end{equation}
Notice that, in the soft limit, $x_{i,\splus\sminus}$ approaches 1, while, in
the collinear limit, i.e.\ $k_i=(1-z)\,\kplus$, it approaches $z$ and that
${\bar K}^\mu = \Lambda^{\mu}_{\;\; \nu}(K,{\bar K}) \, K^\nu$, so that
${\bar K}^2 = K^2$.

Momentum conservation reads
\begin{equation}
\bar{k}_\splus + \bar{k}_\sminus = \sum_{i=1}^n \bar{k}_i\, 
\end{equation}
where
\begin{eqnarray}
\label{eq:bkplus_IS_IS}
\bar{k}_\splus &=& x_{i,\splus\sminus} \,\kplus = \bxplus \Kplus\\
\label{eq:bkminus_IS_IS}
\bar{k}_\sminus &=& \kminus = \bxminus \Kminus
\end{eqnarray}
so that
\begin{eqnarray}
\label{eq:xp_IS_Is}
\bxplus &=& x_{i,\splus\sminus} \; \xplus \\
\label{eq:xm_IS_Is}
\bxminus &=& \xminus\,.
\end{eqnarray}
Introducing the variable
\begin{equation}
{\tilde v}_i = \frac{\kplus \cdot k_i}{\kplus \cdot \kminus} 
\end{equation}
we can write the integral of a generic finite quantity $F$ over the
$(n+1)$-parton phase space as
\begin{eqnarray}
&& \int d\xplus \;d\xminus \;d\PSnpo\(\xplus\Kplus+\xminus\Kminus;
k_1,\ldots,k_{n+1}\)\, 
\Lum(\xplus,\xminus)\nonumber\\
&& \hspace{3cm}  \times \;
F\(\xplus, \xminus;k_1,\ldots,k_{n+1}\) \nonumber\\
&=& \int d\xplus \;d\xminus \; dx \; 
d\PSn\(x\, \xplus  \Kplus +\xminus\Kminus; \bar{k}_1\ldots \bar{k}_n\) 
\lq dk_i\(\kplus,\kminus, x\) \rq \, 
\Lum(\xplus,\xminus) \nonumber\\
&& \hspace{3cm}  \times \;
F\(\xplus, \xminus;k_1,\ldots,k_{n+1}\) \,,
\end{eqnarray}
where $\left[  dk_i\(\kplus,\kminus, x\) \right]$ is given by eq.~(5.151)
of ref.~\cite{Catani:1996vz} in $d=4$ dimensions
%
%
\begin{equation}
\left[ dk_i(\kplus,\kminus, x) \right]
= \frac{(2\kplus\cdot \kminus)}{16\pi^2}
\;\frac{d\phi}{2\pi}
\;d\tilde{v}_i \;dx_{i,\splus\sminus}
\;\stepf\(\tilde{v}_i\)\;\stepf\(1-\frac{\tilde{v}_i}{1-x}\)
\;\stepf\big(x(1-x)\big)\;\delta(x-x_{i,\splus\sminus})\,,
\end{equation}
and $\phi$ is the azimuthal angle of $k_i$ in the centre-of-mass system of
$(\kplus+\kminus)$ or in the laboratory frame.

Performing the integration in $x$, that fixes $x=x_{i,\splus\sminus}$, with
the change of variable of eqs.~(\ref{eq:xp_IS_Is}) and~(\ref{eq:xm_IS_Is}),
we obtain
\begin{eqnarray}
\label{eq:O_IS_Is}
&& \int d\bxplus \;d\bxminus \; \frac{dx_{i,\splus\sminus}}{x_{i,\splus\sminus}}\;
d\PSn\( \bxplus  \Kplus +\bxminus\Kminus; \bar{k}_1\ldots \bar{k}_n\) 
\Lum\(\frac{\bxplus}{x_{i,\splus\sminus}},\bxminus\) \nonumber\\
&& \hspace{2cm}  \times  \;
\frac{(2 \kplus\cdot \kminus)}{16\pi^2}
\;\frac{d\phi}{2\pi}
\;d\tilde{v}_i
\;\stepf\(\tilde{v}_i\)\;\stepf\(1-\frac{\tilde{v}_i}{1-x_{i,\splus\sminus}}\)
\;\stepf\big(x_{i,\splus\sminus}(1-x_{i,\splus\sminus})\big)\nonumber\\
&& \hspace{2cm}  \times \;
\stepf\(x_{i,\splus\sminus} - \bxplus\) 
F\(\xplus, \xminus;k_1,\ldots,k_{n+1}\) \nonumber\\
&=& 
\int d\BKinn \;  d\Rad \;
\Lum\(\frac{\bxplus}{x_{i,\splus\sminus}},\bxminus\) \;  
F\(\xplus, \xminus;k_1,\ldots,k_{n+1}\)\,,
\end{eqnarray}
where
\begin{eqnarray}
\label{eq:ISIS_dphi_rad}
d \Rad &=&
\frac{(2\kplus\cdot \kminus )}{16\pi^2}
\;\frac{d\phi}{2\pi}
\;d\tilde{v}_i\;\frac{dx_{i,\splus\sminus}}{x_{i,\splus\sminus}}
\;\stepf\(\tilde{v}_i\)\;\stepf\(1-\frac{\tilde{v}_i}{1-x_{i,\splus\sminus}}\)
\;\stepf\big(x_{i,\splus\sminus}(1-x_{i,\splus\sminus})\big) \nonumber\\
&&\hspace{1cm}  \times
\stepf\(x_{i,\splus\sminus} - \bxplus\) \;.
\end{eqnarray}

\subsubsection{Inverse construction}
\label{sec:CS_inverse_construction_IS_IS}
To reconstruct the full $(n+1)$-particle final state, given $\bar{k}_\splus$,
$\bar{k}_\sminus$ (i.e.\ $\bxplus$ and $ \bxminus$), the $n$ final-state
momenta $\bar{k}_r$ and the three radiation variables $x_{i,\splus\sminus}$,
${\tilde{v}_i}$ and $\phi$, we follow the same procedure already used in
section~\ref{sec:CS_inverse_construction_FS_FS}.  From
eqs.~(\ref{eq:bkplus_IS_IS}) and~(\ref{eq:bkminus_IS_IS}) we immediately have
\begin{eqnarray}
\label{eq:ISIS_kpkm}
\kplus &=& \frac{\bar{k}_\splus}{x_{i,\splus\sminus}}\\
\kminus &=& \bar{k}_\sminus 
\end{eqnarray}
that is
\begin{equation}
\label{eq:ISIS_xpxm}
\xplus = \frac{\bxplus}{x_{i,\splus\sminus}}\,, \hspace{2cm}
\xminus = \bxminus \,.
\end{equation}
We compute then
\begin{eqnarray}
k_i  \cdot \kplus &=& {\tilde v}_i \, \kplus  \cdot \kminus\\
k_i  \cdot \kminus &=& (1-x_{i,\splus\sminus}-{\tilde v}_i) \, \kplus
\cdot \kminus 
\end{eqnarray}
in the centre-of-mass system of $(\kplus+\kminus)$, and we find that the
energy $E'_i$ and angle $\th'_i$ that the vector $k'_i$ forms with the
$\kplus'$ direction are given by
\begin{eqnarray}
\label{eq:ISIS_energy}
E'_i &=&  \sqrt{\frac{\kplus  \cdot \kminus}{2}}\,(1-x_{i,\splus\sminus})\\
\label{eq:ISIS_angle}
\cos\th'_i &=& \frac{1-2{\tilde v}_i-
  x_{i,\splus\sminus}}{1-x_{i,\splus\sminus}}\,. 
\end{eqnarray}
We boost $k'_i$ back in the laboratory frame and we obtain $k_i$.  Once $k_i$
is known, we can build the two vectors
\begin{eqnarray}
K &=& \kplus + \kminus - k_i \;, \\
{\bar K} &=& \bar{k}_\splus + \kminus \;,
\end{eqnarray}
and the inverse of the boost tensor $\Lambda$
\begin{equation}
\Lambda_{\mu\nu}^{-1}(K,{\bar K}) = g_{\mu \nu} -
\frac{2 (K+{\bar K})_\mu (K+{\bar K})_\nu}{(K+{\bar K})^2}
+\frac{2 {K}_\mu \bar{K}_\nu}{K^2}\;,
\end{equation}
and compute the remaining  $n$ momenta
\begin{equation}
k_r = \Lambda^{-1}(K,{\bar K}) \, \bar{k}_r\,, \qquad\qquad
r=1,\ldots,n\,. 
\end{equation}

\section{Examples}
\label{sec:examples}

In this section we show in detail how to implement the \POWHEG{} formalism in
two simple cases: $e^+e^- \to q \bar{q}$ and vector-boson production at a
hadron-hadron collider, i.e.~$h_{\splus} h_{\sminus} \rightarrow V$.

\subsection{$\boldsymbol{e^+e^- \to q \bar{q}}$ in the \CS{} formalism}
\label{sec:epem_CS_formalism}
\subsubsection*{Born, virtual and real corrections}
We consider the scattering $e^-(p_1) \, e^+(p_2) \,\to\,\gamma^\star\,\to\, q(k_1)
\,\bar{q}(k_2)\, g(k_3)$ where, in the centre-of-mass frame,
\begin{eqnarray}
p_1 &=& \frac{q^0}{2}(1,0,0,1)\\
p_2 &=& \frac{q^0}{2}(1,0,0,-1)\\
k_1 &=& k_1^0\(1,\sin\th_1\cos\phi_1,\sin\th_1\sin\phi_1, \cos\th_1\)\\
k_2 &=& k_2^0\(1,\sin\th_2\cos\phi_2,\sin\th_2\sin\phi_2, \cos\th_2\),
\end{eqnarray}
and define
\begin{equation}
\label{eq:def_x1x2}
 q = p_1+p_2\,,\quad\quad k_3 = q-k_1 - k_2\,,\quad\quad
 x_{\{1, 2, 3\}} = \frac{2\, q\cdot k_{\{1, 2, 3\}}}{q^2} \,. 
\end{equation}
The Born cross section $B$, in arbitrary units, is given by
\begin{equation}
\label{eq:Born_CS}
 B\!\(\bar{k}_1,\bar{k}_2\) \equiv B\!\(\bar{\th}_1\)= 1 + \cos^2 \bar{\th}_1
 \,,  
\end{equation}
where $\bar{k}_1$ and $\bar{k}_2$ are the momenta of the quark and antiquark
at Born level, and are given by
\begin{eqnarray}
\bar{k}_1 &=&\frac{q^0}{2}\(1,\sin\bar{\th}_1\cos\bar{\phi}_1,
\sin\bar{\th}_1\sin\bar{\phi}_1,\cos\bar{\th}_1\)\\  
\bar{k}_2 &=& \frac{q^0}{2}\(1,-\sin\bar{\th}_1\cos\bar{\phi}_1,
-\sin\bar{\th}_1\sin\bar{\phi}_1,-\cos\bar{\th}_1\).
\end{eqnarray}
In this section we drop the flavour index $f_b$ in the Born cross section,
for ease of notation. In virtue of the azimuthal symmetry of the cross section,
we could fix from now on $\bar{\phi}_1=0$, and perform a random azimuthal rotation
at the end of the generation. Here we keep it variable.
The Born two-body phase-space element is given by
\begin{equation}
\label{CS:bPhi2}
d\bar{\Phi}_2 =  \frac{d^3 \bar{k}_1}{2 \bar{k}_1^0 (2 \pi)^3} \,
   \frac{d^3 \bar{k}_2}{2 \bar{k}_2^0 (2 \pi)^3} \,
   (2 \pi)^4 \,\delta^4 (q - \bar{k}_1 - \bar{k}_2) = 
\frac{1}{32 \pi^2} \, d\cos \bar{\th}_1 \, d\bar{\phi}_1\,,
\end{equation}
so that ${\bf \bar{\Phi}}_2 = \lg \bar{\theta}_1,\bar{\phi}_1 \rg$,
while the finite soft-virtual contribution of eq.~(\ref{eq:svcs}) is equal to
\begin{equation}
V(\bar{\th}_1) = \frac{\as}{\pi}\, \CF\,
B(\bar{\th}_1) \,.
\end{equation}
The real-emission term $R$ 
\begin{equation}
\label{eq:epem_real}
 R = \frac{8 \pi \CF \as}{q^2} 
\frac{ x_1^2 \, \(1 +\cos^2 \theta_1\)  + x_2^2\(1 + \cos^2 \, \theta_2\) }
{(1 - x_1) (1 - x_2)} 
\end{equation}
has two soft collinear regions. We call region 1 the $k_1\cdot k_3\,\to\,0$
($x_2\,\to\, 1$) region, and region 2 the $k_2\cdot k_3\,\to\,0$ ($x_1\,\to\,
1$) region.

\subsubsection*{Counterterms}
The counterterms can be computed using eqs.~(5.2) and~(5.7) of
ref.~\cite{Catani:1996vz} and are given by:
\begin{enumerate}
\item Region 1 ($k_1\cdot k_3\,\to\,0$): $\Rad^{(1)}=\lg{y_{13,2},\,\tilde z}_1,\,\phi \rg$
\begin{equation}
C^{(1)}\equiv {\cal D}_{13,2} = \frac{8 \pi \CF \as}{q^2} \frac{1}{y_{13,2}} 
\lq \frac{2}{1-{\tilde z}_1 \(1- y_{13,2}\)} -\(1+{\tilde z}_1\)\rq
B\!\(\bar{k}_1,\bar{k}_2\)
\end{equation}
where
\begin{equation}
\bar{k}_1 = \bar{k}_{13} = q - \frac{1}{1-y_{13,2}}k_2\,, \qquad \qquad
\bar{k}_2 = \frac{1}{1-y_{13,2}}k_2\,,
\end{equation}
so that 
\begin{equation}
\bar{\th}_1=\pi - \th_2\,, \qquad\qquad \bar{\phi}_1 = \(\pi +\phi_2 \){\rm
  mod} \; 2\pi\,, 
\end{equation}
since $\bar{k}_2$ is only rescaled with respect to $k_2$.

We can express the counterterm $C^{(1)}$ in terms of the $x_1$ and $x_2$
variables too, and we have
\begin{equation}
C^{(1)} = \frac{8 \pi \CF \as}{q^2} 
\lq \frac{1}{1-x_2}\(\frac{2}{2-x_1-x_2} - (1+x_1)\) + \frac{1-x_1}{x_2}\rq 
B\(\bar{k}_1,\bar{k}_2\)\,,
\end{equation}
where we have used eqs.~(\ref{eq:y_ijk}) and~(\ref{eq:tz_i})
\begin{equation}
 y_{13,2} = 1-x_2\,,\qquad \qquad {\tilde z}_1 = \frac{x_1+x_2-1}{x_2}\,,
\end{equation}
so that
\begin{equation}
\bar{k}_1 = q - \frac{k_2}{x_2}\,, \qquad \qquad
\bar{k}_2 =  \frac{k_2}{x_2}\,.
\end{equation}
This way, we can use two sets of radiation variables for the first singular
region: the $\lg{y_{13,2},\,\tilde z}_1,\,\phi \rg$ set or $\lg
x_1,\,x_2,\,\phi\rg$ one.

\item Region 2 ($k_2\cdot k_3\,\to\,0$): $\Rad^{(2)}=\lg{y_{23,1},\,\tilde z}_2,\,\phi \rg$

\begin{equation}
C^{(2)}\equiv{\cal D}_{23,1} = \frac{8 \pi \CF \as}{q^2} \frac{1}{y_{23,1}} 
\lq \frac{2}{1-{\tilde z}_2 \(1- y_{23,1}\)} -\(1+{\tilde z}_2\)\rq
B\!\(\bar{k}_1,\bar{k}_2\)
\end{equation}
where
\begin{equation}
\bar{k}_1 = \frac{1}{1-y_{23,1}}k_1\,,\qquad \qquad
\bar{k}_2 = \bar{k}_{23} = q - \frac{1}{1-y_{23,1}}k_1\,,
\end{equation}
so that 
\begin{equation}
\bar{\th}_1=\th_1\,, \qquad\qquad \bar{\phi}_1 = \phi_1\,,
\end{equation}
since $\bar{k}_1$ is only rescaled with respect to $k_1$.

In terms of the $x_1$ and $x_2$
variables, we have
\begin{equation}
C^{(2)} = \frac{8 \pi \CF \as}{q^2} 
\lq \frac{1}{1-x_1}\(\frac{2}{2-x_1-x_2} - (1+x_2)\) + \frac{1-x_2}{x_1}\rq 
B\!\(\bar{k}_1,\bar{k}_2\)
\end{equation}
where we have used eqs.~(\ref{eq:y_ijk}) and~(\ref{eq:tz_i})
\begin{equation}
 y_{23,1} = 1-x_1\,,\qquad \qquad {\tilde z}_2 = \frac{x_1+x_2-1}{x_1}\,,
\end{equation}
so that
\begin{equation}
\bar{k}_1 = \frac{k_1}{x_1}\,,\qquad \qquad
\bar{k}_2 = \bar{k}_{23} = q - \frac{k_1}{x_1}\,.
\end{equation}
This way, we can use two sets of radiation variables for the second singular
region: the $\lg{y_{23,1},\,\tilde z}_2,\,\phi \rg$ set or $\lg
x_1,\,x_2,\,\phi\rg$ one.
\end{enumerate}

\subsubsection*{Radiation phase space and inverse construction}
The radiation phase-space element of eq.~(\ref{eq:Rad_FS_FS}) has the same
functional form in both the two regions:
\begin{eqnarray} 
d\Rad &=&
\frac{q^2}{16\pi^2}\frac{d\phi}{2\pi}\,dz\,dy\,(1-y)\,\stepf\big( z(1-z)\big)
\,\stepf\big( y(1-y)\big)\\
&=& \frac{q^2}{16\pi^2}\frac{d\phi}{2\pi}\,dx_1\,dx_2\,\stepf(1-x_1)
\,\stepf(1-x_2)\,\stepf(x_1+x_2-1)\,,
\end{eqnarray}
and we can use either of the two sets of radiation variables: $\lg y,z,\phi \rg$
or $\lg x_1,x_2,\phi\rg$,
where we identify $y=y_{13,2}$, $z=\tilde{z}_1$ in region 1, or
 $y=y_{23,1}$, $z=\tilde{z}_2$ in region 2.
In each of the two regions, we can express the angles $\th_1$ and $\th_2$ in
the real term~(\ref{eq:epem_real}) in terms of the Born barred variables,
$\bar{\th}_1$ and $\bar{\phi}_1$, and in terms of the radiation variables,
that here we take to be $x_1$, $x_2$ and $\phi$, once all the $3$ final-state
momenta have been constructed, according to the procedure described in
section~\ref{sec:CS_inverse_construction_FS_FS}.  For example, considering
the first singular region and the two Born variables $\bar{\th}_1$ and
$\bar{\phi}_1$ (that fix uniquely $\bar{k}_1$ and $\bar{k}_2$), we have
\begin{equation}
\bar{\th}_2= \pi- \bar{\th}_1\,,\qquad \qquad  
\bar{\phi}_2 = \(\bar{\phi}_1 + \pi\)\,{\rm mod}\, 2\pi\,,
\end{equation}
and $k'_1$ of eq.~(\ref{eq:ki_prime_FS_FS}) is characterized by
\begin{eqnarray}
E'_1 &=& \frac{\sqrt{q^2}}{2} x_1\,,\\
\cos\th'_1 &=& \frac{2 - 2 x_1 - 2 x_2 + x_1 x_2}{x_1\,x_2}\,,
\end{eqnarray}
and by a random angle $\phi$. Observe that, in this case, no boosts are needed
for the inverse construction, since the dipole rest frame coincides with the
$e^+e^-$ centre-of-mass frame.
The vector $k_1$ is simply obtained as follows
\begin{equation}
k_1 = R_z(\bar{\phi}_2)\, R_y(\bar{\th}_2)\, k'_1\,,
\end{equation}
where $R_{z/y}(\alpha)$ is a rotation around the $z/y$ direction of an
angle $\alpha$.  The remaining two momenta are then given by
\begin{equation}
  k_2 = x_2 \, \bar{k}_2\,,    \qquad\qquad  k_3 = q - k_1 - k_2\,.
\end{equation}
The corresponding results for the second region are obtained from the
previous ones with the exchange of indexes $1\,\leftrightarrow \, 2$.

\subsubsection*{Generation of the Born variables}
The $\bar{B}$ function of eq.~(\ref{eq:bbdef}) is then given by
\begin{equation}
\label{eq:epem_bbar}
\bar{B}\!\(\bar{\th}_1\) = B\!\(\bar{\th}_1\) +
V\!\(\bar{\th}_1\) + \int d\Rad \lq R^{(1)} - C^{(1)}\rq +\int d\Rad \lq
R^{(2)} - C^{(2)}\rq,
\end{equation}
where
\begin{equation}
R^{(1)} = R \, \frac{C^{(1)}}{C^{(1)}+C^{(2)}}\,, \qquad \qquad
R^{(2)} = R \, \frac{C^{(2)}}{C^{(1)}+C^{(2)}}\,.
\end{equation}
In order to generate the two Born variables $\bar{\th}_1$ and $\bar{\phi}_1$
distributed according to eq.~(\ref{eq:epem_bbar}), we need to construct the
$\tilde{B}$ function of eq.~(\ref{eq:btdef}).  We then make the following
change of variables
\begin{equation}
x_1 = \Xrad^{(1)}\,, \qquad
x_2 = 1-x_1 + x_1\,\Xrad^{(2)}\, \qquad 
\phi = 2\pi\,\Xrad^{(3)}
\end{equation}
so that
\begin{equation}
d\Rad = \frac{q^2}{16\pi^2} \Xrad^{(1)}\, d\Xrad^{(1)} \,d\Xrad^{(2)}\, d\Xrad^{(3)}
\end{equation}
and we obtain
\begin{equation}
\tilde{B}\(\bar{\th}_1,\Xrad\)=
B\(\bar{\th}_1\) + V\(\bar{\th}_1\) 
+ \frac{q^2}{16\pi^2}\, \Xrad^{(1)}\,\lq \(R^{(1)} - C^{(1)}\) +  
\(R^{(2)} - C^{(2)}\)\rq,
\end{equation}
that satisfies
\begin{equation}
\int_0^1 d\Xrad^{(1)}\int_0^1 d\Xrad^{(2)}\int_0^1 d\Xrad^{(3)}\;
\tilde{B}\!\(\bar{\th}_1,\Xrad\) =
\bar{B}\!\(\bar{\th}_1\) .
\end{equation}
The function $\tilde{B}$ can now be integrated over the full 3-body phase
space, using an integrator that can generate unweighted events, like the
\SPRING-\BASES{} package, so that one can generate the Born configurations
very efficiently.

\subsubsection*{Generation of the radiation variables}
Once we have generated the Born configuration distributed according to the
$\bar{B}$ function, one must generate the hardest radiation. We first define
the $\kt$ of the radiation. The definition is ambiguous to some extent. The
only requirements that it must satisfy is that it must be of the order of the
radiation transverse momentum in the collinear limit, and that it must
coincide with it in the soft-collinear limit. A suitable definition is
\begin{equation}
\label{eq:ktdef} 
  \kt^{2} = q^2 y \, x_3 = q^2 y \lq 1-z(1-y)\rq\,,
\end{equation}
where $x_3$ is the energy fraction of the gluon.

The Sudakov form factor of eq.~(\ref{eq:suddef}) is given by
\begin{eqnarray}
  \Delta\!\(\bar{\Phi}_2,\pt\)\! & = & 
\exp\! \left\{\! - \!\!\!\sum_{\alpha_r \in \{1,2\}}  
\left[ \int d\Rad
  \frac{R\!\(\bar{\theta}_1,\bar{\phi}_1,\phi,z,y\)}
       {B\!\(\bar{\theta}_1\)}     
\, \stepf\!\(\kt^2 - \pt^2\) \right]_{\alpha_r} \right\} \nonumber\\
&=&
\exp\! \left\{\! - \!\!\!\!\!\sum_{\alpha_r \in \{1,2\}}  \!\!
\left[ \frac{q^2}{16\pi^2}\! \int_0^{2\pi}\!\!\frac{d\phi}{2\pi}\!
 \int_0^1 \!\!\!dz \!\int_0^1 \!\!\!dy\,  (1-y) \,
\frac{R\!\(\bar{\theta}_1,\bar{\phi}_1,\phi,z,y\)}
{B\!\(\bar{\theta_1}\)}  
\, \stepf\!\(\kt^2 - \pt^2\) \right]_{\alpha_r} \!\right\}\!. \nonumber\\
\end{eqnarray}
In order to apply the veto method of appendix~\ref{sec:veto_technique} to our
case, following the procedure of section~\ref{sec:gen_rad_var}, we need to
specify the form of an upper bounding function. A suitable choice turns out
to be
\begin{equation}
N\, \as\!\(\kt^2\)\,  \frac{1-y}{y} \frac{1}{1-z(1-y)} \geqslant
\sum_{\alpha_r \in \{1,2\}}  \lq (1-y) \,
\frac{R(\bar{\theta}_1,\bar{\phi}_1,\phi,z,y)}{B(\bar{\theta}_1)}
\rq_{\alpha_r},
\end{equation}
where
\begin{equation}
N = \frac{32\pi\CF}{q^2}\,.
\end{equation}
We denote the integral of the upper bounding function by
\begin{equation}
\Delta_u(\pt) = \exp\lg - \frac{2\CF}{\pi}
\int_0^{2\pi}\frac{d\phi}{2\pi} \int_0^1 dz \int_0^1 dy\,
 \as\!\(\kt^2\)\,  \frac{1-y}{y} \frac{1}{1-z(1-y)}  \stepf\!\(\kt^2 -
\pt^2\)\rg\,.
\end{equation}
The integral in the exponent can now be easily performed
\begin{equation}
\Delta_u(\pt) = \exp\lg - \frac{2\CF}{\pi}
 \int_0^1 dz \int_0^1 dy\,
 \frac{\as\!\(\kt^2\)}{\kt^2/q^2} \, (1-y) \, \stepf\!\( y\(1-z(1-y)\) 
-\frac{\pt^2}{q^2}\)\rg\,,
\end{equation}
and introducing the dimensionless quantities
\begin{equation}
\tildekt^2 = \frac{\kt^2}{q^2}=y\lq 1-z(1-y)\rq,   \qquad \qquad
 \tildept^2 =  \frac{\pt^2}{q^2} \,,
\end{equation}
and the integration over $d\tildekt^2$, we have
\begin{eqnarray}
\Delta_u(\pt) &=& \exp\lg - \frac{2\CF}{\pi}
\int_{\tildept^2}^1 d\tildekt^2 \,\int_0^1 dz \int_0^1 dy \,
 (1-y) \; \delta \left( y\(1-z(1-y)\) 
- \tildekt^2 \right)  \frac{\as\!\(\tildekt^2 q^2\)}{\tildekt^2}\rg
\nonumber\\
 &=&  \exp\lg - \frac{2\CF}{\pi}  \int_{\tildept^2}^1 d\tildekt^2
\frac{\as\!\(\tildekt^2
 q^2\)}{\tildekt^2}\int_{\tildekt^2}^{\tildekt} \frac{dy}{y} \rg
\nonumber\\
&=&  \exp\lg - \frac{2\CF}{\pi}  \int_{\tildept^2}^1 d\tildekt^2\,
\frac{\as\!\(\tildekt^2 q^2\)}{\tildekt^2}\,
\log\frac{1}{\tildekt}\rg.
\end{eqnarray}
The integral can now be easily performed using the one-loop expression
\begin{equation}
\label{eq:as_oneloop}
   \as\!\(\kt^2\) = \frac{1}{b_0 \log \( \kt^2/\Lambda^2 \)}\,,
\end{equation}
and we obtain
\begin{equation}
\label{eq:Delta_u_epem}
\Delta_u(\pt) = \exp\lg - \frac{\CF}{b_0\,\pi} \lq \log\frac{\pt^2}{q^2} 
+ \log\frac{q^2}{\Lambda^2} \,
\log\frac{\log\(q^2/\Lambda^2\)}{\log\(\pt^2/\Lambda^2\)}\rq\rg.
\end{equation}
Observe that we have a lower limit on the acceptable values of $\pt$, since
$\as(\pt)$ must be well defined for both the two-loop and the one-loop
expression of $\as$. We thus introduce a $\pt^{\rm min}\gtrsim \Lambda$.
Summarizing:
\begin{enumerate}
  \item
  Set $\pt^{\max} = \sqrt{q^2}$, where $q^2$ is the maximum value of $\pt^2$,
  such that $\Delta_{u} (\pt^{\max})=1$.
  
  \item \label{item:ptmax} Generate $r$ with $0 < r < 1$, and solve the
  equation \text{$r = \Delta_{u} (\pt)$/$\Delta_{u} (\pt^{\max})$} for
  $\pt$. If no solution is found, or if $\pt<\pt^{\rm min}$ no radiation is
  generated, and the event is returned as a $q\bar{q}$ event.
    
  \item
   If a solution with $\pt>\pt^{\rm min}$ is found,
   the radiation variables, according to the veto technique described in
   appendix~\ref{sec:veto_technique}, are then distributed according to
   \begin{equation}
    \frac{1-y}{y} \frac{1}{1-z(1-y)}\,\delta\big( q^2\,y\(1-z(1-y)\) -\pt^2\big).
   \end{equation}
   The $\delta$ function poses no constraints on the azimuthal angle $\phi$,
   that is then generated uniformly between 0 and $2\pi$, but it fixes $z$ to
   be
   \begin{equation}
   z = \frac{y-\pt^2/q^2}{y(1-y)}\,,
   \end{equation}
   where the probability distribution of $y$ is $dy/y$,
   so that $y$ is uniformly distributed in $\log y$, between the range imposed
   by $0\leq z \leq 1$,
   \begin{equation}
   \log \(\frac{\pt^2}{q^2}\) \leq \log y \leq \frac{1}{2} \log
   \(\frac{\pt^2}{q^2}\)\,. 
   \end{equation}

  \item Generate a uniformly-distributed random number $r'$ in the range
  \begin{equation}
    0 < r' < N\, \as\,  \frac{1-y}{y} \frac{1}{1-z(1-y)}\,. 
  \end{equation}
  If 
  \begin{equation}
    r' < \sum_{\alpha_r \in \{1,2\}}  \lq (1-y) \,
    \frac{R(\bar{\theta}_1,\bar{\phi}_1,\phi,z,y)}
    {B(\bar{\theta}_1)}\rq_{\alpha_r},
  \end{equation}
  accept the event.  Otherwise set $\pt^{\max} = \pt$ and go to
    step \ref{item:ptmax}.

  \item If
  \begin{equation}
   r' \geqslant  \lq (1-y) \,
    \frac{R(\bar{\theta}_1,\bar{\phi}_1,\phi,z,y)}
    {B(\bar{\theta}_1)}\rq_{1} 
   \end{equation}
   then set $\alpha_r=2$. Otherwise set $\alpha_r=1$.
\end{enumerate}
This completes the generation of the radiation variables and of the singular
region $\alpha_r$.

\subsection{$\boldsymbol{e^+ e^- \to q\bar{q}}$ in the \FKS{} formalism}
\subsubsection*{Virtual and real corrections}
Using the notation of section~\ref{sec:epem_CS_formalism},
we write the virtual corrections, normalized as in
eq.~(\ref{eq:FKSoneloop}), as
\begin{equation}
  \mathcal{V}_b = \mathcal{N} \, \frac{\as}{2 \pi} \left[ - 2 \left(
  \frac{\CF}{\epsilon^2} + \frac{3 \CF}{2 \epsilon} \right) + \frac{2
  \CF}{\epsilon} \log \frac{q^2}{Q^2} + \mathcal{V}_{\tmop{fin}}
  \right] B\!\(\bar\th_1\),
\end{equation}
where
\begin{equation}
  \mathcal{V}_{\tmop{fin}} =  \CF \left( \pi^2 - 8 + 3 \log \frac{q^2}{Q^2} -
  \log^2 \frac{q^2}{Q^2} \right) \,
\end{equation}
and
\begin{equation}
 B\!\(\bar\th_1\)=1+\cos^2\bar\th_1\;.
\end{equation}
Setting $\xicut=1$ and $\deltaO=2$,
equations~(\ref{eq:Qdef}) and~(\ref{eq:Iijreg}) become
\begin{eqnarray}
  \mathcal{Q} &=& 2  \CF \left[ \left( \frac{13}{2} - \frac{2 \pi^2}{3}
  \right)  -  \frac{3}{2}\,\log \frac{q^2}{Q^2} \right]\,,
\\
  \mathcal{I}_{12} &=& \frac{1}{2} \log^2 \frac{q^2}{Q^2} -
  \frac{\pi^2}{6}\;,
\end{eqnarray}
and, using eqs.~(\ref{eq:colourcorr}) and~(\ref{eq:FKSsv}) we get
\begin{equation}
 V\!\(\bar\th_1\) = \frac{\as}{2 \pi} \big[ \mathcal{Q} + 2
   \,\CF\,\mathcal{I}_{12}  + 
   \mathcal{V}_{\tmop{fin}} \big] B\!\(\bar\th_1\),
\end{equation}
where the dependence upon the scale $Q^2$ completely cancels.
We now consider the real emission term  $R$.
Following section~\ref{sec:Sfun}, we define the functions
 $d_{ij} = (k_i^0\, k_j^0)^a\, ( 1 -
\cos\theta_{ij})^b$ (see eq.~(\ref{eq:di2})), so that
\begin{eqnarray}
  d_{31} &=& 2^{b-2a} \(q^2\)^a (x_1 x_3)^{a - b} (1 - x_2)^b\,,\\
  d_{32} &=& 2^{b-2a} \(q^2\)^a (x_2 x_3)^{a - b} (1 - x_1)^b\,,
\end{eqnarray}
where $a$ and $b$ are positive real numbers. We introduce the
functions $\Soij$ (see eq.~(\ref{Soijdef}))
\begin{equation}
\Soij=\frac{\frac{1}{d_{ij}}}{\frac{1}{d_{31}} + \frac{1}{d_{32}}}\,,
\quad\quad ij\in \{31,32\}\,.
\end{equation}
Observe that there is no need to
include the $h$ factor of eq.~(\ref{Soijdef}), since only parton 3 has an
associated soft singularity, and therefore we only consider the regions
31 and 32 with $h=1$ (i.e.\ we do not consider the regions 13 and 23).
Furthermore, the region 12 does not correspond to any singularity of $R$,
and thus it is not included.

We define (see eq.~(\ref{eq:csi_yij_def}))
\begin{equation}
  \xi=\frac{2k^0_3}{\sqrt{q^2}}\equiv x_3\,,\qquad\quad
   y_{31} = \frac{\vec{k}_3\cdot\vec{k}_1}{k_3^0\,k_1^0}\,,\qquad\quad
   y_{32} = \frac{\vec{k}_3\cdot\vec{k}_2}{k_3^0\,k_2^0}\,,
\end{equation}
and (see~(\ref{eq:FKSrout}))
\begin{equation}
  \hat{R} = \hat{R}_{31} + \hat{R}_{32}, \qquad\quad
  \hat{R}_{3\{1, 2\}} = \frac{1}{\xi}\left( \frac{1}{\xi}
  \right)_+\! \left( 
  \frac{1}{1 - y_{3\{1, 2\}}} \right)_+ \!\left[ \left( 1 - y_{3\{1,
  2\}} \right) \xi^2 \, R_{3\{1, 2\}} \right],
\end{equation}
where
\begin{equation}
R_{ij}=R\,\Soij\,,\quad\quad ij\in \{31,32\}\,.
\end{equation}
The underlying Born configuration for each
singular region is easily identified as the one that preserve the direction
of the non-collinear parton. Thus, the underlying Born configuration for the
32 region has $\bar{k}_1 \propto k_1$, and for the 31 region has $\bar{k}_2
\propto k_2$.

\subsubsection*{Radiation phase space and inverse construction}
Using eqs.~(\ref{eq:FKS_fac_phinpo}) and~(\ref{eq:FKS_dRad}), we can write
the three-body phase space in terms of the barred and radiation
variables.  We work out explicitly the formulae for the 32 singular
region. The 31 one is treated in the same way.  We have
\begin{equation}
  d \Phi_3 = J_{32} \, d \xi\,  dy_{32}\,  d \phi \, d\bar{\Phi}_2,
\end{equation}
where (see eq.~(\ref{CS:bPhi2}))
\begin{equation}
 d\bar{\Phi}_2=\frac{1}{32\pi^2} d\cos\bar\theta_1\;d\bar{\phi}_1\;.
\end{equation}
In virtue of the azimuthal symmetry of the cross section,
we can fix from now on $\bar{\phi}_1=0$, remembering that, at the
end of the procedure, we should rotate the whole event by a random
azimuthal angle.
From eq.~(\ref{eq:inversefkskn}) we get immediately
\begin{equation}
  x_2 = \frac{2 (1 - \xi)}{2 - \xi (1 - y_{32})} = 1 - \frac{(1 + y_{32})
  \,\xi}{2 - \xi (1 - y_{32})}
\end{equation}
and, using the fact that $x_1 + x_2 +\xi  = 2$, we also get
\begin{equation}
 x_1 = 1 - \frac{(1 - y_{32})\, \xi \,(1 - \xi)}{2 - \xi (1 - y_{32})}\,.
\end{equation}
Given $\bar{\theta}_1$, $\xi$, $y_{32}$ and $\phi$ we can
fully reconstruct $k_1$, $k_2$ and $k_3$. Finally,
from eq.~(\ref{eq:jfsrfks}), we get
\begin{equation}
\label{eq:epemfksjac}
 J_{32} = \frac{q^2 }{(4 \pi)^3}\, \frac{x_2^2}{1 - \xi}\,\xi\,. 
\end{equation}

\subsubsection*{Generation of the Born variables}
The $\bar{B}$ function of eq.~(\ref{eq:bbdef}) is given by
\begin{equation}
  \bar{B}\!\(\bar{\theta}_1\) = B\!\(\bar\theta_1\) +
  V\!\(\bar{\theta}_1\) 
   + \sum_{\alpha_r \in \{32, 31\}} \left[ \int d\phi\, d \xi \, dy \,
 J \,\hat{R}\!\(\bar{\theta}_1, \phi, \xi, y\)
\right]_{\alpha_r}  ,
\end{equation}
where we are using the ``context'' notation introduced in
eq.~(\ref{eq:context}).  We now explicitly show how to deal with the
distributions, by illustrating the computation of the 32 term as an example:
\begin{equation}
 \int d \xi\, d y_{32} \,  d \phi \, J_{32}  \,\hat{R}_{32} = 
\frac{q^2}{(4\pi)^3} \int d \xi \, d y_{32}\, d \phi \left( \frac{1}{\xi}
   \right)_+\!\! \left( \frac{1}{1 - y_{32}} \right)_+\!\!
     \left(1 - y_{32} \right)\, \xi^2 \,R_{32} \,\frac{x_2^2}{1 - \xi}\,. 
\end{equation}
One can easily verify that both $J_{32}$ and $\left( 1 - y_{32} \right)
\xi^2\, R_{32}$ have a finite limit for $y_{32} \rightarrow 1$ or $\xi
\rightarrow 0$. We can thus expand the distributions according to their
definitions~(\ref{eq:uoxidef}), (\ref{eq:uoyipmdef}) and
(\ref{eq:fksplusdef}). Introducing, for
ease of notation, the function
\begin{equation}
  A_{32}\!\(\bar{\theta}_1, \phi, \xi, y_{32}\) = \left( 1 -
  y_{32} \right)\, \xi^2 \, R_{32}\,,
\end{equation}
we have
\begin{eqnarray}
\label{eq:r32int}
  & & \int_0^{2\pi} d \phi \int_0^1 d\xi \int_{-1}^1 dy_{32}
   \,J_{32}\,\hat{R}_{32} =  
   \frac{q^2}{(4\pi)^3}  \int_0^{2\pi} d \phi \int_0^1 d\xi \int_{-1}^1
   dy_{32}  \, 
  \frac{1}{\xi} \, \frac{1}{1 - y_{32}}  \nonumber\\
  && \qquad\qquad
\times \Bigg\{ \!\!\lq \frac{x_2^2}{1 - \xi} \, A_{32}\!\(\bar{\theta}_1,
 \phi, \xi, y_{32}\) - 
  \(1 - \xi\)\,
  A_{32}\!\(\bar{\theta}_1, \phi, \xi, 1\) \rq
 \nonumber\\
  && \qquad\qquad\!\!
{} -  \Big[ A_{32}\!\(\bar{\theta}_1,
  \phi, 0, y_{32}\) - 
   A_{32}\!\(\bar{\theta}_1, \phi, 0, 1\) \Big] \Bigg\}.  
\end{eqnarray}
We observe that the Jacobian of eq.~(\ref{eq:epemfksjac}) has an integrable
divergence at $(\xi,y_{23}) \to (1,-1)$ region.
This happens because the recoiling
system is massless, so that there is no upper bound on $\xi$ (see
eq.~(\ref{eq:xiboundFSR})). In order to clarify the nature of this singularity, we
introduce the parametrization
\begin{equation}
\xi=1-\rho \cos\eta\,,\quad\quad y_{23}=-1+\rho \sin\eta\,,
\end{equation}
and from eq.~(\ref{eq:epemfksjac}) we get
\begin{equation}\label{eq:jacsing}
 J_{32} = \frac{q^2 }{(4 \pi)^3}\frac{4(1-\xi)\xi}{%
\left(2(1-\xi)+\xi(1+y_{32})\right)^2}=\frac{q^2 }{(4 \pi)^3}\,\frac{1}{\rho}
\left[\frac{4 \cos\eta}{%
\left(2\cos\eta+\sin\eta\right)^2} + {\cal O}(\rho)\right],
\end{equation}
diverging as $1/\rho$ in the small $\rho$ limit. This singularity is integrable,
but cannot be integrated using Monte Carlo techniques, since it yields
a divergent error.
On the other hand, $J_{32}\sqrt{1-\xi}\sqrt{1+y_{32}}$ has no divergence,
which suggests the use of square-root importance sampling for both $1-\xi$
and $1+y_{32}$ in order to overcome the problem. We thus parametrize
the integration variables as
\begin{equation}
\xi=1-\left(\Xrad^{(1)}\right)^2,\quad\quad
y_{23}=-1+2\left(\Xrad^{(2)}\right)^2,
\quad\quad \phi=2\pi\,\Xrad^{(3)}\,.
\end{equation}
The generation of the Born variables is performed along the lines
of section~\ref{sec:gen_born_var}.
We define
\begin{equation}
  \tilde{B}\!\(\bar{\theta}_1, \Xrad\) =
  \Big( B\!\(\bar{\theta}_1\) + V\!\(\bar{\theta}_1\) \Big) + 16\pi\,
\Xrad^{(1)}\, \Xrad^{(2)}
\sum_{\alpha_r \in \{32, 31\}} \left[ J \,  \hat{R}\!\(\bar{\theta}_1,
\phi, \xi, y\)\right]_{\alpha_r},   
\end{equation}
so that
\begin{equation}
\int_0^1 d\Xrad^{(1)}\int_0^1d\Xrad^{(2)}\int_0^1d\Xrad^{(3)}\;
 \tilde{B}\!\(\bar{\theta}_1, \Xrad\)  = \bar{B}\!\(\bar{\theta}_1\)\,.
\end{equation}
Then one performs a full integration of $\tilde{B}$, using an integrator that
can generate unweighted events, like the \SPRING-\BASES{} package. After this
step, one can generate Born configurations (i.e.\ $\bar{\theta}_1$)
distributed according to $\bar{B}$.

\subsubsection*{Generation of the radiation variables}
After having generated the Born configuration distributed according to the
$\bar{B}$ function, we must generate the hardest radiation. We first define
the $\kt$ of the radiation.  As we have already pointed out, the definition
is ambiguous to some extent. The only requirements that it must satisfy is
that it must be of the order of the radiation transverse momentum in the
collinear limit, and that it must coincide with it in the soft-collinear
limit. A suitable definition is
\begin{equation}
  \label{eq:ktdefepem} 
  \kt = \frac{q^0}{2} \sqrt{1 - y^2} \,\xi \,.
\end{equation}
The no-radiation probability, according to eq.~(\ref{eq:suddef}), is given
by 
\begin{eqnarray}
\label{eq:FKS_Sud_epem}
  \Delta\!\(\bar{\theta}_1,\pt\) & = & \exp \left\{ -
  \!\!\!\!\sum_{\alpha_r \in \{32, 31\}}
  \left[ \int d \phi \, d \xi \, dy \frac{J \,
  R\!\(\bar{\theta}_1,\phi,\xi,y\)}
  {B\!\(\bar{\theta}_1\)}\stepf\!\(\kt - \pt\) \right]_{\alpha_r}
  \right\}\!.   \phantom{aaa}
\end{eqnarray}
In order to generate the radiation variables distributed according to
eq.~(\ref{eq:FKS_Sud_epem}), we make use of the veto method of
appendix~\ref{sec:veto_technique}. We need then to specify the form of an
upper bounding function. A suitable choice turns out to be
\begin{equation}
  \label{eq:upperbndepem}
U(\xi,y)=\frac{N \as (\kt)}{(1 - y^2) \,\xi} \geqslant
  \sum_{\alpha_r \in \{32, 31\}} \left[ \frac{J \, 
R\!\(\bar{\theta}_1,\phi,\xi,y\)}
    {B\!\(\bar{\theta}_1\)} \right]_{\alpha_r},
\end{equation}
where $y$ stands for the common value of $y_{31}$ and $y_{32}$ on the
right hand side.
Notice that $U(\xi,y)$ is singular also for $y\to -1$. In this the upper
bound properly covers also the singularity in $J$ for $y\to -1$,
(see eq.~(\ref{eq:jacsing})).
The normalization $N$ is evaluated by probing the
$\bar{\theta}_1,\phi,\xi,y$ phase space randomly with an
adequate number of points.
The $\as(\kt)$ factor appearing in eq.~(\ref{eq:upperbndepem})
corresponds to the one-loop expression.\footnote{Since the $e^+e^-\to q\bar{q}$
cross section is of zeroth order in the strong coupling constant, the use
of one-loop $\as$ in the radiative corrections is adequate in this case
also in the full NLO formula.}

We now want to generate \ $\phi, \xi, y$ according to the distribution
\begin{equation}
  \exp \left[ - \int  d \phi' \,d \xi'\,d y'\; U(\xi',y')\,
 \stepf\!\(\kt' - \kt\) \right]  U(\xi,y) \, d \phi \, d \xi \, d y\,,
\end{equation}
where $\kt' = \kt\!\(\xi', y'\)$ and $\kt = \kt\!\(\xi, y\)$, and then use
the veto technique of appendix~\ref{sec:veto_technique}
to correct to the exact distribution. One
generates first the $\kt$ value. Its probability distribution is given by
\begin{equation}
  \int \exp \left[ - \int d \phi'\,d \xi' \, d y' \, U(\xi',y')\,
  \stepf\!\(\kt' - \kt\) \right]  U(\xi,y)
 \, d \phi \,d \xi\,  d y \,
\delta\!\(\kt - \pt\) = - \frac{d \Delta_u (\pt)}{d \pt}\,,
\end{equation}
where
\begin{equation}
  \label{eq:deltau} 
\Delta_u\!\(\pt\) = \exp \left[ - \int d \phi\, d \xi \, d y\;
 U(\xi,y) \,\stepf\!\(\kt - \pt\) \right] .
\end{equation}
The integral in the exponent of eq.~(\ref{eq:deltau}) is easily manipulated
to yield
\begin{equation}
  \int d\phi\, d\xi\, dy \,\frac{N \as (\kt)}{\(1 - y^2\)\, \xi} \,
\stepf\!\(\kt - \pt\) = 2 \pi N \int_{\pt}^{k_{\rm max}} \frac{d \kt}{\kt} \,
\as(\kt)\, \log \frac{1 + \sqrt{1 - (\kt / k_{\rm max})^2}}{1 - 
\sqrt{1 - (\kt/k_{\rm max})^2}}\,,
\end{equation}
with
\begin{equation}\label{eq:kmaxdef}
 k_{\rm max} = q^0 / 2\;.
\end{equation}
So, we want to generate $\pt$ according to the
Sudakov form factor
\begin{equation}
  \label{eq:uktdistr} 
\Delta_u\!\(\pt\) = \exp \left[ - 2 \pi N
  \int_0^{k_{\rm max}} \frac{d \kt}{\kt} \,\as\!\(\kt\) \,\log \frac{1 +
  \sqrt{1 - (\kt / k_{\rm max})^2}}{1 - \sqrt{1 - (\kt / k_{\rm max})^2}}
 \, \stepf\!\(\kt - \pt\) \right],
\end{equation}
which has again the form of eq.~(\ref{eq:deltau}) for a single variable $\kt$
instead of the set of variables $\xi,y$.
The integral in eq.~(\ref{eq:uktdistr}) is
still too complex to be performed analytically, so we resort a second time to
the veto method. Using the inequality
\begin{equation}
  \log \frac{4 k_{\rm max}^2}{\kt^2} \geqslant \log \frac{1 + \sqrt{1 -
  (\kt/k_{\rm max})^2}}{1 - \sqrt{1 - (\kt / k_{\rm max})^2}}\,,
\end{equation}
we generate the $\pt$ distribution according to $d \Delta_{u u} (\pt)$, where
\begin{eqnarray}
  \Delta_{u u}\!\(\pt\) &=& \exp \left[ - 2 \pi N \int_0^{k_{\rm max}} \frac{d
  \kt}{\kt}\, \as (\kt)\, \log \frac{4 k_{\rm max}^2}{\kt^2} \,
\stepf\!\(\kt -  \pt\) \right]\nonumber\\
&=& \exp \left[ - \frac{\pi N}{b_0} \left( \log\frac{4 k_{\rm max}^2}{\Lambda^2}
\log \frac{\log \( k_{\rm max}^2 /\Lambda^2\)}{\log\(\pt^2/\Lambda^2\)} -
  \log \frac{k_{\rm max}^2}{\pt^2} \right) \right].
\end{eqnarray}
Observe that we have a lower limit on the acceptable values of $\pt$, since
$\as(\pt)$ must be well defined for both the two-loop and the one-loop
expression of $\as$. We thus introduce a $\pt^{\rm min}\gtrsim \Lambda$.
Summarizing
\begin{enumerate}
  \item Set $\pt^{\max} = k_{\rm max}$,
  the maximum allowed value according to eq.~(\ref{eq:kmaxdef}).
  We have $\Delta_{u u}(\pt^{\max})=1$.
  
  \item \label{item:first_pt} Generate a uniformly-distributed random number
  $r$ between 0 and 1, and solve the equation $r = \Delta_{u
  u}(\pt)$/$\Delta_{u u}(\pt^{\max})$ for $\pt$. If no solution is found, or
  if $\pt<\pt^{\rm min}$, there is no radiation, and
  a $q\bar{q}$ event is generated.
  
  \item If a solution with $\pt>\pt^{\rm min}$ is found,
  generate $r'$ with $0 \leqslant r' \leqslant \log \frac{4
  k_{\rm max}^2}{\pt^2}$. If $r' \leqslant \log \frac{1 + \sqrt{1 - (\pt /
  k_{\rm max})^2}}{1 - \sqrt{1 - (\pt / k_{\rm max})^2}}$ go to
  step~\ref{item:first_veto}.  Otherwise set $\pt^{\max} = \pt$ and go
  to step~\ref{item:first_pt}.
  
  \item \label{item:first_veto} At this point $\pt$ is generated according to
   eq.~(\ref{eq:uktdistr}).  The radiation variables $\phi, y, \xi$ are then
   distributed with a probability proportional to
  \begin{equation}
  \frac{1}{\xi \(1 - y^2\)} \delta\!\(\kt\!\(\xi, y\) - \pt\)
   d\phi \, dy\, d\xi\,,
  \end{equation}
  where $\kt (\xi, y)$ is given by eq.~(\ref{eq:ktdefepem}).  The $\delta$
  function does not constraint the azimuthal angle $\phi$, that is then
  generated uniformly between 0 and $2\pi$, but it fixes $\xi$ to be
  \begin{equation}
     \xi = \frac{2\pt}{q^0\sqrt{1-y^2}}\,,
  \end{equation}
  and, upon integrating in $\xi$, $y$ is distributed with a probability
  proportional to
  \begin{equation}
   \stepf\!\left( \sqrt{1 - y^2} - \frac{2 \pt}{q^0}
     \right) d \log \frac{1+y}{1 - y}\,.
  \end{equation}
  One thus generates a uniform random number $r_y$ between the minimum and
  maximum value of the logarithm
  allowed by the $\stepf$ function
  \begin{equation}
   - \log \frac{1 + \sqrt{1 - (\pt / k_{\rm max})^2}}{1 - 
\sqrt{1 - (\pt/k_{\rm max})^2}}\, < r_y <
  \log \frac{1 + \sqrt{1 - (\pt / k_{\rm max})^2}}{1 - 
\sqrt{1 - (\pt/k_{\rm max})^2}}
  \end{equation}
  and solves the equation
  \begin{equation}
    r_y=\log\frac{1+y}{1-y}
  \end{equation}
  for $y$.
  \item Now the last veto: generate a random number $r''$ between $0$ and
  $U(\xi,y)$. If
  \begin{equation} 
   r'' \leqslant \sum_{\alpha_r \in \{32, 31\}} \left[ \frac{ J\, 
  R\!\(\bar{\theta}_1,\phi, \xi,
  y\)}{B\!\(\bar{\theta}_1\)} \right]_{\alpha_r} 
  \end{equation}
  accept the event. Otherwise set $\text{$\pt^{\max} = \pt$}$ and go
  to step~\ref{item:first_pt}. 
  \item If
  \[ r'' \geqslant  \left[ \frac{ J \,
   R\!\(\bar{\theta}_1,\phi, \xi,
  y\)}{B\!\(\bar{\theta}_1\)}\right]_{\alpha_r=32} \] 
  set $\alpha_r = 31$. Otherwise
  set $\alpha_r = 32$.
\end{enumerate}
This completes the generation of the radiation variables and of the singular
region $\alpha_r$. Observe that, in our simple case, there is total symmetry
between the regions $31$ and $32$. One could simply double the
contribution of region $32$, and, at the end, choose one of the two
regions with equal probability. The rather pedantic discussion given above
has the only purpose of better clarifying the method in general.

\subsection{$\boldsymbol{h_{\splus} h_{\sminus} \rightarrow V}$ in 
the \CS{} formalism} 
\label{sec:hh_to_gamma_CS}
\subsubsection*{Born term} 
In this section, we consider the production of a massive vector boson $V$ of
mass $M$ in hadron collision, in the CS formalism,
\begin{equation}
  q(\bar{k}_\splus) + \qb(\bar{k}_\sminus) \,\to\, V(\bar{k}_1)\,,
\end{equation}
with $\bar{k}_1^2 = M^2$.  Denoting, as usual, with $S$ the centre-of-mass
energy of the two incoming hadronic beams, $S = (K_{\splus} +
K_{\sminus})^2$, and with $\bar{Y}$ the rapidity of the produced vector
boson,
\begin{equation}
\bar{Y} =
\frac{1}{2}\log\frac{\bar{k}_1^0+\bar{k}_1^3}{\bar{k}_1^0-\bar{k}_1^3}\,, 
\end{equation}
we have, from eq.~(\ref{eq:PSn}),
\begin{equation}
\label{dbarPhi1}
 d{\bf \bar{\Phi}}_1 \equiv d \bar{x}_{\splus} \, d \bar{x}_{\sminus}\,
\, \frac{d^3 \bar{k}_1}{(2 \pi)^3 2 \bar{k}_1^0} \, (2 \pi)^4 
\,\delta^4\(\bar{k}_{\splus} + {\bar k}_{\sminus} - {\bar k}_1\)  = \frac{2
  \pi}{S} d \bar{Y}\,,
\end{equation}
with the constraints
\begin{equation}
\label{eq:CS_VX}
  \bar{x}_{\splus} = \sqrt{\frac{M^2}{S}} e^{\bar{Y}}, \qquad\qquad
\bar{x}_{\sminus} = \sqrt{\frac{M^2}{S}} e^{- \bar{Y}}\,,
\end{equation}
and momentum conservation implies
\begin{equation}
2\, \bar{k}_\splus \cdot \bar{k}_\sminus = \bar{x}_{\splus} \bar{x}_{\sminus}
  S = M^2\,.
\end{equation}
The only Born variable is then $\bar{Y}$, so that
\begin{equation}
\bar{\bf \Phi}_1 = \lg \bar{Y}\rg\,.
\end{equation}
The squared tree-level amplitude, averaged over color and polarization of
incoming partons $(1/(4N_c^2))$, is given by
\begin{equation}
\overline{|{\cal M}_B(\bar{k}_\splus,\bar{k}_\sminus)|^2} =  \frac{1}{N_c}\,
g^2  \(\bar{k}_\splus\cdot \bar{k}_\sminus\)\,,
\end{equation}
where $g$ is the $V q \bar{q}$ coupling.
Introducing the flux factor $1/(2M^2)$, we obtain the differential cross
section 
\begin{equation}
\mathcal{B}_{q\qb}\!\(\bar{\bf \Phi}_1\)\equiv
\mathcal{B}_{q\qb}\!\(\bar{k}_\splus,\bar{k}_\sminus\) =
\frac{1}{4N_c}\,g^2\,, 
\end{equation}
independent from the external momenta.  We denote by $q\bar{q}$
the Born flavour index $f_b$, that, in this example,
carries information on the flavour of the
(massless) incoming partons along the $\splus$ and $\sminus$ direction.
Using a standard convention for the numbering scheme,
we have $q= - 5, \ldots,- 1,1, \ldots, 5$, and $\bar{q}=-q$.

\subsubsection*{Virtual corrections}
With the scale choice $Q^2=M^2$, we have from eqs.~(\ref{eq:FKSoneloop})
and~(\ref{eq:svcs})
\begin{equation}
  \mathcal{V}_{{\rm fin},\, q\qb}\!\(\bar{\bf \Phi}_1\) = \CF \(\pi^2 - 8\) \,
  \mathcal{B}_{q\qb}\!\(\bar{\bf \Phi}_1\)\,,\qquad\qquad 
\mathcal{V}_{q\qb}\!\(\bar{\bf \Phi}_1\) = \frac{\as}{\pi} \, \CF \,
\mathcal{B}_{q\qb}\!\(\bar{\bf \Phi}_1\)\,. 
\end{equation}

\subsubsection*{Real corrections}
Three different sub-processes contribute to the real radiation production
\begin{eqnarray}
\label{eq:Vg}
q(k_\splus) \,\qb(k_\sminus) \,&\to&\, V(k_1) \,g(k_2)\,,\\
\label{eq:Vq}
q(k_\splus) \, g(k_\sminus) \,&\to&\, V(k_1) \,q(k_2)\,,\\
\label{eq:Vqbar}
g(k_\splus) \, \qb(k_\sminus) \,&\to&\, V(k_1) \,\qb(k_2)\,.
\end{eqnarray}
Introducing the usual Mandelstam variables
\begin{eqnarray}
\label{eq:def_s_V}
s &=& (k_\splus+k_\sminus)^2 = (k_1+k_2)^2,\\
t &=& (k_\splus-k_1)^2=(k_\sminus-k_2)^2,\\
\label{eq:def_u_V}
u &=& (k_\splus-k_2)^2= (k_\sminus-k_1)^2,
\end{eqnarray}
we can express them in terms of the two sets of radiation variables (see
section~\ref{sec:CS_IS_IS}): the $\splus$~set,
$\Rad^\splus=\lg\xpm,\vplus,\phi\rg$, that describes the emission from the
$\splus$~parton, and the $\sminus$~set,
$\Rad^\sminus=\lg\xpm,\vminus,\phi\rg$, that describes the emission from the
$\sminus$~parton
\begin{itemize}
\item
$\Rad^\splus$
\begin{equation}
s = \frac{M^2}{\xpm},\qquad\qquad t = -s\(1-\xpm - \vplus \), \qquad\qquad
u = -s \,\vplus,
\end{equation}
\item
$\Rad^\sminus$
\begin{equation}
s = \frac{M^2}{\xpm},\qquad\qquad t = -s \,\vminus, \qquad\qquad
u =  -s\(1-\xpm - \vminus \),
\end{equation}
\end{itemize}
where we have dropped the $i$ subscript, since we have only one radiated
parton, and we have put a $\splusminus$ subscript on the angular $\tilde{v}$
variable, in order to distinguish the two sets.

The squared amplitudes corresponding to the processes in
eqs.~(\ref{eq:Vg})--(\ref{eq:Vqbar}) are:
\begin{enumerate}
\item The squared amplitude for the $q(k_\splus) \,\qb(k_\sminus) \,\to\,
V(k_1) \,g(k_2)$ process, averaged over color and polarization of incoming
partons $(1/(4N_c^2))$, is given by
\begin{eqnarray}
\label{eq:qq_gamglu}
\overline{|{\cal M}_{q\qb}|^2} &=& 2\,\frac{\CF}{N_c} \,g^2\, g_s^2 
\left[\frac{u}{t}+\frac{t}{u}  + \frac{2\,s \,M^2}{tu}\right]\\
 &=& 2\,\frac{\CF}{N_c} \,g^2\, g_s^2 
\left[\frac{\vpm}{1-\xpm - \vpm}
+\frac{1-\xpm - \vpm}{\vpm}
+\frac{2\xpm}{\vpm\(1-\xpm - \vpm\)}\right]. \phantom{aaaaaa}
\end{eqnarray}
This expression has singularities when the gluon is emitted both along the
$\splus$~and the $\sminus$~direction ($\vpm \to 0$).

\item
The squared amplitude for the $q(k_\splus) \, g(k_\sminus) \,\to\, V(k_1)
\,q(k_2)$ process, averaged over color and polarization of incoming partons
$(1/(4N_c(N_c^2-1)))$, is equal to
\begin{eqnarray}
\overline{|{\cal M}_{qg}|^2} &=& 
-2\,\frac{\TF}{N_c}\, g^2\, g_s^2 \left[
\frac{s}{t}+\frac{t}{s}  + \frac{2\,u\,M^2}{st}\rq\\
&=& 2\,\frac{\TF}{N_c}\, g^2\, g_s^2 \,\frac{1}{\vminus} \left[
\vminus^2+1  -2 \xpm \(1-\xpm - \vminus\)\right].  
\end{eqnarray}
\item
The squared amplitude for the $g(k_\splus) \, \qb(k_\sminus) \,\to\, V(k_1)
\,\qb(k_2)$ process, averaged over color and polarization of incoming
partons, is given by
\begin{eqnarray}
\overline{|{\cal M}_{g\qb}|^2} &=& -2\,\frac{\TF}{N_c}\, g^2\, g_s^2 \left[
\frac{u}{s}+\frac{s}{u}  + \frac{2\,t \,M^2}{su}\right]\\
&=& 2\,\frac{\TF}{N_c}\, g^2\, g_s^2 \,\frac{1}{\vplus} \left[
\vplus^2+1  -2 \xpm \(1-\xpm - \vplus\)\right].  
\end{eqnarray}
\end{enumerate}

\subsubsection*{Counterterms}
According to eqs.~(5.136), (5.145) and~(5.146) of ref.~\cite{Catani:1996vz},
the counterterms are given by
\begin{eqnarray}
{\cal D}^{qg,\qb} &=& -\frac{1}{u}\frac{1}{\xpm}
2\,g_s^2\,\CF\lq \frac{2}{1-\xpm} -\(1+\xpm\)\rq
\overline{|{\cal M}_B(\xpm\,k_\splus,k_\sminus )|^2}
\nonumber\\
 &=& 2\,\frac{\CF}{N_c}\, g^2\, g_s^2  \, \frac{1}{\vplus}\lq
\frac{2}{1-\xpm} -\(1+\xpm\)\rq 
\\
{\cal D}^{\qb g,q} &=& -\frac{1}{t}\frac{1}{\xpm}
2\,g_s^2\,\CF\lq \frac{2}{1-\xpm} -\(1+\xpm\)\rq
\overline{|{\cal M}_B(k_\splus,\xpm\,k_\sminus )|^2}
\nonumber\\
 &=& 2\,\frac{\CF}{N_c}\, g^2\, g_s^2  \, \frac{1}{\vminus}\lq \frac{2}{1-\xpm}
-\(1+\xpm\)\rq 
\end{eqnarray}
for gluon radiation from a $q\qb$ initial-state, and
\begin{eqnarray}
{\cal D}^{g\qb,\qb} &=& -\frac{1}{u}\frac{1}{\xpm}
2\,g_s^2\,\TF \lq 1-2\,\xpm\(1-\xpm\)\rq
\overline{|{\cal M}_B(\xpm\,k_\splus,k_\sminus )|^2}
\nonumber\\
 &=& 2\,\frac{\TF}{N_c}\, g^2\, g_s^2  \, \frac{1}{\vplus}
\lq 1-2\,\xpm\(1-\xpm\)\rq 
\\
{\cal D}^{q g,q} &=& -\frac{1}{t}\frac{1}{\xpm}
2\,g_s^2\,\TF\lq1-2\,\xpm\(1-\xpm\)\rq
\overline{|{\cal M}_B(k_\splus,\xpm\,k_\sminus )|^2}
\nonumber\\
 &=& 2\,\frac{\TF}{N_c}\, g^2\, g_s^2  \, \frac{1}{\vminus}
\lq 1-2\,\xpm\(1-\xpm\)\rq 
\end{eqnarray}
for the $g\qb$ and $qg$ initiated processes, respectively.

\subsubsection*{Radiation variables and inverse construction}
Following section~\ref{sec:CS_inverse_construction_IS_IS}, given the rapidity
$\bar{Y}$, the invariant squared mass of the produced boson $M^2$ and the
three radiation variables ($\xpm$, $\vpm$ and $\phi$), we can construct the
full 2-body event.

For definiteness, we assume that the emitting parton is the $\splus$~one.  A
similar procedure can be followed if the emitting parton is the
$\sminus$~one, with the exchange of the roles of the $\splus$ and
$\sminus$~directions.  Using eq.~(\ref{eq:CS_VX}), we can compute
$\bar{x}_{\splus}$ and $\bar{x}_{\sminus}$, and from
eqs.~(\ref{eq:ISIS_kpkm})--(\ref{eq:ISIS_xpxm}), we have
\begin{equation}
k_{\splus} = x_{\splus} K= \frac{\bar{x}_{\splus}}{\xpm}  K\,,
\qquad\qquad 
k_{\sminus} = x_{\sminus} K = \bar{x}_{\sminus} K\,.
\end{equation}
In the centre-of-mass system of $(\kplus+\kminus)$, the energy $k_2^{\prime
0}$ of the final-state parton and its angle $\theta'_2$ with respect to the
$k_{\splus}$ direction are given by (see eqs.~(\ref{eq:ISIS_energy})
and~(\ref{eq:ISIS_angle}))
\begin{equation}
\label{eq:VX_E_costh}
k_2^{\prime 0} = \sqrt{\frac{\kplus \cdot\kminus}{2}}\,(1-\xpm)\,,
\qquad
\cos\th'_2 = \frac{1-2\vplus-  \xpm}{1-\xpm}\,,
\end{equation}
so that 
\begin{equation}
k_2^{\prime 1} = k_2^{\prime 0} \, \sin\th'_2\, \cos\phi\,,\qquad
k_2^{\prime 2} = k_2^{\prime 0} \, \sin\th'_2\, \sin\phi\,,\qquad
k_2^{\prime 3} = k_2^{\prime 0} \, \cos\th'_2\,.
\end{equation}
We can now boost back in the laboratory frame the momentum $k_2'$, with a
boost parallel to the $\splus$ direction, with velocity given by
\begin{equation}
\beta = \frac{x_{\splus}-x_{\sminus}}{x_{\splus}+x_{\sminus}}\,,
\end{equation}
and obtain the radiated parton momentum $k_2$. There is no need of further
Lorentz boosts to obtain $k_1$, since we can simply use momentum conservation
\begin{equation}
k_1 = k_{\splus} + k_{\sminus} - k_2\,.
\end{equation}
According to eq.~(\ref{eq:ISIS_dphi_rad}), the infinitesimal phase-space
volume, $d{\bf \Phi}_2$, can be written as
\begin{equation}
d\xplus \, d\xminus \,d\Phi_2 = d{\bf\bar{\Phi}}_1 \,  d\Rad^\splus\,
\end{equation}
where
\begin{equation}
\label{eq:dPhi_rad_plus}
d \Rad^\splus =
\frac{(2\kplus\cdot \kminus )}{16\pi^2}
\;\frac{d\phi}{2\pi}
\;d\vplus\;\frac{d\xpm}{\xpm}
\;\stepf\!\(\vplus\)\,\stepf\!\(1-\frac{\vplus}{1-\xpm}\)
\stepf\big(\xpm(1-\xpm)\big)\, \stepf\!\(\xpm -\bxplus\). 
\end{equation}
The expression for the infinitesimal phase-space radiation volume, in the
case of emission from the $\sminus$~parton, $d \Rad^\sminus$, can be obtained
with the interchange $\vplus\leftrightarrow\vminus$,
$\bar{x}_\splus\leftrightarrow\bar{x}_\sminus$.
The $\stepf$ functions appearing in eq.~(\ref{eq:dPhi_rad_plus}) have a
simple physical interpretation: they constrain the energy in
eq.~(\ref{eq:VX_E_costh}) to be positive, the cosine of the $\th'_2$ angle
(see eq.~(\ref{eq:VX_E_costh})) to be in the appropriate range and force
$\xplus$ and $\xminus$ to be in the physical range.

\subsubsection*{Collinear remnants}
The collinear remnants of eq.~(\ref{eq:cremcs}) are given by
\begin{eqnarray}
\mathcal{G}_{\splus}^{q\qb}(\bar{\bf \Phi}_{1,\splus}) 
&=& \frac{\as}{2\pi}\CF\!\!\lq 
\( \frac{2}{1-z}\log\frac{(1-z)^2}{z}\)_+ 
- (1+z)\log\frac{(1-z)^2}{z} +(1-z) \right.
\nonumber\\
&&\phantom{\frac{\as}{2\pi}\CF\Big[\!\!}
\left.+\(\frac{2}{3}\pi^2-5\) \delta(1-z)+
\(\frac{1+z^2}{1-z}\)_+\log\frac{M^2}{\muF^2}\rq 
\mathcal{B}_{q\qb}\!\(z k_\splus,k_\sminus\)\,,
\\
\mathcal{G}_{\splus}^{g\qb}(\bar{\bf \Phi}_{1,\splus})
 &=& \frac{\as}{2\pi}\TF\Bigg\{
\lq z^2+(1-z)^2\rq\lq \log\frac{(1-z)^2}{z} +\log\frac{M^2}{\muF^2}\rq
+2z(1-z) \Bigg\}\,  \mathcal{B}_{q\qb}\!\(z k_\splus,k_\sminus\)\,.\nonumber\\
\end{eqnarray}
The other two collinear remnants $\mathcal{G}_{\sminus}^{q\qb}(\bar{\bf
\Phi}_{1,\sminus})$ and $\mathcal{G}_{\sminus}^{qg}(\bar{\bf
\Phi}_{1,\sminus})$ are equal to $\mathcal{G}_{\splus}^{q\qb}(\bar{\bf
\Phi}_{1,\splus})$ and $\mathcal{G}_{\splus}^{g\qb}(\bar{\bf
\Phi}_{1,\splus})$ respectively, with
$\mathcal{B}_{q\qb}\(zk_\splus,k_\sminus\)$ replaced by
$\mathcal{B}_{q\qb}\(k_\splus,zk_\sminus\)$.

\subsubsection*{\POWHEG{} ingredients}
We define the contributions to the real differential cross section, and
corresponding counterterms, according to the three singular regions, by
attaching the flux factor $1/(2s)=\xpm/(2M^2)$, to the squared matrix
elements
\begin{eqnarray}
\label{eq:Rqqbar}
&&{\displaystyle
\mathcal{R}_{q\qb}(\bar{\bf \Phi}_1,\Rad^\splusminus) = \frac{\xpm}{2M^2}\, 
\overline{|{\cal M}_{q\qb}|^2}}\,,
\\[2mm]
&&{\displaystyle \mathcal{R}_{qg}(\bar{\bf \Phi}_1,\Rad^\sminus) =
  \frac{\xpm}{2M^2}\, 
\overline{|{\cal M}_{qg}|^2}}\,,\\[2mm]
\label{eq:Rgqbar}
&&{\displaystyle\mathcal{R}_{g\qb}(\bar{\bf \Phi}_1,\Rad^\splus) =
  \frac{\xpm}{2M^2}\, 
\overline{|{\cal M}_{g\qb}|^2}}\,,
\end{eqnarray}
and
\begin{eqnarray}
&&{\displaystyle
\mathcal{C}_{q\qb}^\splus(\bar{\bf \Phi}_1,\Rad^\splus) = \frac{\xpm}{2M^2}
\, {\cal D}^{qg,\qb}}\,,
\\[2mm]
&&{\displaystyle
\mathcal{C}_{q\qb}^\sminus(\bar{\bf \Phi}_1,\Rad^\sminus) = \frac{\xpm}{2M^2}
\, {\cal D}^{q g,q}}\,,
\\[2mm]
&&
{\displaystyle\mathcal{C}_{g\qb}(\bar{\bf \Phi}_1,\Rad^\splus) =
  \frac{\xpm}{2M^2}\, {\cal D}^{q g,q}} \,,
\\[2mm]
&&
{\displaystyle \mathcal{C}_{qg}(\bar{\bf \Phi}_1,\Rad^\sminus) =
  \frac{\xpm}{2M^2}\, 
{\cal D}^{g\qb,\qb}} \,.
\end{eqnarray}
We also define
\begin{eqnarray}
\mathcal{R}_{q\qb}^\splus\(\bar{\bf \Phi}_1,\Rad^\splus\) &=&
\mathcal{R}_{q\qb}\(\bar{\bf \Phi}_1,\Rad^\splus\)
\,\lq \frac{\mathcal{C}_{q\qb}^\splus}{\mathcal{C}_{q\qb}^\splus 
+ \mathcal{C}_{q\qb}^\sminus}\rq_{\Rad^\splus},\\
\mathcal{R}_{q\qb}^\sminus\(\bar{\bf \Phi}_1,\Rad^\sminus\) &=&
\mathcal{R}_{q\qb}\(\bar{\bf \Phi}_1,\Rad^\sminus\)\,
\lq \frac{\mathcal{C}_{q\qb}^\sminus}{\mathcal{C}_{q\qb}^\splus +
  \mathcal{C}_{q\qb}^\sminus}\rq_{\Rad^\sminus},
\end{eqnarray}
where the counterterms are evaluated at the values specified by the radiation
variables.

The expression for $\bar{B}_{q\qb}\!\(\bar{\bf \Phi}_1\)$ of
eq.~(\ref{eq:bbdef}) is then given by
\begin{eqnarray}
\label{eq:VX_CS_Bbar}
 \bar{B}_{q\qb}\!\(\bar{\bf \Phi}_1\) &=& B_{q\qb}\!\(\bar{\bf \Phi}_1\) + 
V_{q\qb}\!\(\bar{\bf \Phi}_1\)  \nonumber\\
&& + \int d\Rad^\splus \lq  
R_{q\qb}^\splus(\bar{\bf \Phi}_1,\Rad^\splus)
 - C_{q\qb}^\splus(\bar{\bf\Phi}_1,\Rad^\splus) \rq \nonumber\\
&& + \int d\Rad^\sminus \lq  
R_{q\qb}^\sminus(\bar{\bf \Phi}_1,\Rad^\sminus)- 
C_{q\qb}^\sminus(\bar{\bf \Phi}_1,\Rad^\sminus) \rq \nonumber\\
&& +\int d\Rad^\splus \lq  R_{g\qb}(\bar{\bf \Phi}_1,\Rad^\splus) -
C_{g\qb}(\bar{\bf \Phi}_1,\Rad^\splus) \rq \nonumber\\
&& +\int d\Rad^\sminus \lq  R_{qg}(\bar{\bf \Phi}_1,\Rad^\sminus) -
C_{qg}(\bar{\bf \Phi}_1,\Rad^\sminus) \rq \nonumber\\
&&+\int_{\bar{x}_{\splus}}^1 \frac{dz}{z}
\lq G_{\splus}^{q\qb}(\bar{\bf \Phi}_{1,\splus})+G_{\splus}^{g\qb}(\bar{\bf
  \Phi}_{1,\splus})\rq   
+  \int_{\bar{x}_{\sminus}}^1 \frac{dz}{z} \lq G_{\sminus}^{q\qb}(\bar{\bf
  \Phi}_{1,\sminus})+ 
G_{\sminus}^{qg}(\bar{\bf \Phi}_{1,\sminus})  \rq,\nonumber\\
\end{eqnarray}
where
\begin{equation}
B_{q\qb}\!\(\bar{\bf \Phi}_1\) = \mathcal{B}_{q\qb}\!\(\bar{\bf \Phi}_1\) 
\, \Lum_{q\qb}\(\bar{x}_{\splus},\bar{x}_{\sminus}\),\quad\quad
V_{q\qb}\!\(\bar{\bf \Phi}_1\) = \mathcal{V}_{q\qb}\!\(\bar{\bf \Phi}_1\)
\, \Lum_{q\qb}\(\bar{x}_{\splus},\bar{x}_{\sminus}\),
\end{equation}
and the quantities $R$ and $C$ are obtained multiplying the corresponding
quantities ${\cal R}$ and ${\cal C}$  by the corresponding factor
$\Lum\(x_{\splus},x_{\sminus}\)$. Furthermore
\begin{eqnarray}
G_{\splus}^{q\qb}(\bar{\bf \Phi}_{1,\splus}) &=& 
\mathcal{G}_{\splus}^{q\qb}(\bar{\bf \Phi}_{1,\splus}) 
\, \Lum_{q\qb}\( \frac{\bar{x}_{\splus}}{z},\bar{x}_{\sminus}\),\\
G_{\splus}^{g\qb}(\bar{\bf \Phi}_{1,\splus}) &=&
\mathcal{G}_{\splus}^{g\qb}(\bar{\bf \Phi}_{1,\splus})  
\, \Lum_{g\qb}\( \frac{\bar{x}_{\splus}}{z},\bar{x}_{\sminus}\),\\
G_{\sminus}^{q\qb}(\bar{\bf \Phi}_{1,\sminus}) &=&
\mathcal{G}_{\sminus}^{q\qb}(\bar{\bf \Phi}_{1,\sminus})  
\, \Lum_{q\qb}\(\bar{x}_{\splus},\frac{\bar{x}_{\sminus}}{z}\),\\
G_{\sminus}^{qg}(\bar{\bf \Phi}_{1,\sminus}) &=&
\mathcal{G}_{\sminus}^{qg}(\bar{\bf   \Phi}_{1,\sminus})  
\, \Lum_{qg}\(\bar{x}_{\splus},\frac{\bar{x}_{\sminus}}{z}\),
\end{eqnarray}
with
\begin{equation}
\label{eq:luminosity}
\Lum_{ff'}\(\xplus,\xminus\) =  f_f^{\splus}(\xplus, \muF^2)\;
  f_{f'}^{\sminus} (\xminus, \muF^2)\,.
\end{equation}
where $f^f_{\splusminus}(x, \muF^2)$ is the parton density function of the
parton $f$ in the hadron $\splusminus$, and $\muF^2$ is the factorization
scale.

\subsubsection*{Generation of the Born variable $\boldsymbol{\bar{Y}}$}
\label{sec:gen_Born_V_CS}
From
eq.~(\ref{eq:dPhi_rad_plus}), we can write
\begin{equation}
d \Rad^\splusminus = \frac{M^2}{16\pi^2}
\int_0^{2\pi}\frac{d\phi}{2\pi}
\int_{\bar{x}_\splusminus}^1 d\xpm
\int_0^{1-\xpm} d\vpm\; \frac{1}{\(\xpm\)^2}\,,
\end{equation}
and with the change of variables
\begin{equation}
\label{eq:VH_CS_changevar}
\phi = 2\,\pi\,\varphi\,,\qquad\qquad
\vpm = \(1-\xpm\) \tilde{v}\,,\qquad\qquad
\xpm = \bar{x}_\splusminus + \(1-\bar{x}_\splusminus\) x\,,
\end{equation}
we have
\begin{equation}
d \Rad^\splusminus = \frac{M^2}{16\pi^2} \(1-\bar{x}_\splusminus\)^2
\int_0^{1} d\varphi
\int_{0}^1 dx \, (1-x) 
\int_0^{1} d\tilde{v}\;\, \frac{1}{\(\bar{x}_\splusminus +
  \(1-\bar{x}_\splusminus\) x\)^2}\,,
\end{equation}
so that now both $d \Rad^\splus$ and $d \Rad^\sminus$ are expressed in terms
of the same integration variables and integration ranges.

The set of variables $\lg \varphi,\, x,\,\tilde{v}\rg$ are equivalent to the
three $\Xrad$ variables, introduced in section~\ref{sec:gen_born_var}.  We
can now insert the process-dependent part of the Jacobian into the integrand,
and define $d^3\Xrad$ as
\begin{equation}
\int d^3\Xrad = \int_0^{1}d\varphi \int_{0}^1 dx \int_0^{1} d\tilde{v}\,,
\end{equation}
so that
\begin{equation}
\label{eq:VX_CS_Phirad}
d \Rad^\splusminus = d^3\Xrad \, J^\splusminus\,,\qquad\qquad
J^\splusminus = \frac{M^2}{16\pi^2} \(1-\bar{x}_\splusminus\)^2 (1-x) 
\frac{1}{\(\bar{x}_\splusminus + \(1-\bar{x}_\splusminus\) x\)^2}\,.
\end{equation}
In addition, we make a change of variable in the integrals of the two
collinear remnants of eq.~(\ref{eq:VX_CS_Bbar}),
\begin{equation}
z = \bar{x}_\splusminus + \(1- \bar{x}_\splusminus \)x,\qquad{\rm and}\qquad
\frac{dz}{dx} = 1-\bar{x}_\splusminus\,,
\end{equation}
 so that the limits of $x$ are consistently between 0 and 1.
The $\tilde{B}$ function of eq.~(\ref{eq:btdef}) is then given by
\begin{eqnarray}
 \tilde{B}_{q\qb}\!\(\bar{\bf \Phi}_1,\Xrad\) &=& 
B_{q\qb}\!\(\bar{\bf \Phi}_1\) + V_{q\qb}\!\(\bar{\bf \Phi}_1\)  \nonumber\\
&+& J^\splus \lq  
R_{q\qb}^\splus(\bar{\bf \Phi}_1,\Rad^\splus)
- C_{q\qb}^\splus(\bar{\bf
  \Phi}_1,\Rad^\splus) \rq 
+  J^\sminus \lq  
R_{q\qb}^\sminus(\bar{\bf \Phi}_1,\Rad^\sminus)- 
C_{q\qb}^\sminus(\bar{\bf \Phi}_1,\Rad^\sminus) \rq \nonumber\\
&+& J^\splus \lq  R_{g\qb}(\bar{\bf \Phi}_1,\Rad^\splus) -
C_{g\qb}(\bar{\bf \Phi}_1,\Rad^\splus) \rq 
+J^\sminus \lq  R_{qg}(\bar{\bf \Phi}_1,\Rad^\sminus) -
C_{qg}(\bar{\bf \Phi}_1,\Rad^\sminus) \rq \nonumber\\
&+& \frac{1-\bar{x}_\splus}{z}
\lq G_{\splus}^{q\qb}(\bar{\bf \Phi}_{1,\splus})+G_{\splus}^{g\qb}(\bar{\bf
  \Phi}_{1,\splus})\rq   
+ \frac{1-\bar{x}_\sminus}{z}\lq G_{\sminus}^{q\qb}(\bar{\bf
  \Phi}_{1,\sminus})+ 
G_{\sminus}^{qg}(\bar{\bf \Phi}_{1,\sminus})  \rq,\nonumber\\
\end{eqnarray}
so that
\begin{equation}
\int d^3\Xrad\, \tilde{B}_{q\qb}\!\(\bar{\bf \Phi}_1,\Xrad\) =
  \bar{B}_{q\qb}\!\(\bar{\bf \Phi}_1\)\,.
\end{equation}
The generation of the Born variable $\bar{Y}$ is performed using the method
illustrated in section~\ref{sec:gen_born_var}.  We use an
integrator-unweighter program (like for example the \BASES-\SPRING{}
package), that, after a single 4-dimensional integration of the function
\begin{equation}
\tilde{B}\!\(\bar{\bf \Phi}_1,\Xrad\) = \sum_q\tilde{B}_{q\qb}\!\(\bar{\bf
  \Phi}_1,\Xrad\)\,, 
\end{equation}
can generate 4-tuples of $\lg\bar{Y},\,\varphi,\, x, \, \tilde{v}\rg$ values,
distributed according to $\tilde{B}\!\(\bar{\bf \Phi}_1,\Xrad\)$. For each
generated 4-tuple, we generate $q$ with a probability proportional to
$\tilde{B}_{q\qb}\!\(\bar{\bf\Phi}_1,\Xrad\)$.\footnote{The standard
procedure to generate an index $0 < j \leqslant n$ with a probability
proportional to $a_j$ is to generate a random number $0 < r < 1$, and then
choose j so that $\sum_{j' = 1}^{j - 1} a_{j'} < r \sum_{j' = 1}^n a_{j'} <
\sum_{j' = 1}^j a_{j'}$.} We only keep the $\bar{Y}$ and $q$ generated value,
and neglect $\varphi, \, x,\, \tilde{v}$, which corresponds to integrate over
them.

Observe that $\tilde{B}\!\(\bar{\bf \Phi}_1,\Xrad\)$ is positive, unless the
$\mathcal{O}(\as)$ terms overcome the Born term. If this happens, the whole
perturbative approach breaks down. This can happen if $M$ is too small, or if
we are in extreme regions of phase space, like the threshold region (i.e.\
when the values of $\bar{Y}$ and $M$ are forcing $\bar{x}_{\splus}$ or
$\bar{x}_{\sminus}$ to 1). In practical applications, it is wise to check
that the fraction of negative weights in the integral of
$\tilde{B}\!\(\bar{\bf \Phi}_1,\Xrad\)$ is small, and can be safely neglected.

\subsubsection*{Generation of the radiation variables}
The transverse momentum $\kt$ of the radiation, with respect to the incoming
beams, in the two singular regions is given by
\begin{equation}
{\kt^2}_{\splusminus} = s\(1-\xpm\)\vpm = M^2\,\frac{1-\xpm}{\xpm} \,\vpm\,,
\end{equation}
where $\vplus$ is used when the emitter is the $\splus$~parton, and $\vminus$
when it is the $\sminus$~one, and the Sudakov form factor of
eq.~(\ref{eq:suddef}) is
\begin{eqnarray}
\label{eq:DY_sudakov}
\Delta_{q}\(\bar{\bf \Phi}_1,\pt\) &=& 
\exp\lg -\!\!\int d\Rad^\splus 
\frac{R_{q\qb}^\splus(\bar{\bf\Phi}_1,\Rad^\splus) + 
R_{g\qb}(\bar{\bf \Phi}_1,\Rad^\splus)}{B_{q\qb}\!\(\bar{\bf \Phi}_1\)}
\,\stepf\!\({\kt^2}_{\splus} - \pt^2\)\right. \nonumber\\
 &&\phantom{\exp\Big\{} \left.\!
-\!\!\int d\Rad^\sminus 
\frac{R_{q\qb}^\sminus(\bar{\bf\Phi}_1,\Rad^\sminus) + 
R_{qg}(\bar{\bf \Phi}_1,\Rad^\sminus)}{B_{q\qb}\!\(\bar{\bf \Phi}_1\)}
\,\stepf\!\({\kt^2}_{\sminus} - \pt^2\)\rg \,.\phantom{aaa}
\end{eqnarray}
In order to generate the radiation variables, we use the hit-and-miss
technique introduced in section~\ref{sec:gen_rad_var}. Suitable upper
bounding functions are given by
\begin{eqnarray}
\frac{R_{q\qb}^\splus(\bar{\bf\Phi}_1,\Rad^\splus)}{B_{q\qb}\!\(\bar{\bf
    \Phi}_1\)} 
&\lesssim& 8\,\CF\,g_s^2\, \frac{\xpm}{2M^2}\,\frac{1}{\vplus} \,
\frac{2}{1-\xpm}\,
\frac{f^q_\splus(\bar{x}_\splus/\xpm,\muF^2)}{
f^q_\splus(\bar{x}_\splus,\muF^2)}\,,\\  
\frac{R_{q\qb}^\sminus(\bar{\bf\Phi}_1,\Rad^\sminus)}{B_{q\qb}\!\(\bar{\bf
    \Phi}_1\)} 
&\lesssim& 8\,\CF\,g_s^2\, \frac{\xpm}{2M^2}\,\frac{1}{\vminus}
\, \frac{2}{1-\xpm} \,\frac{f_\sminus^{\qb}(\bar{x}_\sminus/\xpm,\muF^2)}{
f_\sminus^{\qb}(\bar{x}_\sminus,\muF^2)}\,,\\      
\frac{R_{g\qb}(\bar{\bf\Phi}_1,\Rad^\splus)}{B_{q\qb}\!\(\bar{\bf \Phi}_1\)}
&\lesssim& 8\,\TF\,g_s^2\, \frac{\xpm}{2M^2} \, \frac{1}{\vplus} \,
\frac{f_\splus^g(\bar{x}_\splus/\xpm,\muF^2)}{
f_\splus^q(\bar{x}_\splus,\muF^2)}\,,\\
\frac{R_{qg}(\bar{\bf\Phi}_1,\Rad^\sminus)}{B_{q\qb}\!\(\bar{\bf \Phi}_1\)}
&\lesssim& 8\,\TF\,g_s^2\,  \frac{\xpm}{2M^2}\, \frac{1}{\vminus} \,
\frac{f_\sminus^g(\bar{x}_\sminus/\xpm,\muF^2)}{
  f_\sminus^{\qb}(\bar{x}_\sminus,\muF^2)}\,. 
\end{eqnarray}
Following appendix~\ref{sec:pdf_upper_bounds}, we can put an upper bound on
the ratios of the parton distribution functions (pdf) too, and we obtain
\begin{eqnarray}
\frac{R_{q\qb}^\splus(\bar{\bf\Phi}_1,\Rad^\splus)}{B_{q\qb}\!\(\bar{\bf
    \Phi}_1\)} 
&\leq& N^\splus_{q\qb}\,\frac{\as}{\vplus} \,
\frac{\xpm}{1-\xpm}\,,\\  
\frac{R_{q\qb}^\sminus(\bar{\bf\Phi}_1,\Rad^\sminus)}{B_{q\qb}\!\(\bar{\bf
    \Phi}_1\)} 
&\leq&  N^\sminus_{q\qb}\,\frac{\as}{\vminus}
\, \frac{\xpm}{1-\xpm} \,,\\      
\frac{R_{g\qb}(\bar{\bf\Phi}_1,\Rad^\splus)}{B_{q\qb}\!\(\bar{\bf \Phi}_1\)}
&\leq& N_{g\qb}\,\frac{\as}{\vplus} \,
\frac{\xpm}{1-\xpm}\,,\\  
\frac{R_{qg}(\bar{\bf\Phi}_1,\Rad^\sminus)}{B_{q\qb}\!\(\bar{\bf \Phi}_1\)}
&\leq& N_{qg}\, \frac{\as}{\vminus}
\, \frac{\xpm}{1-\xpm}\,,
\end{eqnarray}
where $N^\splusminus_{q\qb}$, $N_{g\qb}$ and $N_{qg}$ are upper-bound
constants determined by sampling the whole phase-space of the $R/B$ ratios.

We can then write an upper bound for the integrand function in the Sudakov
form factor of eq.~(\ref{eq:DY_sudakov})
\begin{eqnarray}
\label{eq:DY_sud_upper}
&& \int d\Rad^\splus 
\frac{R_{q\qb}^\splus + R_{g\qb}}{B_{q\qb}}
\,\stepf\!\({\kt^2}_{\splus} - \pt^2\)
+ \int d\Rad^\sminus \frac{R_{q\qb}^\sminus + R_{qg}}{B_{q\qb}}
\,\stepf\!\({\kt^2}_{\sminus} - \pt^2\)
\nonumber\\
&&\hspace{1cm}\leq
N_\splus\int_{\bar{x}_\splus}^1 d\xpm
\int_0^{1-\xpm} d\vplus\;
\frac{1}{\xpm\(1-\xpm\)}\frac{1}{\vplus} \,\as\({\kt^2}_{\splus}\)\,
\stepf\!\({\kt^2}_{\splus} - \pt^2\)
\nonumber\\
&&\hspace{1cm}+\, 
N_\sminus\int_{\bar{x}_\sminus}^1 d\xpm
\int_0^{1-\xpm} d\vminus\;
\frac{1}{\xpm\(1-\xpm\)}\frac{1}{\vminus} \,\as\({\kt^2}_{\sminus}\)\,
\stepf\!\({\kt^2}_{\sminus} -\pt^2\) ,\phantom{aaa}
\end{eqnarray}
where $N_\splusminus$ are constant with respect to the radiation variables.
The two integrals have the same functional form, so that we just need to
generate radiation variables using the highest-$\pt$ bid method of
appendix~\ref{sec:highest_pt} and the veto technique of
appendix~\ref{sec:veto_technique}, where the $\pt$ is generated uniformly in
$\Delta_u(\pt)$
\begin{equation}
\Delta_u(\pt) = \exp\lg
-N\int_{\bar{x}}^1 dx \int_0^{1-x} dv\;
\frac{1}{x\(1-x\)}\frac{1}{v} \,\as\({\kt^2}\)\,
\stepf\!\({\kt^2} -\pt^2\) \rg,
\end{equation}
where $\kt$ must be of the order of the radiation transverse momentum in the
collinear limit, and it must coincide with it in the soft-collinear limit. A
suitable definition is given by
\begin{equation}
{\kt^2} = M^2\,(1-x)\,v\,.
\end{equation}
Introducing the dimensionless quantities
\begin{equation}
\tildekt^2 = \frac{\kt^2}{M^2}=(1-x)\,v,   \qquad \qquad
 \tildept^2 =  \frac{\pt^2}{M^2} \,,
\end{equation}
and the integration over $d\tildekt^2$, we have
\begin{eqnarray}
\Delta_u(\pt) &=& \exp\lg
-N \int_{\tildept^2}^{\tildektmaxsq} d\tildekt^2
\int_{\bar{x}}^1 dx \int_0^{1-x} dv\;
\frac{1}{x\(1-x\)}\frac{1}{v} \,\delta\!\big( v(1-x) -\tildept^2\big)
\as\({\kt^2}\) \rg
\nonumber\\
&=& \exp\lg
-N \int_{\tildept^2}^{\tildektmaxsq} d\tildekt^2
\int_{\bar{x}}^{1-\tildekt} dx \,
\frac{1}{x\(1-x\)}\,\frac{\as\({\kt^2}\)}{\tildekt^2}\,,
\rg
 \nonumber\\
&=& \exp\lg
-N \int_{\tildept^2}^{\tildektmaxsq} d\tildekt^2
\frac{\as\({\kt^2}\)}{\tildekt^2} 
\lq \log\frac{1-\tildekt}{\tildekt} +
\log\frac{1-\bar{x}}{\bar{x}}
\rq \rg,
\end{eqnarray}
where
\begin{equation}
\tildektmaxsq = \(1-\bar{x}\)^2\,,
\end{equation}
and $\as\({\kt^2}\)$ is given by eq.~(\ref{eq:as_oneloop}).  Since the
generation of $\pt$ according to
\begin{equation}
\label{eq:dDeltau_DY}
d\Delta_u(\pt) =  d\exp\lg
-N \int_{0}^{\tildektmaxsq} d\tildekt^2
\frac{\as\({\kt^2}\)}{\tildekt^2} 
\lq \log\frac{1-\tildekt}{\tildekt} +
\log\frac{1-\bar{x}}{\bar{x}}
\rq \stepf\!\( \kt^2 - \pt^2\)
\rg
\end{equation}
is still too complex to be performed analytically, we use a second time
the veto method, using the upper bounding function
\begin{equation}
 \log\frac{1}{\tildekt} 
\geq
\log\frac{1-\tildekt}{\tildekt} 
\,,
\end{equation}
and we then generate the $\pt$ distribution according to
$d\Delta_{uu}\(\pt\)$, where
\begin{eqnarray}
\Delta_{uu}\(\pt\) &=& \exp\lg
-N \int_{0}^{\tildektmaxsq} d\tildekt^2\,
\frac{\as\({\kt^2}\)}{\tildekt^2} 
\lq \log\frac{1}{\tildekt} +
\log\frac{1-\bar{x}}{\bar{x}}
\rq \stepf\!\( \kt^2 - \pt^2\)
\rg
\nonumber\\
&=& \exp\Bigg\{
-\frac{N}{b_0} 
\Bigg[ -\frac{1}{2}\log\frac{M^2\(1-\bar{x}\)^2}{\pt^2}
\nonumber\\
&&\hspace{1cm} +\log\frac{\log\big(\(1-\bar{x}\)^2 
M^2/\Lambda^2\big)}{\log \(\pt^2/\Lambda^2\)}  
\( \log\frac{1-\bar{x}}{\bar{x}} +\frac{1}{2}\log\frac{M^2}{\Lambda^2}
\) \Bigg]\Bigg\}.
\end{eqnarray}
In order to apply the highest-$\pt$ bid technique to
eq.~(\ref{eq:DY_sudakov}), we need to repeat the following steps for both the
last two integrals in eq.~(\ref{eq:DY_sud_upper}).  We follow in detail the
first one. The second integral is treated in the same way.

Observe that we have a lower limit on the acceptable values of $\pt$, since
$\as(\pt)$ must be well defined for both the two-loop and the one-loop
expression of $\as$. We thus introduce a $\pt^{\rm min}\gtrsim \Lambda$.

When dealing with the first integral, we set $N=N_\splus$ and
$\bar{x}=\bar{x}_\splus$. 

\begin{enumerate}
  \item Set $\pt^{\max} =M\(1-\bar{x}\)$, where $M^2\(1-\bar{x}\)^2$ is the
  maximum value of $\pt^2$, such that $\Delta_{uu} (\pt^{\max})=1$.
  
  \item \label{item:delta_uu} Generate $r$ with $0 < r < 1$, and solve the
  equation $r = \Delta_{uu} (\pt)$/$\Delta_{u u} (\pt^{\max})$ for $\pt$.

  If no solution is found, or if $\pt<\pt^{\rm min}$, no radiation is
  produced, and a Born event is generated.
    
  \item If a solution with $\pt>\pt^{\rm min}$ is found,
  generate $r'$ with $0 \leqslant r' \leqslant \log(M/\pt)$.  If $r'
  \leqslant \log(M/\pt-1)$ go to step~\ref{item:rad_variables}.  Otherwise
  set $\pt^{\max} = \pt$ and go to step~\ref{item:delta_uu}.

  \item \label{item:rad_variables} At this point $\pt$ is generated according
  to eq.~(\ref{eq:dDeltau_DY}).  We need to generate $x$, $v$ and $\phi$ with
  a probability proportional to
  \begin{equation}
    \frac{1}{x\(1-x\)}\,\frac{1}{v} \,\delta\big( M^2 v(1-x) -\pt^2\big)
  dv\,dx\,d\phi. 
  \end{equation}
  This is done as follows: integrate the above in $v$ using the $\delta$
  function, so that
  \begin{equation}
    v = \frac{\pt^2}{M^2}\,\frac{1}{1-x}\,,
  \end{equation}
  and generate $x$ according to
  \begin{equation}
    \stepf\!\(x-\bar{x}\) \, \stepf\!\(1-\frac{\pt}{M}-x \)\,
  d\log\frac{x}{1-x}\,,
  \end{equation}
  i.e., generate a uniformly-distributed random number $r''$ with
  \begin{equation}
  \log\frac{\bar{x}}{1-\bar{x}} \leqslant r'' \leqslant 
  \log\( \frac{M}{\pt}-1\)\,
  \end{equation}
  and solve $r''=\log\big( x/(1-x)\big)$ for $x$, that is
  \begin{equation}
    x = \frac{1}{\exp(-r'')+1}\,.
  \end{equation}
  One also generates uniformly a random value of $\phi$ between $0$ and $2
  \pi$. 
  
  \item \label{item:rad_var_qq} Set $\xpm=x$ and $\vplus=v$. Now we need to
   apply the last veto. Generate a random number $r_\splus$ in the range
  \begin{equation}
   0 \leq r_\splus \leq \(N_{q\qb}^\splus + N_{g\qb}\) \frac{\as}{\vplus} \,
   \frac{\xpm}{1-\xpm}\,.
  \end{equation}
  If 
  \begin{equation}
    r_\splus \leq \frac{R_{q\qb}^\splus\!\(\bar{\bf\Phi}_1,\Rad^\splus\) +
  R_{g\qb}\!\(\bar{\bf\Phi}_1,\Rad^\splus\)}{B_{q\qb}\!\(\bar{\bf \Phi}_1\)} 
  \end{equation}
  then accept the event. Otherwise, set $\pt^{\max} = \pt$ and go to
  step~\ref{item:delta_uu}.

  \item
  Call the $\pt$ generated this way $\pt^\splus$. Set $\pt^\splus=0$ in case
  the event is returned as a Born one.  Repeat the previous steps for the
  second integral, that is set $N=N_\sminus$ and $\bar{x}=\bar{x}_\sminus$,
  with the obvious changes in step~\ref{item:rad_var_qq}, that now becomes

  \begin{itemize}
  \item[\ref{item:rad_var_qq}$'$.] Set $\xpm=x$ and $\vminus=v$ and generate
   a random number $r_\sminus$ in the range
   \begin{equation}
   0 \leq r_\sminus \leq \(N_{q\qb}^\sminus + N_{qg}\) \frac{\as}{\vminus} \,
   \frac{\xpm}{1-\xpm}\,.
  \end{equation}
  If 
  \begin{equation}
    r_\sminus \leq \frac{R_{q\qb}^\sminus\!\(\bar{\bf\Phi}_1,\Rad^\sminus\) +
  R_{qg}\!\(\bar{\bf\Phi}_1,\Rad^\sminus\)}{B_{q\qb}\!\(\bar{\bf \Phi}_1\)} 
  \end{equation}
  then accept the event. Otherwise, set $\pt^{\max} = \pt$ and go to
   step~\ref{item:delta_uu}.  Call the $\pt$ generated this way
   $\pt^\sminus$. Set $\pt^\sminus=0$ in case the event is returned as a Born
   one.
  \end{itemize}

  \item According to the highest-$\pt$ bid method of
  appendix~\ref{sec:highest_pt}, choose the highest value between
  $\pt^\splus$ and $\pt^\sminus$. If both $\pt^\splus$ and $\pt^\sminus$
  are zero, then return the event as a Born-like one.

  \begin{itemize}
  \item  
  In case the $\pt$ chosen is $\pt^\splus$,  if
  \begin{equation}
    r_\splus >  \frac{R_{q\qb}^\splus\!\(\bar{\bf\Phi}_1,\Rad^\splus\)}
    {B_{q\qb}\!\(\bar{\bf\Phi}_1\)}\,,
  \end{equation}
  then generate a $g\qb$ event. Otherwise, generate a $q\qb$ event with
  radiation emitted by the $\splus$ incoming quark.

  \item  
  In  case the $\pt$ chosen is $\pt^\sminus$,  if
  \begin{equation}
    r_\sminus >  \frac{R_{q\qb}^\sminus\!\(\bar{\bf\Phi}_1,\Rad^\sminus\)}
    {B_{q\qb}\!\(\bar{\bf\Phi}_1\)} \,,
  \end{equation}
  then generate a $qg$ event. Otherwise, generate a $q\qb$ event with
  radiation emitted by the $\sminus$ incoming quark.
  \end{itemize}

\end{enumerate}

\subsection{$\boldsymbol{h_{\splus} h_{\sminus} \rightarrow V}$ in the \FKS{}
  formalism} 
\label{sec:hh_to_gamma_FKS}
We now consider the case of the production of a massive vector $V$ of mass
$M$ in hadron-hadron collisions, in the FKS formalism.

\subsubsection*{Radiation variables and inverse construction}
We use the kinematics
notation introduced in section~\ref{sec:hh_to_gamma_CS}, that is, the Born 
kinematics is characterized by the rapidity $\bar{Y}$ of the produced vector
boson, and the phase space and momentum fractions are given by
eqs.~(\ref{dbarPhi1}) and~(\ref{eq:CS_VX})
\begin{equation}
d{\bf \bar{\Phi}}_1 = \frac{2
  \pi}{S} d \bar{Y}\,, \qquad \bar{x}_{\splus} = \sqrt{\frac{M^2}{S}}
e^{\bar{Y}}, \qquad 
\bar{x}_{\sminus} = \sqrt{\frac{M^2}{S}} e^{- \bar{Y}}\,.
\end{equation} 
According to eqs.~(\ref{eq:Vg})--(\ref{eq:Vqbar}), the real-emission
processes are characterized by two final-state momenta, $k_1$ (the $V$
momentum) and $k_2$, the momentum of the radiated parton. The (only) radiated
parton is the FKS parton.  Following the prescriptions of
section~\ref{sec:powhegfksiss}, we define
\begin{equation}
  k^0_2 = \frac{\sqrt{s}}{2}\, \xi\,,
\qquad\qquad
 k_2 = k^0_2 \(1,\sin \theta \sin \phi,
  \sin \theta \cos \phi, \cos \theta\),
\end{equation}
and we take $\xi$, $y = \cos \theta$ and $\phi$ as the radiation variables.
The real-emission phase space is given by (see eqs.~(\ref{eq:dPhinpo_IS_FKS})
and~(\ref{eq:dRad_IS_FKS}))
\begin{equation}
d{\bf \Phi}_2 = d{\bf \bar{\Phi}}_1 \, d\Rad\,, \qquad\qquad
d\Rad=\frac{s}{(4\pi)^3}\,\frac{\xi}{1-\xi}\,d\xi\,dy\, d\phi\,,
\end{equation}
and the procedure described in section~\ref{sec:inv_construction_FKS} allows
one to fully reconstruct the two-body phase space, give the Born barred
variable $\bar{Y}$ and the three radiation variables.  In particular, the
requirement that both $x_{\splus}$ and $x_{\sminus}$ of
eq.~(\ref{eq:xp_xm_IS_FKS}) be less than 1 forces $\xi$ to be in the range
$0\leq\xi \leq \xi_{\rm max}$, where $\xi_{\rm max}$ is given by
eq.~(\ref{eq:xi_max}).

\subsubsection*{Virtual corrections}
The virtual corrections to the process can be found in
ref.~\cite{Altarelli:1979ub}, eq.~(82) 
\begin{equation}
  \mathcal{V}_{b,\,q\qb} = \left( \frac{4 \pi \mu^2}{M^2}
  \right)^{\epsilon} \frac{\Gamma (1 + \epsilon) \, \Gamma (1 -
  \epsilon)^2}{\Gamma (1 - 2 \epsilon)}  \frac{\as}{2 \pi}\, \CF \left[ -
  \frac{2}{\epsilon^2} - \frac{3}{\epsilon} - 8 + \pi^2
  \right]\mathcal{B}_{\,q\qb},
\end{equation}
where we drop the argument ${\bf\bar{\Phi}_1}$ of the virtual and Born term,
for ease of notation.  Following section~\ref{sec:hh_to_gamma_CS}, we
indicate with $q$ the Born flavour index $f_b$, that carries information on
the flavour of the (massless) incoming parton, along the $\splus$
direction. Using a standard convention for the numbering scheme, we have
$q = - 5, \ldots, - 1, 1, \ldots, 5$, and $\qb=-q$. We notice that
\begin{equation}
  \frac{\Gamma (1 + \epsilon) \,\Gamma (1 - \epsilon)^2}{\Gamma (1 - 2
  \epsilon)} = \frac{1}{\Gamma (1 - \epsilon)} +\mathcal{O}\(\epsilon^3\),
\end{equation}
so that, choosing $Q^2 = M^2$ in eq.~(\ref{eq:virtnorm}), we get
\begin{equation}
  \mathcal{V}_{b,\,q\qb} =\mathcal{N} \,\frac{ \as}{2 \pi}\,\CF
  \left[ - \frac{2}{\epsilon^2} - \frac{3}{\epsilon} - 8 + \pi^2 \right]
\mathcal{B}_{q\qb}.
\end{equation}
From eq.~(\ref{eq:FKSoneloop}) we thus get
\begin{equation}
  \mathcal{V}_{\tmop{fin},\,q\qb} = \CF \(\pi^2 - 8\)\mathcal{B}_{q\qb}\, .
\end{equation}
Notice that the $\mathcal{B}_{i j}$ terms in (\ref{eq:FKSoneloop}) have only
the $\mathcal{B}_{\splus \sminus}$ and $\mathcal{B}_{\sminus \splus}$
components in this case, and their contribution vanishes because we have
chosen $Q^2 = M^2 = 2 \, (k_{\splus}\cdot k_{\sminus})$.  According to
eqs.~(\ref{eq:colourcorr}), (\ref{eq:Qdef}) and~(\ref{eq:Iijreg}) we have
\begin{equation}
  \mathcal{B}_{\splus \sminus} =\mathcal{B}_{\sminus \splus} = \CF
  \mathcal{B}_{q\qb}\,,
\end{equation}
\begin{equation}
  \sum_{\overset{i, j \in \mathcal{I}}{i \ne j}}
  \mathcal{I}_{ij}\,\mathcal{B}_{ij} = \(\mathcal{B}_{\splus
  \sminus} +\mathcal{B}_{\sminus \splus}\) \mathcal{I}_{\splus \sminus} = -
 (\mathcal{B}_{\splus \sminus} +\mathcal{B}_{\sminus \splus}) \,\tmop{Li}_2
  (1)=  -\frac{\pi^2}{6} \(\mathcal{B}_{\splus \sminus} +\mathcal{B}_{\sminus
  \splus}\) ,
\end{equation}
\begin{equation}
\mathcal{Q} =  - 2 \,\CF\,\log \frac{\muF^2}{M^2} \,,
\end{equation}
where we have chosen  $\xi_c = 1$.
Finally, eq.~(\ref{eq:FKSsv}) gives
\begin{equation}
  \mathcal{V}_{q\qb}= \frac{\as}{2 \pi} \,\CF \left( - 2 \, \log
  \frac{\muF^2}{M^2} + \frac{2}{3}\, \pi^2 - 8 \right)\mathcal{B}_{q\qb}\,.
\end{equation}

\subsubsection*{Collinear remnants}
From eq.~(\ref{eq:FKScrp}) we immediately get the collinear remnants
\begin{eqnarray}
  \mathcal{G}_{\splus}^{q\qb} &=& \frac{ \as}{2 \pi} \,\CF\left\{ 
\(1 + z^2 \) \left[ \left( \frac{1}{1 - z} \right)_+ \log \frac{M^2}{z
  \muF^2} + 2 \left( \frac{\log (1 - z)}{1 - z} \right)_+ \right] + 1 - z
  \right\} \mathcal{B}_{q\qb}\!\(z k_\splus,k_\sminus\)\,,\nonumber
\\\\
  \mathcal{G}_{\splus}^{g \bar{q}} &=& \frac{ \as}{2 \pi}\,\TF \left\{ \left[
  z^2 + (1 - z)^2 \right] \left[ \log \frac{M^2}{z \muF^2} + 2 \log (1 - z)
  \right] + 2 z (1 - z) \right\} \mathcal{B}_{q\qb}\!\(z
  k_\splus,k_\sminus\), \phantom{aa}
\end{eqnarray}
and the corresponding ones for the $\sminus$ collinear direction, and where
we have set $\deltaI=2$.

\subsubsection*{Real corrections}
The real production contributions are given by
eqs.~(\ref{eq:Rqqbar})--(\ref{eq:Rgqbar}) (see also~\cite{Altarelli:1979ub})
\begin{eqnarray}
  \mathcal{R}_{q\qb} &= &
 2\,\frac{\CF}{N_c} \,g^2\, g_s^2 \,\frac{1}{2\,s}
\left[\frac{u}{t}+\frac{t}{u}  + \frac{2\,s \,M^2}{tu}\right],\\
  \mathcal{R}_{g\qb} &=& 
-2\,\frac{\TF}{N_c}\, g^2\, g_s^2  \,\frac{1}{2\,s}
\left[\frac{u}{s}+\frac{s}{u}  + \frac{2\,t \,M^2}{su}\right] ,
\end{eqnarray}
(and the analogous one for $\mathcal{R}_{qg}$) where, according to the
definitions given in eqs.~(\ref{eq:def_s_V})--(\ref{eq:def_u_V}),
\begin{equation}
  s = \frac{M^2}{1-\xi}\,, \qquad
  t = - \frac{s}{2} \,\xi\, (1 + y)\,, \qquad
  u = - \frac{s}{2}\, \xi \,(1 - y) \,.
\end{equation}
In terms of the FKS variables we have
\begin{eqnarray}
  \mathcal{R}_{q\qb} &=& 2\,\frac{\CF}{N_c} \,g^2\, g_s^2 \,\frac{1}{2\,s}
\left\{2\,\frac{1 + y^2}{1 - y^2} + \frac{8 (1 - \xi)}{\xi^2 \(1 - y^2\)}
\right\}, 
\\
  \mathcal{R}_{g\qb} &=& 2\,\frac{\TF}{N_c}\, g^2\, g_s^2\,\frac{1}{2\,s}
\left\{\frac{2 [1 - \xi (1 - \xi) (1 + y)]}{\xi (1 - y)} + \frac{\xi (1 -
  y)}{2}   \right\},
\end{eqnarray}
so that $(1 - y^2)\, \xi^2\, \mathcal{R}$ are regular functions. Using
eq.~(\ref{eq:FKSrin}) we obtain immediately 
\begin{eqnarray}
  \hat{\mathcal{R}}_{q\qb} &=& 2\,\frac{\CF}{N_c} \,g^2\, g_s^2
  \,\frac{1}{2\,s}
\lq 2 \(1 + y^2\) \xi^2 + 8 (1 - \xi) \rq \left\{ \frac{1}{2} \left(
  \frac{1}{\xi} \right)_{\!\!+} \! \left[ \left( \frac{1}{1 - y} \right)_+ +
  \!\!\left( 
  \frac{1}{1 + y} \right)_+ \right] \right\} \frac{1}{\xi}\,,\phantom{aa}
\nonumber\\
\\
  \hat{\mathcal{R}}_{g\qb} &=& 2\,\frac{\TF}{N_c}\, g^2\, g_s^2\,\frac{1}{2\,s}
\left\{
  2 [1 - \xi (1 - \xi) (1 + y)] + \frac{\xi^2 (1 - y)^2}{2} \right\} 
\left( \frac{1}{1 - y} \right)_+  \frac{1}{\xi}\,,
\end{eqnarray}
with the $+$ distributions defined in eq.~(\ref{eq:fksplusdef}).

\subsubsection*{Generation of the Born variable $\boldsymbol{\bar{Y}}$}
The expression for $\bar{B}_{q\qb}\!\(\bar{\bf \Phi}_1\)$ of
eq.~(\ref{eq:bbdef}) is then given by
\begin{eqnarray}
\bar{B}_{q\qb}\!\(\bar{\bf \Phi}_1\) \!&=& B_{q\qb}\!\(\bar{\bf \Phi}_1\) + 
V_{q\qb}\!\(\bar{\bf \Phi}_1\)  \nonumber\\
  &  & + \int d \Phi_{\tmop{rad}} \left[ \hat{R}_{q\qb}\! \(
  {\bf\bar{\Phi}_1}, \Phi_{\tmop{rad}}\) + \hat{R}_{qg}\! \(
  {\bf\bar{\Phi}_1}, \Phi_{\tmop{rad}}\) + \hat{R}_{g\qb}\! \(
  {\bf\bar{\Phi}_1}, \Phi_{\tmop{rad}}\) \right] \nonumber\\
&&+\int_{\bar{x}_{\splus}}^1 \frac{dz}{z}
\lq G_{\splus}^{q\qb}(\bar{\bf \Phi}_{1,\splus})+G_{\splus}^{g\qb}(\bar{\bf
  \Phi}_{1,\splus})\rq   
+  \int_{\bar{x}_{\sminus}}^1 \frac{dz}{z} \lq G_{\sminus}^{q\qb}(\bar{\bf
  \Phi}_{1,\sminus})+ 
G_{\sminus}^{qg}(\bar{\bf \Phi}_{1,\sminus})  \rq,\phantom{aaaaaa} 
\end{eqnarray}
where
\begin{eqnarray}
B_{q\qb}\!\(\bar{\bf \Phi}_1\) &=& \mathcal{B}_{q\qb}\!\(\bar{\bf \Phi}_1\) 
\, \Lum_{q\qb}\(\bar{x}_{\splus},\bar{x}_{\sminus}\),\\
V_{q\qb}\!\(\bar{\bf \Phi}_1\) &=& \mathcal{V}_{q\qb}\!\(\bar{\bf \Phi}_1\)
\, \Lum_{q\qb}\(\bar{x}_{\splus},\bar{x}_{\sminus}\),\\
\hat{R}_{q\qb}(\bar{\bf \Phi}_1,\Rad) &=&
  \hat{\mathcal{R}}_{q\qb}(\bar{\bf \Phi}_1,\Rad)  
\, \Lum_{q\qb}\(x_{\splus},x_{\sminus}\),\\
\hat{R}_{qg}(\bar{\bf \Phi}_1,\Rad) &=& \hat{\mathcal{R}}_{qg}(\bar{\bf
  \Phi}_1,\Rad) 
\, \Lum_{qg}\(x_{\splus},x_{\sminus}\),\\
\hat{R}_{g\qb}(\bar{\bf \Phi}_1,\Rad)  &=&
\hat{\mathcal{R}}_{g\qb}(\bar{\bf\Phi}_1,\Rad)  
\, \Lum_{g\qb}\(x_{\splus},x_{\sminus}\),\\
G_{\splus}^{q\qb}(\bar{\bf \Phi}_{1,\splus}) &=& 
\mathcal{G}_{\splus}^{q\qb}(\bar{\bf \Phi}_{1,\splus}) 
\, \Lum_{q\qb}\( \frac{\bar{x}_{\splus}}{z},\bar{x}_{\sminus}\),\\
G_{\splus}^{g\qb}(\bar{\bf \Phi}_{1,\splus}) &=&
\mathcal{G}_{\splus}^{g\qb}(\bar{\bf \Phi}_{1,\splus})  
\, \Lum_{g\qb}\( \frac{\bar{x}_{\splus}}{z},\bar{x}_{\sminus}\),\\
G_{\sminus}^{q\qb}(\bar{\bf \Phi}_{1,\sminus}) &=&
\mathcal{G}_{\sminus}^{q\qb}(\bar{\bf \Phi}_{1,\sminus})  
\, \Lum_{q\qb}\(\bar{x}_{\splus},\frac{\bar{x}_{\sminus}}{z}\),\\
G_{\sminus}^{qg}(\bar{\bf \Phi}_{1,\sminus}) &=&
\mathcal{G}_{\sminus}^{qg}(\bar{\bf   \Phi}_{1,\sminus})  
\, \Lum_{qg}\(\bar{x}_{\splus},\frac{\bar{x}_{\sminus}}{z}\),
\end{eqnarray}
with $x_{\splus}, x_{\sminus}$ given in eq.~(\ref{eq:xp_xm_IS_FKS}) and the
luminosity $\mathcal{L}$ defined in eq.~(\ref{eq:luminosity}).

As explained in section~\ref{sec:FKS_IS}, the integration range in the
radiation variable $\xi$ is limited by the condition~(\ref{eq:xi_limits})
\begin{equation}
0 \leq \xi \leq \xi_{\rm max}\,.
\end{equation}
Making the change of variable
\begin{equation}
  1 - \xi = z\,,
\end{equation}
the integration limits become
\begin{equation}
 z_{\min}(y) \leq z \leq 1\,,
\end{equation}
where
\begin{eqnarray}
z_{\min}(y) &=& {\rm max}\lg
\frac{2(1+y)\,\bar{x}_\splus^2}{\sqrt{(1+\bar{x}_\splus^2)^2(1-y)^2 +
 16\,y\,\bar{x}_\splus^2}+(1-y)(1-\bar{x}_\splus^2)},\right.
\nonumber\\
&&\phantom{aaaaa}\left.
\frac{2(1-y)\,\bar{x}_\sminus ^2}{\sqrt{(1+\bar{x}_\sminus ^2)^2(1+y)^2 -
    16\,y\,\bar{x}_\sminus ^2} +(1+y)(1-\bar{x}_\sminus ^2)}\rg.
\end{eqnarray}
It is convenient to introduce a rescaled variable $\tilde{z}$
\begin{equation}
  \tilde{z} = \frac{z - z_{\min}(y)}{1 - z_{\min}(y)}\,, \qquad\qquad
z = z_{\min}(y) + \tilde{z}\, \(1 - z_{\min}(y)\)\,,
\end{equation}
and use the identity
\begin{equation}
 \label{eq:xscaled} 
  \left( \frac{1}{\xi} \right)_+ = \left( \frac{1}{1 - z}
  \right)_+ = \frac{1}{1-z_{\min}(y)}\lg\left( \frac{1}{1 - \tilde{z}}
  \right)_+ + \delta\!\(1 - \tilde{z}\)\, \log\!\lq 1 - z_{\min}(y)\rq\rg\,,
\end{equation}
in order to simplify the treatment of the $+$ distribution in the integration
of $\hat{R}$.

The integral of the distributions requires some care. We follow, as an example,
the integral of the $1 / (1 - y)_+$ contribution in the $\hat{R}_{q\qb}$
term.  The integral has the form
\begin{equation}
  I = \int_{- 1}^1 d y \int_{1 - z_{\min}(y)}^1 d \xi \left( \frac{1}{1 - y}
  \right)_+  \left( \frac{1}{\xi} \right)_+ f (\xi, y)\, .
\end{equation}
We change variable from $\xi$ to $\tilde{z}$ using eq.~(\ref{eq:xscaled}), and
obtain
\begin{eqnarray}
  I & = & \int_{- 1}^1 d y \int_0^1 d \tilde{z}  \left( \frac{1}{1 - y}
  \right)_+  \left( \frac{1}{1 - \tilde{z}} \right)_+ f\!\(\xi\!\(\tilde{z},
  y\),y\)
\nonumber \\
&& + \int_{- 1}^1 d y \left( \frac{1}{1 - y} \right)_+ \log \lq
  1-z_{\min}(y) \rq f (0, y) \nonumber\\ 
  & = & \int_{-1}^1 d y \int_0^1 d \tilde{z} \, \frac{1}{1 - y} \left[ \frac{f
  (\xi ( \tilde{z}, y), y) - f (0, y)}{1 - \tilde{z}} - \frac{f (\xi (
  \tilde{z}, 1), 1) - f (0, 1)}{1 - \tilde{z}} \right]
  \nonumber\\
  & & +\int_{-1}^1 d y \,\frac{1}{1 - y} \,\Big\{ \log\lq 1 - z_{\min} (y)\rq f
  (0, y)  - \log\lq 1 - z_{\min} (1) \rq f (0, 1) \Big\} \,, 
\end{eqnarray}
that is manifestly finite and can be computed numerically.

The collinear contributions also have a restricted range in the $z$ variable.
Considering for simplicity only the $\mathcal{G}_{\splus}^{q\qb}$ term, we
have $\bar{x}_{\splus}\leq z \leq 1$, so that, as usual, we need to make use
of the identities
\begin{eqnarray}
  \int_{\bar{x}_{\splus}}^1 d z \left( \frac{1}{1 - z} \right)_+ f (z) & = & 
  \log (1 - \bar{x}_{\splus}) f (1) + \int_{\bar{x}_{\splus}}^1 d z\, \frac{f
  (z) - f (1)}{1 - z}\,,\\
  \int_{\bar{x}_{\splus}}^1 d z \left( \frac{\log (1 - z)}{1 - z} \right)_+ f
  (z) & = & \frac{1}{2} \log^2 (1 - \bar{x}_{\splus}) f (1) +
  \int_{\bar{x}_{\splus}}^1 d z \,\frac{\log (1 - z) [f (z) - f (1)]}{1 - z},
\phantom{aaaaaa}
\end{eqnarray}
and also in this case we define a rescaled variable $0 \leq \tilde{z} \leq 1$
\begin{equation}
  \tilde{z} = \frac{z - \bar{x}_{\splus}}{1 - \bar{x}_{\splus}}\,, \qquad\qquad
z =  \bar{x}_{\splus} + \tilde{z} (1 - \bar{x}_{\splus})\,,
\end{equation}
and rewrite the $z$ integrations as $\tilde{z}$ integrations. At the end of
this procedure, the most general form one can obtain for $\bar{B}$ is
\begin{eqnarray}
  \bar{B}_{q\qb}\!\(\bar{\bf \Phi}_1\) & = & B^a_{q\qb}\!\( \bar{\bf
  \Phi}_1\) +   \int_0^1  
d\tilde{z}\, B^b_{q\qb}\!\(\bar{\bf \Phi}_1,\tilde{z}\) + \int_{- 1}^1 d y
\int_0^{2 \pi} d \phi \, B^c_{q\qb}\!\(\bar{\bf \Phi}_1,y, \phi\) \nonumber\\
  & & +\int_{-1}^1 d y \int_0^1 d \tilde{z} \int_0^{2 \pi} d \phi \,
B^d_{q\qb}\!\(\bar{\bf \Phi}_1,\tilde{z}, y, \phi\)\,.
\end{eqnarray}
In order to perform numerically the generation of the Born term, we introduce
the function (see eq.~(\ref{eq:btdef}))
\begin{eqnarray}
  \tilde{B}_{q\qb}\!\(\bar{\bf \Phi}_1,\tilde{z}, y, \phi\) & = & \frac{1}{4
  \pi} 
  B^a_{q\qb}\!\(\bar{\bf \Phi}_1\) + \frac{1}{4 \pi} \int_0^1 d \tilde{z} \, 
B^b_{q\qb}\!\(\bar{\bf \Phi}_1, \tilde{z}\)
  + \int_{- 1}^1 d y \int_0^{2 \pi} d \phi \, B^c_{q\qb}\!\(\bar{\bf
  \Phi}_1,y, \phi\) 
  \nonumber\\
  &  & +\int_{- 1}^1 d y \int_0^1 d \tilde{z} \int_0^{2 \pi} d \phi \,
  B^d_{q\qb}\!\(\bar{\bf \Phi}_1, \tilde{z}, y, \phi\),  
\end{eqnarray}
so that
\begin{equation}
  \bar{B}_{q\qb}\!\( \bar{\bf \Phi}_1\) = \int_{- 1}^1 d y \int_0^1 d \tilde{z}
  \int_0^{2 \pi} 
  d \phi \, \tilde{B}_{q\qb}\!\(\bar{\bf \Phi}_1, \tilde{z}, y, \phi\)\, .
\end{equation}
The generation of the Born variable $\bar{\bf \Phi}_1=\{\bar{Y}\}$ is
performed by using an integrator-unweighter program (like for example the
\BASES-\SPRING{} program), that after a single four-dimensional integration
of the function
\begin{equation}  
\tilde{B}\!\(\bar{\bf \Phi}_1, \tilde{z}, y, \phi\) = \sum_q 
\tilde{B}_{q\qb}\!\(\bar{\bf \Phi}_1, \tilde{z}, y, \phi\) 
\end{equation}
in the variables $\bar{Y},\tilde{z}, y, \phi$, can generate 4-tuples of
$\bar{Y},\tilde{z}, y, \phi$ values, distributed according to the weight
$\tilde{B}\!\(\bar{Y}, \tilde{z}, y, \phi\)$. For each generated 4-tuple, we
also generate $q$ with a probability proportional to
$\tilde{B}_{q\qb}\!\(\bar{Y},\tilde{z}, y, \phi\)$.  We only keep the
$\bar{Y}$ and $q$ generated value, and neglect $y, \tilde{z}, \phi$, which
corresponds to integrating over them.

As already noticed at the end of section~\ref{sec:gen_Born_V_CS}, observe
that $\tilde{B}\!\(\bar{Y}, \tilde{z}, y, \phi\)$ is positive, unless the
$\mathcal{O}(\as)$ terms overcome the Born term. If this happens, the whole
perturbative approach breaks down. This can happen if $M$ is too small, or if
we are in extreme regions of phase space, like the threshold region (i.e.\
when the values of $\bar{\bf \Phi}_1$ and $M$ are forcing $\bar{x}_{\splus}$
or $\bar{x}_{\sminus}$ to 1). In practical applications it is wise to check
that the fraction of negative weight in the integral of
$\tilde{B}\!\(\bar{Y}, \tilde{z}, y, \phi\)$ is small, and can be safely
neglected.

\subsubsection*{Generation of the radiation variables}
We assume that we have generated $\bar{\bf \Phi}_1$ and $q$ according to the
procedure given above. The Sudakov form factor for the generation of
radiation, according to eq.~(\ref{eq:suddef}) is given by
\begin{equation}
  \label{eq:Vsuda} 
  \Delta_q\!\(\bar{\bf \Phi}_1, \pt\) \!= \exp\!
  \left\{\! -\!\! \int \! d\Rad \, \frac{
  R_{q\qb}\!\(\bar{\bf \Phi}_1,\Rad\) + 
  R_{qg}\!\(\bar{\bf \Phi}_1,\Rad\) +  
  R_{g \qb}\!\(\bar{\bf \Phi}_1,\Rad\)}{B_{q\qb}\!\(\bar{\bf \Phi}_1\)}
  \,\stepf\!\(\kt - \pt\)\! \right\}\!, 
\end{equation}
where
\begin{equation}
  \kt^2 = \frac{s}{4}\, \xi^2 \(1 - y^2\) =
 \frac{M^2}{4 (1 - \xi)} \,\xi^2 \(1 -  y^2\)
\end{equation}
is the exact squared transverse momentum of the radiated parton. The
factorization and renormalization scales in eq.~(\ref{eq:Vsuda}) should be
taken equal to $\kt^2$.  In order to generate the radiation variables, we use
the hit-and-miss technique introduced in section~\ref{sec:gen_rad_var}.  We
need to find a simple upper bound for the expression
\begin{equation}
  \frac{s}{(4 \pi)^3}\, \frac{\xi}{1 - \xi} \,
\frac{
  R_{q\qb}\!\(\bar{\bf \Phi}_1,\Rad\) + 
  R_{qg}\!\(\bar{\bf \Phi}_1,\Rad\) +  
  R_{g \qb}\!\(\bar{\bf \Phi}_1,\Rad\)}{B_{q\qb}\!\(\bar{\bf \Phi}_1\)}\,,
\end{equation}
in order to perform the generation of radiation using the veto method.  With
the same considerations of section~\ref{sec:gen_rad_var} and using the upper
bounds for the parton distribution functions found in
appendix~\ref{sec:pdf_upper_bounds}, it is easy to show that the upper
bounding function found in ref.~{\cite{Nason:2006hf}}
\begin{equation}
  \label{eq:ubound} U_q = N_q \,\frac{\as\!\(\kt^2\)}{\xi\(1 - y^2\)}
\end{equation}
is suited also to the present case.

When implementing the bound~(\ref{eq:ubound}) in practice, the whole phase
space is searched randomly in order to determine the bound normalization
$N_q$. At this stage, problems can arise, due to the fact that available
parton-density parametrization do not necessarily comply with the
bound~(\ref{eq:gqbound}). One simple example is the case of the $b$ quark
density, that is set arbitrarily to zero above a given value of $x$,
generally smaller than the value of $x$ at which the gluon density is set to
zero. These regions give very small contribution to the cross section, and
must be carefully cut away, unless one is using a pdf parametrization that is
fully consistent with evolution also in the very large-$x$ region.

The bound~(\ref{eq:ubound}) has been obtained assuming that we are not near
the small-$x$ region. If this is the case, the bound may become inefficient.
We observe that, in case the parton distribution functions obey Feynman
scaling (i.e.\ $f(x) \approx 1 / x$ at small $x$), the bound is still
adequate in the small-$x$ region. Modern pdf parametrizations, however,
prefer the faster rising behavior \ $f (x) \approx 1 / x^{\delta}$, with
$\delta > 1$. In this case, the bound may become inefficient at small $x$,
and numerical tricks (like, for example, dividing the integration range in
small intervals and adopting different $N_q$ values in each interval) should
be used to overcome this problem.

The generation of the event according to the bound (\ref{eq:ubound}) is
documented in great detail in appendix~D of ref.~{\cite{Nason:2006hf}}, and
we do not repeat it here.


\section{Conclusions}
In this work we have presented in full details the \POWHEG{} method
for interfacing NLO calculations to parton shower. The purpose of this
paper, besides providing all the technical details needed for the
implementation of the method, was also to demonstrate its full
generality.  We have given detailed formulae in the case of cross
sections that do not involve massive coloured partons.  The extension
to this case is straightforward both in the FKS and in the CS
framework. We have not tried to include it in our discussion, in order
not to increase the complexity of the presentation.

We believe that the method presented here can be applied to a large
number of interesting LHC processes, and that authors of QCD calculation
should be able to build their own implementation with only a modest
effort. In order to ease this task, we have collected \POWHEG{}
software in a repository at the location:\\
\centerline{{\tt http://moby.mib.infn.it/\!$\boldsymbol{\sim}$\,nason/POWHEG/FNOpaper/\ }.} 
Among other things, 
the software for the implementation of the inverse mapping
in FKS, the program {\tt mint} (see ref.~\cite{mint})
for the integration and generation
of unweighted events, and the full implementation of the $e^+e^-$
example in the FKS approach can be found there.

\vspace{1cm}
{\bf \noindent Acknowledgments}

We wish to thank S.~Catani and T.~Sj\"ostrand for useful discussions.
S.~F.~wishes to thank the CERN Theory Division for its kind hospitality
during this work.


\appendix

\section{The veto technique}
\label{sec:veto_technique}
In this section we describe a method to generate a set of $d$ variables $x$,
distributed according to
\begin{equation}
\label{eq:prob_distrib}
f(x)\; \Delta(h(x))
\end{equation} 
where
\begin{equation}
\Delta(h)=\exp\left\{-\int d^d x^\prime\,
f\!\(x^\prime\)\,\stepf\!\(h\!\(x^\prime\)-h\)\right\}. 
\end{equation}
We assume, as usual, that $f$ and $h$ are non-negative functions, and that
the unrestricted integral of $f$ is divergent, that is
\begin{equation}
\Delta(0) = \exp\left\{-\int d^dx^\prime\,
f\!\(x^\prime\)\right\} = 0\,.
\end{equation}
With these assumptions, upon multiplying the infinitesimal probability
$f(x)\, \Delta(h(x))\,d^dx$ by $\delta(h-h(x))\, dh$, we can integrate over
$d^dx$ and interpret it as infinitesimal probability for the variable $h$
\begin{equation}
dh \int d^dx\, \delta(h-h(x))\, f(x) \exp\left\{-\int d^d
x^\prime\, 
f\!\(x^\prime\)\,\stepf\!\(h\!\(x^\prime\)-h\)\right\} 
=  dh\, \frac{d\Delta(h)}{dh} = d\Delta(h)\,,
\end{equation}
that shows that the probability is uniform in $\Delta(h)$.  In principle, the
generation of events is therefore straightforward: one generates a uniform
random number $r$ between 0 and 1, and solves the equation $\Delta(H)=r$ for
$H$ (here we have used the fact that $\Delta(0)=0$).  At this point, the
variables $x$ have distribution function equal to
\begin{equation}
\delta(H-h(x))\, f(x) \exp\left\{-\int d^d x^\prime\,
f\!\(x^\prime\)\,\stepf\!\(h\!\(x^\prime\)-H\)\right\},
\end{equation}
where the exponent is now just a number (a normalization factor), so that the
variables $x$ are on the surface $\delta(H-h(x))$ with a distribution
function proportional to $f(x)\,\delta(H-h(x))$.  The generation of these
variables can be done, for example, with a hit-and-miss technique, or, if the
integration can be performed analytically, by generating $(d-1)$ random
numbers $r_i$, uniformly distributed between 0 and 1, and solving
\begin{equation}
\int^{X_i}_{x_{i0}} dx_i\, \prod_{k\neq i} \int dx_k\, \delta(H-h(x))\, f(x)
= r_i 
\end{equation}
for $X_i$, where $x_{i0}$ is the lower limit for the variable $x_i$, and all
the other variables are integrated over their full range of validity.

In practise, however, the solution of the equation $\Delta(H)=r$ is, in most
cases, very heavy, from a numerical point of view.  This difficulty can be
overcome by means of the so-called veto method. We assume that there is a
function $F(x)\geq f(x)$ for all $x$ values, and that
\begin{equation}
\Delta_F(h)=\exp\left[-\int d^dx^\prime\, F\!\(x^\prime\)\,
\stepf\!\(h\(x^\prime\)-h\)\right]\;
\end{equation}
has a simple form, so that the solution of the equation $\Delta_F(H)=r$ and
the generation of the distribution $F(x)\,\delta(H-h(x))$ are reasonably
simple. Then, we implement the following procedure:
\begin{enumerate}
\item Set $H_{\max}$ equal to the maximum allowed value, such that
  $\Delta_F(H_{\max}) =1$.
\item Generate a flat random number $0<r<1$ and solve the equation
\begin{equation}
\frac{\Delta_F(H)}{\Delta_F(H_{\max})}=r
\end{equation}
for $H$ (a solution with $0<H<H_{\max}$ always exists for $0<r<1$).
\item Generate $x$ according to $F(x)\,\delta(h(x)-H)$.
\item Generate a new random number $r^\prime$.
\item
If $r^\prime>f(x)/F(x)$ then the event is vetoed, we set $H_{\max}=H$, go to
step~2 and continue.  Otherwise the event is accepted, and the procedure
stops.
\end{enumerate}
The resulting events are distributed according to
eq.~(\ref{eq:prob_distrib}).  The proof of this statement is simple but
non-trivial, and is given in appendix~C of ref.~\cite{Nason:2006hf}.

\section{The highest-$\boldsymbol{\pt}$ bid procedure}
\label{sec:highest_pt}
Our aim is to generate $\(k,\,x_k\)$ pairs with a probability
\begin{equation}\label{eq:initialdistr}
 f_k(x_k) \prod_i \Delta_i\!\(h_k(x_k)\) \,d^d x_k\,,
\end{equation}
where
\begin{equation}
\Delta_i(h)=\exp\left\{-\int d^d x_i^\prime\,
f_i\!\(x_i^\prime\)\,\stepf\!\(h_i\!\(x_i^\prime\)-h\)\right\}. 
\end{equation}
We assume, as usual, that the $f_i$ and $h_i$ are non-negative functions, and
that the unrestricted integral of the $f_i$ is divergent, that is
\begin{equation}
\label{eq:delta_zero}
\Delta_i(0) = \exp\left\{-\int d^d x_i^\prime\,
f_i\!\(x_i^\prime\)\right\} = 0\,.
\end{equation}
Under these conditions, we have the identity
\begin{equation}
\label{eq:exact_diff}
\int d^d x_k \,\Delta_k\!\(h_k(x_k)\) \, f_k(x_k) \,
\delta\!\(h_k(x_k)-h\)=\frac{d}{dh} \Delta_k(h)\,,
\end{equation}
so that
\begin{eqnarray}
\label{eq:noemission}
&&\int d^d x_k\, \Delta_k(h_k(x_k)) \,f_k(x_k)\, \stepf(h-h_k(x_k)) 
\nonumber\\
&& \hspace{2cm}
=\int_0^\infty \!\! dh^\prime \!\int d^d x_k\,\Delta_k(h_k(x_k)) \,f_k(x_k)
\, \delta(h^\prime-h_k(x_k)) \; \stepf\!\(h-h^\prime\)\nonumber \\
&& \hspace{2cm}
=\int_0^\infty\!\!  dh^\prime \,\frac{d}{dh^\prime} \Delta_k\!\(h^\prime\)
\stepf\!\(h-h^\prime\)= 
\int_0^h \!\! dh^\prime\,
\frac{d}{dh^\prime}\Delta_k\!\(h^\prime\)=\Delta_k(h)\,, 
\end{eqnarray}
where we have used the fact that $\Delta_k(0)=0$.  If we interpret $h$ and
$h_k$ as transverse momenta, then $\Delta_k(h)$ in eq.~(\ref{eq:noemission})
corresponds to the probability of not emitting a parton with transverse
momentum bigger than $h$.

The procedure to generate the distribution in eq.~(\ref{eq:initialdistr}),
using the highest-bid method, is the following. For each $k$, we generate an
$x_k$ value with probability
\begin{equation}
\label{eq:gener_probab}
\Delta_k(h_k(x_k))\, f_k(x_k) \, d^d x_k\,,
\end{equation}
as described in appendix~\ref{sec:veto_technique}, and then we pick the $k$
value with the largest $h_k(x_k)$. In fact, the probability that the
generated $(k,\,x_k)$ has the largest $h_k(x_k)$ is precisely given by the
product of its generation probability (eq.~(\ref{eq:gener_probab})) times the
probability that all the other $h_i(x_i)$ are less then $h_k(x_k)$, which is
given by the product
\begin{equation}
\prod_{i\neq k} \Delta_i(h_k(x_k))\,,
\end{equation}
and together they reconstruct eq.~(\ref{eq:initialdistr}).

\section{Soft angular integrals}\label{app:Softints}
The soft angular integrals eqs.~(\ref{eq:softint1}) and~(\ref{eq:softint2})
can all be obtained from the basic integral
\begin{equation}
I\!\(k_i,v,k^0\)=\int d\Omega \frac{k_i\cdot v}{(k_i\cdot k)(k\cdot v)},
\end{equation}
where $v$ is an arbitrary timelike vector and $d\Omega$ is the element of the
solid angle of $k$ with respect to a reference direction.  First we notice
that, since the integrand scales like $(k^0)^{-2}$,
\begin{equation}
\int d\Omega \int_{f(\Omega)}^{e f(\Omega)} k^0 dk^0
\frac{k_i\cdot v}{(k_i\cdot k)(k\cdot v)}= I(k_i,v,1),
\end{equation}
(where $e$ is the Euler constant) independent of $f$, for any arbitrary
function $f(\Omega)$.
We can also write
\begin{equation}
I(k_i,v,1)=\int \frac{d^3 k}{k^0} \frac{k_i\cdot v}{(k_i\cdot k)(k\cdot v)}
\stepf\!\(k^0-f(\Omega)\)\,\stepf\!\(ef(\Omega)-k^0\),
\end{equation}
and perform a change of variables, that is also a Lorentz transformation, $k
\to \Lambda k$, to obtain
\begin{equation}
I(k_i,v,1)=\int \frac{d^3 k}{k^0} \frac{k_i^\prime\cdot
  v^\prime}{(k_i^\prime\cdot k)(k\cdot v^\prime)} 
\stepf\!\(k^0-f^\prime(\Omega)\)\,\stepf\!\(ef^\prime(\Omega)-k^0\),
\end{equation}
where $v^\prime=\Lambda^{-1} v$ and $k_i^\prime=\Lambda^{-1}k_i$. The
function $f$ also undergoes a change, and becomes $f^\prime$. But we have
just shown that the integral does not depend upon $f$, so that we conclude
\begin{equation}
I(k_i,v,1)=I(k_i^\prime,v^\prime,1).
\end{equation}
We thus perform the integral in a frame where $v$ has only the time
component. We get
\begin{equation}\label{eq:basicsoft}
I(k_i,v,k^0)=\frac{2\pi}{ k_0^2} \int d\cos\theta
\,\frac{1}{1-\beta\cos\theta}= \frac{2\pi}{\beta \, k_0^2}
\log\frac{1+\beta}{1-\beta}\,,\qquad\quad \beta=\sqrt{1-\frac{k_i^2\,
    v^2}{(k\cdot v)^2}}\,. 
\end{equation}
Setting $v=k_{ij}$ in eq.~(\ref{eq:basicsoft}) yields immediately
eq.~(\ref{eq:softint2}).  For small $k_i^2$, eq.~(\ref{eq:basicsoft}) becomes
\begin{equation}
I(k_i,v,k^0)=\frac{2\pi}{k_0^2}
\log\frac{4 (k_i\cdot v)^2}{k_i^2 \, v^2}\,.
\end{equation}
Taking the difference of the above formula with $v=k_{ij}$ and $v=n$ we
get~(\ref{eq:softint1}).

\section{Parton-distribution-function upper bounds}
\label{sec:pdf_upper_bounds}
In this section, we discuss the bounds on the parton-distribution-function
(pdf) ratios of the form
\begin{equation}
\frac{f_g(x/z,\muF^2)}{f_g(x,\muF^2)}\,,\qquad
\frac{f_q(x/z,\muF^2)}{f_q(x,\muF^2)}\,,\qquad
\frac{f_g(x/z,\muF^2)}{f_q(x,\muF^2)}\,,\qquad
\frac{f_q(x/z,\muF^2)}{f_g(x,\muF^2)}\,,
\end{equation}
where $q$ stands for a quark or an antiquark, and $x\leq z \leq 1$. We begin
by noticing that the first two ratios are always bounded by a number of order
1, because the pdf's are never fast growing functions of their argument. In
fact, only the pdf of the valence quarks grows mildly for moderate $x$
values. In the remaining two cases, the ratio of parton densities is between
different parton species, and must be discussed with care.  We consider first
the $g/q$ case.  First of all, since the gluon pdf is a monotonic decreasing
function
\begin{equation}
f_g(x/z,\muF^2) < f_g(x,\muF^2)\,,
\end{equation}
so that we only need a bound for
\begin{equation}
\frac{f_g(x,\muF^2)}{f_q(x,\muF^2)}\,.
\end{equation}
At small values of $x$, this ratio is always bounded, because the gluon and
quark densities have similar behavior in the small-$x$ limit. For large $x$,
if $q$ is a valence quark, the ratio is also bounded, since the gluon is
generated by valence quarks through evolution.  In case $q$ is a sea quark,
the corresponding density may be softer than the gluon density. In the worst
case, however, the sea quark is generated by the gluon through evolution.
Assuming that parton densities behave as a power of $1-x$ at large $x$,
\begin{equation}
f_g(x,\muF^2) \sim (1-x)^\delta\,,
\end{equation}
the Altarelli-Parisi equation in the large-$x$ limit yields
\begin{equation}
\mu^2 \frac{d f_q(x,\muF^2)}{d\muF^2} \sim \frac{T_F \as}{2\pi}
\int_x^1\frac{dz}{z}\, (1-z)^\delta
\sim \frac{T_F \as}{2\pi} \, (1-x)^{\delta+1}\;,
\end{equation}
and therefore
\begin{equation}
\frac{f_g(x,\muF^2)}{f_q(x,\muF^2)}
\lesssim \frac{C}{1-x}\,.
\end{equation}
Since
\begin{equation}
\frac{1}{1-x}<\frac{1}{1-z}\;,
\end{equation}
we conclude that
\begin{equation}
\label{eq:gqbound}
\frac{f_g(x/z,\muF^2)}{f_q(x,\muF^2)}\leq \frac{C}{1-z}\,.
\end{equation}
For the $q/g$ case, since
\begin{equation}\label{eq:valenceless}
f_q(x/z,\muF^2) \lesssim f_q(x,\muF^2)\,,
\end{equation}
we need a bound for
\begin{equation}\label{eq:qoverg}
\frac{f_q(x,\muF^2)}{f_g(x,\muF^2)}\,.
\end{equation}
The symbol $\lesssim$ in eq.~(\ref{eq:valenceless}) is to be interpreted as
``almost everywhere smaller than''. In fact, if $q$ is a valence quark, its
pdf can grow mildly in an intermediate $x$ range. If $q$ is not a valence
quark, the ratio~(\ref{eq:qoverg}) is certainly small. We should only worry
about the case when $q$ is a valence quark. In this case, we assume for the
quark a large $x$ behavior of the form
\begin{equation}
f_q(x,\muF^2) \sim (1-x)^\delta\,,
\end{equation}
and, with a reasoning similar to the one used before, we can conclude that
\begin{equation}
\frac{f_q(x/z,\muF^2)}{f_g(x,\muF^2)}\leq \frac{C}{1-z}\,.
\end{equation}

\end{document}